\documentclass[11pt]{article}

\usepackage{stolenstyle}

\usepackage{amsmath,epsf,amssymb,latexsym,amsthm,setspace,bbm,array,pifont,multirow}

\usepackage[matrix,arrow,frame,import,curve,color]{xy}

\DeclareFontFamily{U}{rsf}{}
\DeclareFontShape{U}{rsf}{m}{n}{
  <5> <6> rsfs5 <7> <8> <9> rsfs7 <10-> rsfs10}{}
\DeclareMathAlphabet\Scr{U}{rsf}{m}{n}

\def\CO#1#2{{[#1,#2]}}
\def\AC#1#2{{\{#1,#2\}}}

\def\rep#1{{{\boldsymbol{#1}}}}

\def\cDb{{\overline{\cD}}}
\def\cQb{{\overline{\cQ}}}

\def\C{{\mathbb C}}

\def\P{{\mathbb P}}
\def\R{{\mathbb R}}
\def\Z{{\mathbb Z}}

\def\End{\operatorname{End}}

\def\Pic{\operatorname{Pic}}

\def\ch{\operatorname{ch}}

\def\rank{\operatorname{rank}}

\def\SO{\operatorname{SO}}

\def\GL{\operatorname{GL}}

\def\SU{\operatorname{SU}}
\def\GU{\operatorname{U{}}}

\def\GE{\operatorname{E}}

\def\so{\operatorname{\mathfrak{so}}}

\def\su{\operatorname{\mathfrak{su}}}

\def\p{\partial}
\def\pb{\bar{\partial}}

\def\la{\langle}
\def\ra{\rangle}

\def\ff#1#2{{\textstyle\frac{#1}{#2}}}
\def\half{\frac{1}{2}}

\def\cA{{\cal A}}

\def\cC{{\cal C}}
\def\cD{{\cal D}}
\def\cE{{\cal E}}
\def\cF{{\cal F}}

\def\cH{{\cal H}}

\def\cJ{{\cal J}}
\def\cK{{\cal K}}
\def\cL{{\cal L}}

\def\cO{{\cal O}}

\def\cQ{{\cal Q}}

\def\cT{{\cal T}}

\def\cV{{\cal V}}
\def\cW{{\cal W}}
\def\cX{{\cal X}}

\def\ep{{\epsilon}}

\newcommand\alphah{\widehat{\alpha}}
\newcommand\betah{\widehat{\beta}}
\newcommand\gammah{\widehat{\gamma}}

\newcommand\alphab{\overline{\alpha}}
\newcommand\betab{\overline{\beta}}
\newcommand\gammab{\overline{\gamma}}

\newcommand\epb{\overline{\ep}}

\newcommand\etab{\overline{\eta}}
\newcommand\thetab{\overline{\theta}}

\newcommand\mub{\overline{\mu}}
\newcommand\nub{\overline{\nu}}

\newcommand\phib{\overline{\phi}}
\newcommand\chib{\overline{\chi}}

\newcommand\nut{\widetilde{\nu}}

\newcommand\Gammah{\widehat{\Gamma}}

\newcommand\Gammat{\widetilde{\Gamma}}

\newcommand\cb{\overline{c}}

\newcommand\qb{\overline{q}}

\newcommand\ub{\overline{u}}

\newcommand\xb{\overline{x}}
\newcommand\yb{\overline{y}}
\newcommand\zb{\overline{z}}

\newcommand\Ab{\overline{A}}
\newcommand\Bb{\overline{B}}

\newcommand\Hb{\overline{H}}

\newcommand\Jb{\overline{J}}

\newcommand\Lb{\overline{L}}
\newcommand\Mb{\overline{M}}

\newcommand\Qb{\overline{Q}}

\newcommand\Yb{\overline{Y}}

\def\bY{{{\boldsymbol{Y}}}}
\def\bR{{{\boldsymbol{R}}}}
\def\br{{{\boldsymbol{r}}}}

\def\bx{{{\boldsymbol{x}}}}
\def\bq{{{{\boldsymbol{q}}}}}
\def\bqb{{{\overline{\bq}}}}

\def\bQ{{{\boldsymbol{Q}}}}
\def\bQb{{{\boldsymbol{\Qb}}}}

\def\mfU{{{\mathfrak{U}}}}

\def\Gammah{{{\widehat{\Gamma}}}}
\def\Gammat{{{\tilde{\Gamma}}}}

\def\htt{{{\tilde{h}}}}

\def\pbz{{{\pb_{\!\zb}}}}

\def\e{{{\mathfrak{e}}}}
\def\mfU{{{\mathfrak{U}}}}

\def\cXb{{\overline{\cX}}}

\def\coker{{{\text{coker}}}}

\def\tot{{{\text{tot}}}}
\def\lcm{{{\text{lcm}}}}
\def\Sym{{{\text{Sym}}}}
\def\obs{{{\text{obs}}}}
\def\obst{{{\widetilde{\text{obs}}}}}

\def\preprint{IPMU17-0170}

\title{(0,2) hybrid models}
\author[a] {Marco Bertolini}
\author[b] {and M.~Ronen Plesser}
\affiliation[a]{Kavli Institute for the Physics and Mathematics of the Universe (WPI),\\
The University of Tokyo, Kashiwa, Chiba 277-8583, Japan}
\affiliation[b]{Center for Geometry and Theoretical Physics, Box 90318 \\
Duke University, Durham, NC 27708-0318, USA}
\emailAdd{marco.bertolini@ipmu.jp}
\emailAdd{plesser@cgtp.duke.edu}

\abstract{We introduce a class of (0,2) superconformal field theories based on hybrid geometries,
generalizing various known constructions.
We develop techniques for the computation of the complete massless
spectrum when the theory can be interpreted as 
determining a
perturbative heterotic string compactification.
We provide evidence for surprising properties regarding RG flows and IR accidental symmetries in (0,2) hybrid CFTs.
We also study the conditions for embedding a hybrid theory in a particular class of gauged linear sigma models.
This perspective suggests that our construction generates models which cannot be realized or analyzed by previously known methods.}

\begin{document}

\maketitle

\section{Introduction}
\label{s:intro}

Despite the renewed interest that conformal field theories (CFTs) with (0,2) supersymmetry 
have reclaimed in the last decade
\cite{McOrist:2010ae,Sharpe:2015vza}, little is still known about the
moduli space of deformations of such theories (the conformal manifold)
in general.  In the context of heterotic string theory, this is the
target space in which the massless scalars of the spacetime theory
take values.
Moreover, much of the work concerning (0,2) theories has focused on deformations of theories with (2,2) supersymmetry. 
While there are certainly notable challenges even 
in this tamer setting \cite{McOrist:2008ji,Melnikov:2010sa,Aspinwall:2010ve,Melnikov:2012hk,Aspinwall:2014ava,Aspinwall:2011vp,Closset:2015ohf}, 
in many instances features of (2,2) theories
are more or less straightforwardly inherited by this larger space of deformations.
Thus, it is of most interest to pursue the study of (0,2) models which do not admit a (2,2) locus. 

In the geometric setting of nonlinear sigma models (NLSMs), a (0,2)
model is determined by a (conformally invariant) Calabi--Yau target space $M$ along
with a polystable holomorphic bundle $\cE$ over it, subject to some
conditions (for some recent work on these see
\cite{Bouchard:2006dn,Donagi:2006yf}).  Examples with a (2,2) locus
are obtained when $\cE$ is a deformation of $TM$.  NLSMs are UV free
theories and for suitable choices of the geometric data are expected
to flow to nontrivial SCFTs in the IR.  A first
approximation to the moduli space of deformations of the SCFT 
is given by deformations of the
complex structure of $M$, deformations of its K\"ahler structure, and
deformations of $\cE$ (there are some caveats to this associated to
Atiyah classes).  The NLSM in principle contains in the UV theory
models that flow to the entire conformal manifold.
In general, however, worldsheet instantons are
expected to lift some of these vacua \cite{Greene:1990du}, so that the
moduli space is a quantum corrected version of the classical space.
In general we lack the tools to compute these corrections.  Since the
large-radius limit in which these corrections vanish is not a point in
the moduli space, NLSM calculations are strictly valid only where such
instanton corrections are absent.

Landau-Ginzburg orbifold (LGO) theories are another class of UV free
models expected to flow to nontrivial SCFTs
\cite{Distler:1993mk,Melnikov:2009nh}.  For these theories we have
computational techniques
\cite{Kawai:1994np,Melnikov:2007xi,Gadde:2016khg}  
that produce exact results.  These are valid
for a locus in the moduli space (the LG locus); deformations along
this (realized as deformations of the LG potential) are described
exactly.  Deformations away from this locus (generically of high
codimension) are not well described, restricting the kind of
computations possible.  Even in this rather manageable class of models, the (0,2)
story is not as simple as the (2,2) situation.  In general, accidental
Abelian global symmetries of the IR theory mix with the R-symmetry,
invalidating calculations based on the UV symmetry \cite{Benini:2013cda,Melnikov:2016dnx}.
Some of these ``accidents'' in LG models can be explicitly exhibited
in terms of the UV theory \cite{Bertolini:2014ela}, although in general accidental IR
symmetries can be difficult to see in the UV.

Gauged linear sigma models (GLSMs) \cite{Witten:1993yc} make explicit the possibility that
the LGO locus is a subspace of the NLSM moduli space, although
disjoint from the large radius limit.  In the GLSM set-up, the large
radius limit and the LGO arise as ``phase limits'', limiting values of
the UV parameters in which a simplified description of the IR physics
is applicable.   In the GLSM framework, worldsheet instanton effects
take the form of gauge instanton effects, and in some cases
\cite{Beasley:2003fx,Bertolini:2014dna} one can 
show that they do not lift GLSM vacua.  The space of GLSMs flows to a
subspace of the moduli space.  LGO phases appear to be relatively
rare.  More common, although far from generic, are hybrid phases in
which the space of classical vacua is compact and there are, in
addition to the NLSM on this, massless fields interacting via a
superpotential.

The goal of this paper is to introduce a novel class of (0,2) models
based on hybrid geometries, extending the construction of (2,2) hybrid
models in \cite{Bertolini:2013xga}, 
and describe a technique to compute the complete massless spectrum.\footnote{Aspect of hybrid theories with (2,2) supersymmetry have recently received attention 
also from a mathematical perspective \cite{Babalic:2016mbw,Babalic:2016hjh,2016arXiv160808962B,2015arXiv150204872S}.} 
Hybrid theories 
are a natural generalization of both LG models and NLSMs, and while it is fair to say
that a generic (0,2) SCFT will not be described by a hybrid model, we are nonetheless enlarging the set of
points where exact computations can be carried out. Indeed, under favorable circumstances, a hybrid model,
while carrying non-trivial geometric structure, is simple enough that
explicit computations can be performed. We offer evidence that the
problem of accidental symmetries is ameliorated by hybrids in the
sense that ``accidental'' LG models can produce well-behaved theories
and suggest that many of the models we construct do not, in fact,
arise as GLSM phase limits.

Although our main interest lies in models suitable for heterotic compactifications, our construction can be straightforwardly 
generalized to yield more general (0,2) SCFTs. In the first part of this work we will take the more general approach,
and we will point out which conditions should be further imposed for string applications. In the second part of this note,
in which we tackle the computation of the massless spectrum and we apply our techniques to solve explicit examples,
we will restrict our attention to hybrid SCFTs suitable for heterotic compactifications.

The rest of this note is organized as follows. We begin in section \ref{s:model} by describing the 
construction of our class of models.
In section \ref{s:anomalies} we proceed by deriving the constraints from anomaly cancellation and exploring the basic low-energy properties. 
We then turn in section \ref{s:spectrum} to a generalization of the methods for computing
the massless spectrum for heterotic compactifications based on our models, and apply these techniques to
nontrivial examples in section \ref{s:examples}. Here, we will
provide evidence for using hybrids as a ``cure" for resolving LG accidents. In section \ref{s:GLSMembed} we present a criterion 
for a hybrid to arise as a phase in a GLSM exhibiting a Calabi-Yau
large radius limit and show that even for a toric base this is nontrivial. 

\acknowledgments The authors would like to thank N.~Addington, P.~Aspinwall, K.~Hori and M.~Romo for useful discussions.
It is a pleasure to thank I.~Melnikov for many discussions
that lead to some of the results presented in this work. 
We would also like to thank I.~Melnikov and M.~Romo for helpful comments on the manuscript.
We thank the organizers of the conference {\it Current problems in theoretical physics} in Vietri sul Mare, as well as D.~Isra\"el and the theory group at LPTHE (Paris)
for allowing MB to present a preliminary version of this work.
MB thanks the Perimeter Institute for Theoretical Physics and the physics department at Duke University for hospitality while some of this work was completed.
MRP thanks the theory group at LPTHE, the mathematics department at
UCSB,  and especially the particle
theory group at the Weizmann Institute of Science for their gracious
hospitality during several stages of this work.
MB and MRP are supported by NSF Grant PHY-1521053.
This work was supported by World Premier International Research Center Initiative (WPI Initiative), MEXT, Japan.
Any opinions, findings, and conclusions or
recommendations expressed in this material are those of the authors
and do not necessarily reflect the views of the National Science Foundation.

\section{The model}
\label{s:model} 

In this section we will introduce the geometric ingredients suitable
to build a (0,2) hybrid model, we will construct the action for the
corresponding NLSM and analyze its symmetries.  

\subsection*{(0,2) superspace conventions}

Throughout this work we work in Euclidean signature and (0,2) superspace with bosonic coordinates $(z,\zb)$ and fermionic coordinates $(\theta,\thetab)$.
The supercharges are
\begin{align}
\label{eq:supercharges}
\cQ &= -\frac{\p}{\p\theta} + \thetab \pbz~,
&\cQb&= -\frac{\p}{\p\thetab} +\theta \pbz~,
\end{align}
where $\pb_{\!\zb} \equiv \p/\p \zb$, and satisfy the algebra
\begin{align}
\cQ^2 &= \cQb^2 =  0~, &\AC{\cQ}{\cQb} &= -2\pbz~. 
\end{align}
The supercovariant derivatives anticommute with \eqref{eq:supercharges} and are defined as
\begin{align}
\cD &= \frac{\p}{\p\theta} + \thetab \pbz~,
&\cDb &= \frac{\p}{\p\thetab} +\theta \pbz~,
\end{align}
realizing the corresponding algebra
\begin{align}
\cD^2 &= \cDb^2 =  0~, &\AC{\cD}{\cDb} &= 2\pbz~.
\end{align}

\subsection{A peek at hybrids}

The models we construct in this work provide a description of suitable
limiting loci in the moduli space of (0,2) SCFTs.  Hybrid limits have
been discussed in the context of (0,2) GLSMs \cite{Chiang:1997kt} 
or of (0,2) NLSMs with superpotential \cite{Guffin:2008pi}.
Our approach is closer in spirit to the latter, so we begin with a quick
review of this.

In general, to write a (0,2) NLSM with superpotential\footnote{The perspective of the construction of hybrid limits in \cite{Guffin:2008pi} 
is however different from the one developed in the present work. For instance, while the author in \cite{Guffin:2008pi} do consider 
NSLMs on non-compact spaces with cokernel bundles,
the classical vacuum of the theory is not the base of the fibration.}  \cite{Guffin:2008pi} one
requires a $D$-dimensional K\"ahler target space $\bY_0$ equipped with a K\"ahler
metric $g_{\alpha\betab}$ determined by the K\"ahler potential
$K(y,\yb)$, a rank-$R$ holomorphic vector bundle $\cE_0\to \bY_0$
equipped with a Hermitian inner product $h_{A\Bb}$, as well as a holomorphic
section $J\in \Omega^0(\cE_0^\ast)$.   The model is then constructed
using $D$ bosonic chiral multiplets $Y^\alpha$ and $R$ fermionic
chiral multiplets $\cX^A$, with expansions
\begin{align}
\label{eq:fieldcontent}
Y^\alpha & = y^\alpha + \sqrt{2}\theta \eta^\alpha +\theta\thetab \pbz y^{\alpha}~,&&&
\Yb^{\alphab} & = {\yb}^{\alphab} - \sqrt{2}\thetab \etab^{\alphab} -\theta\thetab \pbz \yb^{\alphab}~, \nonumber\\
\cX^A & = \chi^A + \sqrt{2}\theta H^A +\theta\thetab \pbz \chi^{A}~,&&&
\cXb^{\Ab} & = {\chib}^{\Ab} + \sqrt{2}\thetab \Hb^{\Ab} -\theta\thetab \pbz \chib^{\Ab}~.
\end{align}
Here, $y^\alpha$, $\alpha=1,\dots, D$, are coordinates on $\bY_0$, $\eta^\alpha$ are right-moving fermions on the worldsheet that transform as sections of the tangent bundle $T_{\bY_0}$, 
$\chi^A$, $A=1,\dots,R$, are left-moving fermions on the worldsheet which transform as sections of the bundle $\cE_0$
and $H^A$ are non-propagating auxiliary fields.

In terms of these we can write the classical Lagrange density as a sum
of a kinetic term $\cL_{\text{kin}}$ and a superpotential term
$\cL_W$ where
\begin{align}
\cL_{\text{kin}} &= \int\! d^2\theta \half (K_\alpha \p_z Y^\alpha - K_{\alphab}\p_z
                   \Yb^{\alphab}) + \cH_{A\Bb}\cX^A\cXb^{\Bb}~,	\nonumber\\
\cL_W &= m\int\! d\theta \cX^A J_A + {\rm h.c.}~,
\end{align}
with $K_\alpha \equiv \p_\alpha K$, and $m$ is a coupling with
dimensions of mass.

With $J=0$, or at energies well above $m$, this is just a NLSM with
target space $\bY_0$ and left-moving fermions coupled to $\cE_0$.  If
$c_1(T_{\bY_0}) = c_1(\cE_0)=0$ and $\ch_2(T_{\bY_0}) = \ch_2(\cE_0)$, this will
determine a superconformally invariant theory to all orders in
perturbation theory.  We typically consider this SCFT as describing
the IR dynamics of the theory.  The details of the various metrics
introduced are associated to irrelevant deformations of this; the space of
SCFTs will be locally parameterized by the complex structure of
$\bY_0$, the K\"ahler class of $g$, and
the deformations of $\cE_0$ as a holomorphic bundle.
Non-perturbative
effects (worldsheet instantons) can modify this moduli space, 
and will be discussed below.  The NLSM is
weakly coupled when the K\"ahler class is deep within the K\"ahler
cone of $\bY_0$.  An important role in understanding the low-energy
dynamics is played by the chiral $\GU(1)_L\times\GU(1)_R$ symmetry
under which $\chi$, resp.~$\psi$ are charged.  The topological
conditions mentioned ensure that these are nonanomalous.
Strictly speaking these results are well known when $\bY_0$ is
compact; this will not be the case in our examples but we expect no
subtleties to arise due to this.

Given values of the parameters, however, we can consider the SCFT they
determine as a UV fixed point and study the low-energy dynamics
obtained when we add a superpotential, which is a relevant coupling.
In general, this breaks the $\GU(1)_L\times\GU(1)_R$ symmetry. If
$\bY_0$ admits a holomorphic Killing vector $V$ and $\End (\cE_0)$ a
global holomorphic section $Q$ such that 
\begin{align}
Q\chi \cdot J = \chi\cdot {\cal L}_V J\ ,
\end{align}
we have an unbroken  $\GU(1)_L\times\GU(1)_R$ under which 
\begin{align} 
\delta\chi &= -i\epsilon_LQ\chi - i\epsilon_R(Q + 1)\chi~,
&\delta y &= i\epsilon_L V(y) + i\epsilon_R V(y)\ .
\end{align}
For sufficiently generic $J$, there will be a unique such $V$.  Since
the action differs by a non-chiral correction from the na\"\i ve
symmetry, this is nonanomalous under the same topological conditions
mentioned above.

The superpotential is a relevant deformation and at low energies the
UV theory will flow, under suitable conditions, to a family of
low-energy SCFTs parameterized in addition by the section $J$.  
The superpotential interaction induces a potential for $y$ given by 
\begin{align}
U(y) = m^2|J(y)|^2~,
\end{align}
so low-energy dynamics will be determined by the structure of $\bY_0$
and $\cE_0$ in the vicinity of $B = J^{-1}(0)$.   Our interest will be
in models in which a generic $J$ vanishes on a {\sl compact\/} $B$.
This is the {\sl potential condition\/}.
In this limit, we expand the K\"ahler potential $K$ to quadratic order about $B$ as 
\begin{align}\label{eq:Kexp}
K = \widehat{K}(x,\xb) + \phi^{\dagger} h \phi + \cdots~,
\end{align}
where $\widehat K$ is a K\"ahler potential for a metric on $B$, with local coordinates $x$, and $h$
is a hermitian metric on the normal bundle $NB$, with local coordinates $\phi$. This corresponds to
approximating $\bY_0$ as the total space of $\bY = NB\to B$,
and the bundle $\cE_0$ as a bundle $\cE\to\bY$. 
In the large-radius limit one might think of the theory as described 
fiberwise over $B$ by a LG model, adiabatically varying over the
base.  In order for this description to be useful, it is important
that the isometry $V$ be {\sl vertical\/} fixing the base pointwise.
Models satisfying this criterion are called {\sl good hybrids\/}.

The models obtained by this method will inherit the property 
that $c_1(\cE)=c_1(T_{\bY})=0$.  However, it is possible that the 
resulting structure contains 
{\sl spectator fields\/} as in \cite{Distler:1995mi,Guffin:2008pi}.  Consider a model 
in which we have a chiral field $S$ taking values in a line bundle 
over $\bY$ and a Fermi field $\Xi$ taking values in the dual line
bundle.  These admit a superpotential coupling 
\begin{align}
\cW_S = m_S S\Xi\ .
\end{align}
These fields introduce no massless modes, and at energies well below
$m_S$ can be integrated out.   The quotient bundle $\cE'$ obtained by
dropping $\xi$ and restricting to $\bY'=\{s=0\}$ will then be 
described as above, but we will in general find after reduction to low
energy the weaker condition $c_1(\cE')+c_1(T_{\bY'})=0$.  

In this work we will take a bottom-up approach.  We will construct a
hybrid model determined by 
the quadruple $(\bY,\cE,V,J)$,
where $\bY$ is the total space of a vector bundle over a compact
K\"ahler base $B$, and $\cE$ is a holomorphic vector bundle over $\bY$
\begin{align}
\label{eq:geomstructure}
\cE &\longrightarrow \bY~,		&\bY=\tot\left( X\overset{\pi}{\longrightarrow} B\right)~.
\end{align}
We denote
\begin{align}
d&=\dim_\C B~,	&n&=\dim_\C X~,		 &R&=\rank \cE~.
\end{align}
$V$ determines a vertical $\GU(1)$-action on $\bY$, 
and a lift of this action to $\cE$.
The last piece of data is given by the (0,2) superpotential $J\in
\Gamma(\cE^\ast)$. 
We assume that this satisfies the  {\it potential condition\/} of
\cite{Bertolini:2013xga}, i.e., $J^{-1}(0) = B$.

In the remainder of this section we will construct a (0,2) theory from
these ingredients, and argue that under suitable conditions the IR
dynamics will be determined by a (0,2) superconformal theory which is
our real interest.    
The $V$ action will endow our theory with a
global $\GU(1)_L$ symmetry which will be essential in controlling IR
physics \cite{Distler:1995mi}.

Before delving into the details of the construction two remarks are in
order.  In contrast to the situation in (2,2) models, in (0,2)
theories the relation between UV and IR physics exhibits various
subtleties.  In particular, a unique $\GU(1)_L\times \GU(1)_R$ symmetry in
the UV (which we will assume) does not exclude mixing with accidental
symmetries in the IR, rendering predictions based on the UV symmetry incorrect.
As demonstrated 
in the simple setting of LG models in \cite{Bertolini:2014ela}, even
the central charge of the IR symmetry is in general not manifest in
the UV.  Presumably analogous phenomena occur in more elaborate models
such as those considered here, and we do not know how to resolve the
question definitively.  
We will provide some evidence here that the
hybrid construction in fact resolves some of the problems found there,
in the sense that some LG models suffering from ``accidents'' can be used
to construct hybrid models that do not exhibit the expected
pathologies.
Furthermore, computations of physically
interesting quantities\footnote{A study of the topological ring in
B- and B/2-twisted hybrid models will be presented in \cite{Bertolini:2018now}.}
like the ones presented here yield a good deal of information about
the model, and can provide evidence in favor of, or against, a
putative model.

A second remark concerns the extent to which our construction produces
a well-defined theory.  To write
a (0,2) NLSM on $\bY$ we require a K\"ahler metric on $\bY$ such that
$V$ acts as a vertical isometry.  As noted in \cite{Bertolini:2013xga},
for sufficiently ``negative'' bundles (see section \ref{ss:Y}) we can use $h$ 
to construct such a metric.  In general, when this is not the case, 
we can imagine a space $\bY_0$ 
with an immersion $B\to\bY_0$ such that $T_{\bY_0}|_B = T_B\oplus X$ and 
equipped with a K\"ahler metric reducing near $B$ to the form
(\ref{eq:Kexp}).   The formulation we provide will give an effective
description of the dynamics for energies $E\ll m$, and will be
valid deep enough in the K\"ahler cone that the nonlinear sigma model
is weakly coupled at the scale $E$.

\subsection{The target space}
\label{ss:Y}

The construction begins with a (0,2) NLSM on the target space
$\bY=\tot\left( X\overset{\pi}{\longrightarrow} B\right)$.
Coordinates on $\bY$ will be denoted $y^\alpha$ and following the
bundle structure we split these into coordinates on the base and the
fiber as $y^\alpha=(y^\mu,\phi)$, with $\mu=1,\dots,d$ and
$\phi\in\Gamma(X)$.

We equip $\bY$ with a K\"ahler metric induced by a K\"ahler potential
$K$.  The low-energy physics of a well-behaved hybrid will be
described by small fluctuations of the fiber fields around the base,
so we expand $K$ to quadratic order in $\phi$ as
\begin{align}
\label{eq:Kquadrexp}
K = \widehat{K}(y^\mu,\yb^{\mub}) + \phi^{\dagger} h \phi + \cdots~,
\end{align}
where $\widehat K$ is a K\"ahler potential for a metric on $B$ and
$h$ is a hermitian metric on $X$.

As usual the detailed form of the metric will be irrelevant to
the IR physics, but the K\"ahler class will appear as a parameter in
the low-energy physics in the models we discuss.  Because $\bY$
retracts to $B$ this is determined by a class in $h^2(B)$.  Our NLSM
will be weakly coupled, and the IR physics simply connected to the UV
data, when this takes values deep into the K\"ahler cone of $B$.

The $V$ action on $\bY$ is generated by a holomorphic Killing vector
field of the metric.  As shown in \cite{Bertolini:2013xga}, this is
determined by a covariantly constant section $A\in\Gamma(X\otimes
X^*)$ such that $h A^{\dagger} = - A h$.  The bundle $X$ then has an 
orthogonal decomposition $X=\oplus_i X_i$ into eigenspaces of $A$ with
eigenvalue $-iq_i$.  Reasonable models will emerge for $0 < q_i< 1$.
We will assume here that $X_i$ can be taken to
be line bundles.  This is not an
essential restriction but it will simplify notation and provide
sufficiently varied examples.  We can then represent the field
$\phi = \sum_i \phi^i$ where $\phi^i\in\Gamma(X_i)$; we will refer to
$q_i$ as the charge of $\phi^i$. 
Our considerations apply with no real modification to the case in which
$X = \widehat X/\Gamma$ is an orbi-bundle given by a discrete group
quotient.

\subsection{The left-moving bundle}
\label{ss:leftbundle}

As mentioned above, the choice of a holomorphic vector bundle $\cE$
over $\bY$ is part of the defining data of our construction.  Our
construction will work if $\cE$ respects the bundle structure of $\bY$
and admits a suitable lift of the $V$ action.  We will in the
following describe a particular class of such bundles, for which we
have a relatively straightforward method to 
explicitly compute the complete massless spectrum following \cite{Kachru:1993pg,Bertolini:2013xga}.
The calculations are in fact simplest for bundles over $\bY$ that pull back
from bundles over $B$, but this excludes, for example, $(2,2)$
models; the class of models we consider is large enough to contain
these while still amenable to our computational method.

We start our construction with an auxiliary collection of irreducible holomorphic bundles $\cF_\Gamma\rightarrow B$,
$\Gamma=1,\dots R+m$, for some $m\geq0$.
As in the case of the bundle $X$ above,  
we choose to restrict ourselves to the case where $\cF_\Gamma$ are
line bundles to simplify notation.  Again, this restriction is not essential.
In terms of these, we define $\cE$ 
by the SES
\begin{align}
\label{eq:Ebundle}
\xymatrix@R=0mm@C=10mm{
0\ar[r] &\cO_{\bY}^{\oplus m} \ar[r]^-{E} &\oplus_\Gamma \pi^\ast (\cF_\Gamma) \ar[r]^-{F} & \cE \ar[r] &0~,
}
\end{align}
where $\pi$ is the projection map defined in \eqref{eq:geomstructure}.

The existence of a lift of the $V$ action to $\cE$ is imposed on this
structure by assigning 
charge $Q_\Gamma$ to the summand
$\cF_\Gamma$ and charge $-1$ to the first term, and requiring that the
maps are all equivariant, ensuring that the cokernel in fact carries a
$V$ action.  
We order $\Gamma$ 
so that $Q_1\leq Q_2\leq\cdots$.
Reasonable IR theories will emerge for $-1\leq Q_\Gamma < 0$.
The map $E$ is then represented by a collection of sections
$E^\Gamma\in\Gamma(\pi^\ast \cF_\Gamma)$ of charge
\begin{align}
\label{eq:VactionEs}
V(E^\Gamma)&=(Q_\Gamma+1)E^\Gamma~.
\end{align}
We assume that for sufficiently generic maps this defines a smooth
bundle (or an orbi-bundle).
In particular, since $q_i> 0$, $Q_\Gamma=-1$ implies that 
$E^\Gamma$ is constant along fibers of $X$, while for $Q_\Gamma > -1$,
$E^\Gamma$ is quasi-homogeneous of positive degree on the fibers and
vanishes on the base.
Let $-1 = Q_P < Q_{P+1}$ determining $0\le P\le R+m$.

If we consider the restriction of \eqref{eq:Ebundle} to the base
\begin{align}
\xymatrix@R=0mm@C=10mm{
0\ar[r] &\cO_B^{\oplus m}\ar[r]^-{E|_B} &\oplus_\Gamma \cF_\Gamma \ar[r]^-{F|_B} & \cE|_B \ar[r] &0~,
}
\end{align}
the vanishing of the last $R+m-P$ maps shows that 
the bundle splits as $\cE|_B=\oplus_{\Gammat=P+1}^{R+m}\cF_{\Gammat}\oplus \cE_B$~, where $\cE_B\rightarrow B$ is defined by
\begin{align}
\label{eq:EBbundle}
\xymatrix@R=0mm@C=10mm{
0\ar[r] &\cO_B^{\oplus m} \ar[r]^-{E_B} &\oplus_{\Gammah=1}^P \cF_{\Gammah} \ar[r]^-{F_B} & \cE_B \ar[r] &0~.
}
\end{align}
The map $E_B$ ($F_B$) is obtained from the restriction map $E|_B$ ($F|_B$) simply by ignoring the last $R+m-P$ rows (columns).
In other words, \eqref{eq:EBbundle} is the projection to $B$ of the
restriction of \eqref{eq:Ebundle} to degree $-1$ under the $V$-action. 
Of course, in (2,2) theories $\cE_B=T_B$.
From \eqref{eq:Ebundle} and \eqref{eq:EBbundle} it then follows that the bundle $\cE$ can be equivalently defined by the following SES
\begin{align}
\label{eq:EasExt}
\xymatrix@R=0mm@C=10mm{
0\ar[r] &\oplus_{\Gammat=P+1}^{R+m} \pi^\ast \cF_{\Gammat}    \ar[r] &\cE   \ar[r] & \pi^\ast \cE_B \ar[r] &0~.
}
\end{align}
A proof of the above statement can be found in appendix
\ref{app:proofbundles}.  

In other words, given a collection of $\cF_\Gamma$ and maps $E_B$, 
the bundles $\cE$ we can obtain are classified by extensions of
$\pi^\ast \cE_B$ by $\oplus_{P+1}^{R+m} \pi^\ast\cF_{\Gammat}$. 
For our purposes, we can take \eqref{eq:EasExt} to be our definition of the bundle for our models, 
and keep the discussion above as motivation for this particular form. 
It is worth noting that a nonsingular bundle $\cE$ might still give
rise to a singular model by failing to satisfy the potential
condition.  In some cases we can obtain both smooth and singular
models depending on the extension we choose.

We can make this dependence on the fiber coordinates explicit.  We can
write the map $E^\Gammat$ as 
\begin{align}
\label{eq:Eexplexpr}
E^\Gammat &= \sum_{\br\in\Delta_\Gammat} \mathsf{M}_{\br}S_\Gammat^{\br}~,	&\mathsf{M}_{\br} &= \prod_i (\phi^i)^{r_i}~,
\end{align}
where 
\begin{align}\label{eq:DeltaIdef}
\Delta_\Gammat = \Big\{ \br\in\Z_{\ge 0}^{\oplus n} \Big| \sum_i q_i r_i = Q_\Gammat+1 \Big\}
\end{align}
indicates the monomials of suitable charge for inclusion in
$E^\Gammat$ and $S_\Gammat^{\br}$ is a section of 
$\cL_\Gammat^{\br} = \cF_\Gammat\otimes_i \left(X_i^\ast\right)^{\otimes r_i}$.

Now, let $\mathfrak{U_{a,b}}$ be two patches of $B$,
$\mfU_a\cap\mfU_b\neq\emptyset$, parametrized by local coordinates
$y^\mu_{a,b}$.
Let $g_{ab}^{\Gammat}(y)$ be the transition function for the bundle
$\cF_{\Gammat}$ and $T_{ab}(y)$ the transition matrix for the bundle $\cE_B$.
Then, a section $\lambda$ of $\cE$ transforms as $\lambda_b(y^\alpha_b)=\lambda_a(y^\alpha_a)G_{ab}(y^\alpha_a)$,  where
\begin{align}
\label{eq:Etransfuncts}
G_{ab}(y_a)=
\begin{pmatrix}
  T_{ab} 	& C^1 		                &C^2                               & \cdots 		\\
0 		& g^{P+1}_{ab} 			& 0 					& \cdots 		\\
0 		& 0 					& g^{P+2}_{ab} 			& \cdots 		\\
\vdots 	& \vdots 				& \vdots 				& \ddots 
\end{pmatrix}~,
\end{align} 
and
\begin{align}
C^\Gammat = \sum_{\br \in \Delta_\Gammat} \mathsf{M}_\br (\phi_a)
  f^\Gammat_{ab,\br} (y^\mu_a) ~,
\end{align}
where the collection of functions $f^\Gammat_{ab,\br} (y^\mu_a)$ is
given by the restriction of a section of $\cE_B^\ast\otimes \cL_\Gammat^{\bf r}$.

To illustrate this construction and its connection to \eqref{eq:Ebundle}, we provide an example in appendix \ref{app:proofbundles}.

\subsection*{Field content}

The field content of the theory consists of a set of $d+n$ (0,2) bosonic chiral multiplets $Y^\alpha$ and $R$ (0,2) fermionic chiral multiplets $\cX^A$ together with their 
conjugate anti-chiral multiplets
\begin{align}
\label{eq:fieldcontent}
Y^\alpha & = y^\alpha + \sqrt{2}\theta \eta^\alpha +\theta\thetab \pbz y^{\alpha}~,&&&
\Yb^{\alphab} & = {\yb}^{\alphab} - \sqrt{2}\thetab \etab^{\alphab} -\theta\thetab \pbz \yb^{\alphab}~, \nonumber\\
\cX^A & = \chi^A + \sqrt{2}\theta H^A +\theta\thetab \pbz \chi^{A}~,&&&
\cXb^{\Ab} & = {\chib}^{\Ab} + \sqrt{2}\thetab \Hb^{\Ab} -\theta\thetab \pbz \chib^{\Ab}~.
\end{align}
Here, $y^\alpha$, $\alpha=1,\dots,n+d$, are coordinates on $\bY$, $\eta^\alpha$ are right-moving fermions on the worldsheet that transform as sections of the tangent bundle $T_{\bY}$, 
$\chi^A$, $A=1,\dots,R$, are left-moving fermions on the worldsheet which transform as sections of the bundle $\cE$
and $H^A$ are non-propagating auxiliary fields.

Given the bundle structure for $\bY$ in \eqref{eq:geomstructure},
it is often convenient to split the coordinates as $y^\alpha=(y^\mu,\phi^i)$, $\mu=1,\dots,d$, $i=1,\dots,n$, which represents the splitting into base versus fiber coordinates. 
Following the construction of the bundle $\cE$ in the previous section, it is natural to introduce a similar splitting for the left-moving fermions $\chi^A$. 
We introduce two sets of indices $A=(M,I)$, where $M=1,\dots,R_B$, and $I=R_B+1,\dots,R_B+N$,
where for convenience we have relabeled $R_B = P-m$ and $N = R - R_B$.  The field $\cX^I$ takes values in $\pi^\ast \cF_{I+m}$.  
Locally we can take $\cX^M$ to be a section of $\pi^\ast\cE_B$.  
The distinction between the two sets of Fermi fields is not valid globally when \eqref{eq:EasExt} is a nontrivial extension.
In particular, when $\cE$ is a deformation of $T_{\bY}$ this represents  the fiber/base splitting.

\subsection{The action}

Let us denote by $K$ the K\"ahler potential
for a K\"ahler metric on $\bY$ and by $\cH_{A\Bb}$ a Hermitian metric on $\cE\rightarrow \bY$. 
We approximate $K$ for small $\phi$ as in \eqref{eq:Kquadrexp}
and neglect the terms of higher powers in $\phi,\phib$ resulting in $K$ sesquilinear in $\phi$.

The action in (0,2) superspace for our models consists of D-terms for the kinetic part and F-terms specifying the (0,2) superpotential.
Requiring our action to possess a global $\GU(1)_L$ symmetry, and
$\bY$ to be K\"ahler with a torsion-free connection\footnote{More
  general kinetic terms are compatible with $(0,2)$ supersymmetry; we
  will not investigate this here.}
forces the kinetic term to be of the form \cite{Melnikov:2011ez} 
\begin{align}
\label{eq:lagrangian}
\cL_{\text{kin}} &= \half (K_\alpha \p_z Y^\alpha - K_{\alphab}\p_z \Yb^{\alphab}) + \cH_{A\Bb}\cX^A\cXb^{\Bb}~,
\end{align}
where $K_\alpha \equiv \p_\alpha K$.
The (0,2) superpotential will be assumed of the form $\cW = \cX^A J_A$ with $J\in\Gamma(\cE^\ast)$.

It is easy to derive the equations of motion in superspace. Up to boundary terms we obtain
\begin{align}
\label{eq:eomsuperspace}
\cDb\left[\cH_{A\Bb} \cXb^{\Bb} \right] & = \sqrt2 J_A~,	
&\cDb \left[  K_{\alpha\betab}\p \Yb^{\betab} +\cH_{A\Bb,\alpha}  \cXb^{\Bb} \cX^A \right] &= \sqrt2 \cX^A J_{A,\alpha}~.
\end{align}
These equations suggest 
the field redefinitions
\begin{align}
\label{eq:newfermion}
\cXb_{A} &\equiv \cH_{A\Bb} \cXb^{\Bb}~,		&P_\alpha&\equiv K_{\alpha\betab}\p \Yb^{\betab} +\Gamma^C_{A\alpha}\cXb_C \cX^A~,
\end{align}
where we introduced the connection $\Gamma^C_{A\alpha}\equiv \cH_{A\Bb,\alpha} \cH^{\Bb C}$ on $\cE\rightarrow\bY$. The corresponding curvature is given by 
$\cF_{I\alphab\beta \Jb}\equiv \Gamma^K_{I\beta,\alphab} \cH_{K\Jb}$.
The component action then reads
\begin{align}
\label{eq:compaction}
\half \cL &=\rho_\alpha  \pb y^\alpha + K_{\alphab\beta}\etab^{\alphab} D^\text{K} \eta^\beta
+\chib_A D^\text{H} \chi^A - \cF_{A\alphab\beta}{}^{B}\etab^{\alphab}\eta^\beta\chib_{B}\chi^A  \nonumber\\
&\quad -\chi^A\eta^\alpha D_\alpha J_A + \chib^{\Ab}\etab^{\alphab} D_{\alphab} \Jb_{\Ab} +\cH^{A\Bb} \Jb_{\Bb}J_A~,
\end{align}
where we implemented the field redefinitions \eqref{eq:newfermion} on the lowest components of the corresponding superfields
\begin{align}
\chib_A&\equiv \cH_{A\Bb}\chib^{\Bb}~,		&\rho_\alpha&\equiv K_{\alpha\betab}\p \yb^{\betab} + \Gamma^B_{A\alpha} \chib_B\chi^A~.
\end{align}
In writing the action \eqref{eq:compaction} we have made use of 
the covariant derivatives \footnote{The superscripts on $D^\text{K}$ and $D^\text{H}$ stand for ``K\"ahler" and ``Hermitian" respectively.}
\begin{align}
D^\text{K} \eta^\beta&=\p\eta^\beta + \Omega^\beta_{\alpha\gamma}\p y^\alpha \eta^\gamma~,		&D^\text{H} \chi^A&=\p\chi^A + \Gamma^A_{B\alpha}\p y^\alpha \chi^B~,		
&D_\alpha J_A&=\p_\alpha J_A +\Gamma_{A\alpha}^B J_B~,
\end{align}
where $\Omega^\delta_{\alpha\beta}\equiv K_{\alpha\gammab,\beta} K^{\gammab\delta }$ is the K\"ahler connection on $\bY$. 

\subsection*{Symmetries}

The action we described in the previous section is invariant under (0,2) supersymmetry,
implemented by the operators $\bQ$ and $\bQb$.\footnote{We use the notation $\epb\bQb\cdot\bullet\equiv\CO{\epb\bQb}{\bullet}$, where 
$\epb$ is an anti-commuting parameter
and $\bQb$ is the operator corresponding to the supercharge $\cQb$, and similarly for $\bQ$.}
In later sections we will study in detail the cohomology of $\bQb$, thus it is convenient to introduce its action on the component fields.
This is given by
\begin{align}
\label{eq:Qbaractioncomp}
\bQb\cdot \yb^{\alphab}&=-\etab^{\alphab}~,	&\bQb\cdot \eta^{\alpha}&=-\pb y^{\alpha}~,	&\bQb\cdot \chib_A&=J_A~,	&\bQb\cdot \rho_{\alpha}&=\chi^AJ_{A\alpha}~,
\end{align}
where $J_{A\alpha}\equiv \p/ \p y^\alpha J_A$.

As we mentioned above, chiral symmetries play a fundamental role in our construction.
In addition to the right-moving $\GU(1)_R$ R-symmetry, under which 
the multiplets \eqref{eq:fieldcontent} are invariant and $\theta$ has charge 1, the action \eqref{eq:compaction} at $J=0$ exhibits a global $\GU(1)_L$ symmetry, 
under which $\cX^A$ have charge $-1$ and everything else is invariant.
The inclusion of the superpotential generically breaks this symmetry, but
our construction of $\cE$ in section \ref{ss:leftbundle} ensures this
is preserved since
the superpotential $J\in\Gamma(\cE^\ast)$ satisfies the quasi-homogeneity condition 
\begin{align}
\label{eq:quasihom}
-Q_AJ_A = \sum_\alpha q_\alpha y^\alpha \p_\alpha J_A~,
\end{align}
for some $0\leq q_\alpha<1$ and $-1\leq Q_A <0$.\footnote{The upper
  bounds for the charges are determined by the potential condition and
  unitarity constraints.}
The hybrid action is then invariant under a $\GU(1)_L\times\GU(1)_R$ symmetry under which the superfields have charges
\begin{align}
\label{eq:symclasact}
\xymatrix@R=0mm@C=8mm{
		&Y^\mu		&\Phi^i		&\cX^A		&\theta \\
\bq		&0			&q_i			&Q_A 		&0\\
\bqb		&0			&q_i			&Q_A+1 		&1
}
\end{align}
where we denote by $\bq$ and $\bqb$ the charges under 
$\GU(1)_L$ and $\GU(1)_R$, respectively. 
Note that in \eqref{eq:symclasact} we assumed that the base
coordinates $y^\mu$ have $q_\mu=0$ ($V$ acts vertically). 
A parallel discussion concerns the fermionic fields.
Following our former distinction we have $Q_M=-1$ and $Q_I>-1$, where, with some abuse of notation,
we denote $Q_I=Q_{\Gammat=I+m}$. 
Again, the distinction is valid locally over a patch in $B$ but the form of the transition functions ensures that the charge assignments hold in any patch.

\section{Anomalies and IR physics}
\label{s:anomalies}

In this section we are going to derive the constraints for anomaly
cancellation that our models need to satisfy in order for their IR
behavior to be reliably described in terms of the UV data.  Our action
enjoys a $\GU(1)_L\times\GU(1)_R$ action, and we are interested in
models for which $\GU(1)_R$ is the current algebra in the right-moving
superconformal algebra of the low-energy SCFT while $\GU(1)_L$ is a
left-moving conserved current.

The NLSM on $\bY$ from which we began our discussion is well-defined
provided the sigma model anomaly condition
\begin{align}
\label{eq:hetanom}
\ch_2(\cE) = \ch_2(T_{\bY})~
\end{align}
holds.
It is possible to relate $\ch_2(T_{\bY})$ to $\ch_2(T_B)$ and the Chern classes $c_1(X_i)$ as follows
\begin{align}
\ch_2(T_{\bY}) &= \ch_2(T_B) + \half \sum_i c_1(X_i)^2~.
\end{align}
Similarly, we can express $\ch_2(\cE)$ in two equivalent ways according to our definitions \eqref{eq:Ebundle} and \eqref{eq:EasExt} as
\begin{align}
\ch_2(\cE) &= \ch_2(\cE_B) + \half \sum_{\Gammat} c_1(\cF_{\Gammat})^2=\half \sum_{\Gammah} c_1(\cF_{\Gammah})^2+ \half \sum_{\Gammat} c_1(\cF_{\Gammat})^2~.
\end{align}
Of course, the above equalities are statements about pullbacks of classes in $H^4(B,\Z)$, which means that \eqref{eq:hetanom} 
is trivially satisfied for $d\leq1$. While this fact is obvious for $d=0$, i.e., LG theories, it is somewhat more surprising for $d=1$ models.
In the case where the hybrid describes a (0,2) NLSM with target space
a CY manifold, \eqref{eq:hetanom} reduces to the usual anomaly
cancellation condition.

If the superconformal theory is to determine a  heterotic string
vacuum, \eqref{eq:hetanom} is the condition that four-dimensional
gauge and gravitational anomalies can be cancelled by the
Green-Schwarz mechanism.

\subsection{Worldsheet instantons in hybrid models}

In fact, the $\GU(1)_L\times\GU(1)_R$ symmetry of the classical action is generically broken 
in the presence of worldsheet instantons.
This will remain a symmetry of the quantum theory if the path-integral
measure is invariant in the background of an instanton.  
As usual, the charge of this is determined by the charges of the
fermionic zero modes in the presence of the instanton.
Thus, we start with a description of worldsheet instantons in hybrid models.

In a (0,2) NLSM with target space $\bY$ worldsheet
instantons correspond to topologically nontrivial maps of
the worldsheet $\Sigma=\P^1$ into $\bY$.  For a simply connected space these are
classified by $H^2(\bY,\Z)$.   Understanding the spaces of such maps
for a CY target space is in general highly nontrivial.  In LG models,
the low-energy theory localizes to a point, and there are no worldsheet
instantons.  In hybrid models, worldsheet instantons are associated
with nontrivial holomorphic maps of the worldsheet to $B$.  Since the
space of such maps can be rather simple, these models provide an
opportunity to study the effects of worldsheet instantons in detail.

In the background of an instanton,
the Fermi fields will take values in bundles $\overline{K}_\Sigma^{\frac{1}{2}}
\otimes y^\ast(T_{\bY})$ and
$K_\Sigma^{\frac{1}{2}} \otimes y^\ast(\cE)$ for right-movers and left-movers,
respectively, where $K_\Sigma$ is the anti-canonical bundle of the
worldsheet $\Sigma=\P^1$.  Because the instanton background satisfies
$y(\Sigma)\subset B$ we have  
\begin{align}
y^\ast(T_{\bY}) &= y^\ast(T_B)_{(0,-1)}\oplus_i y^\ast(X_i)_{(q_i,q_i-1)}~,  \nonumber\\
y^\ast(\cE) &=y^\ast(\cE_B)_{(-1,0)} \oplus_{\Gammat=P+1}^{R+m} y^\ast(\cF_{\Gammat})_{(Q_\Gammat,Q_\Gammat +1)}~,
\end{align}
where the subscripts indicate the charges under
$\GU(1)_L\times\GU(1)_R$.

Left-moving zero modes will correspond to
holomorphic sections of $K_\Sigma^{\frac{1}{2}} \otimes y^\ast(\cE)$ with the
charges indicated, and of the conjugate bundle with the charges
reversed,
while right-moving zero modes to antiholomorphic sections of 
$\overline{K}_\Sigma^{\frac{1}{2}} \otimes y^\ast(T_{\bY})$ and of the conjugate.

\subsection{$\GU(1)_L\times\GU(1)_R$ anomaly}

At this point we posses all the tools to derive the constraints from
anomaly cancellation.  
For a left-moving fermion $\chi$ taking values in 
$K_\Sigma^{\frac{1}{2}}\otimes Z_{(q,\qb)}$ where $Z\to\P^1$ is a holomorphic
bundle, the net charge carried by
the measure is given by $-c_1(Z)(q,\qb)$.\footnote{This can be seen for
example by using the splitting principle to decompose $Z\to\P^1$ as
a sum of line bundles and applying Riemann-Roch to find the
difference between the number of zero modes of $\chi$ and its
conjugate field, since these are holomorphic sections of the
appropriate bundles.} For
right-moving fermions we can use a hermitian metric on the fibers to
replace a holomorphic bundle with an antiholomorphic one
\cite{Katz:2004nn}, and find that for a right-moving fermion coupled
to $\overline{K}_\Sigma^{\frac{1}{2}}\otimes Z_{(q,\qb)}$ 
the net charge carried by the measure is $c_1(Z)(q,\qb)$.

Under $\GU(1)_L$ the measure in the background
of an instanton will be neutral if 
\begin{align}
\sum_i q_i c_1\left( y^\ast(X_i)\right) + c_1\left(y^\ast(\cE_B)\right)
   -\sum_\Gammat Q_\Gammat c_1\left(y^\ast(\cF_\Gammat)\right) = 0~.
\end{align}
This must hold for arbitrary holomorphic $y:\Sigma\to B$, which by the
Lefschetz theorem for $(1,1)$ forms implies
\begin{align}\label{eq:anomaly1}
\sum_i q_i c_1(X_i) + c_1(\cE_B) -\sum_\Gammat Q_\Gammat c_1(\cF_\Gammat)=0~.
\end{align}
The condition for $\GU(1)_R$ similarly implies
\begin{align}\label{eq:anomaly2}
\sum_i(q_i-1) c_1(X_i) - c_1(TB) - \sum_\Gammat (Q_\Gammat+1) c_1(\cF_\Gammat)=0~.
\end{align}
Subtracting \eqref{eq:anomaly1} from \eqref{eq:anomaly2} we get
\begin{align}
\label{eq:anomaly3}
c_1(\cE_B) + \sum_\Gammat c_1(\cF_\Gammat) + \sum_ic_1(X_i) + c_1(T_B) = c_1(\cE) + c_1(T_{\bY})=0~.
\end{align}
This last equation implies that $\cE\rightarrow \bY$ is Calabi-Yau,
but as noted above, $\bY$ need not be. 
However, it is comforting to note that for the case of a pure (0,2)
NLSM this yields the standard condition.
Indeed, in this case $\bY=B$, while 
$X$ and the superpotential are trivial. In particular, $Q_A=-1$ for all left-moving fermions.
It then follows from \eqref{eq:anomaly1} that $c_1(\cE)=0$, which together with \eqref{eq:anomaly3} yields $c_1(T_{\bY})=0$, as expected.

We can interpret the above constraints as follows: \eqref{eq:anomaly3} is a pure geometric condition 
on the admissible fibration structures; \eqref{eq:anomaly2} instead involves a non-trivial interplay between geometry 
and the charges \eqref{eq:symclasact}, regulating the compatibility
between the fiber LG theory and the underlying geometry.

The last condition corresponds to the mixed $\GU(1)_L-\GU(1)_R$
anomaly, which must vanish if these are to descend to chiral currents
in the IR. This reads
\begin{align}
\label{eq:mixanom}
\sum_A Q_A^2-\sum_i q_i^2=-\sum_A Q_A - \sum_i q_i~.
\end{align}
This is a pure fiber condition and it corresponds to the anomaly
condition of the fiber LG theory\cite{Distler:1993mk,Kawai:1994np}.  In pure LG theories this condition
serves merely to fix the normalization of the charges \cite{Bertolini:2014ela}
but note that in hybrid models \eqref{eq:anomaly2} is not invariant
under rescaling $q,Q$.

\subsection{Left-moving algebras in cohomology and IR physics}
\label{s:IRphys}

In this section we show the existence of representatives for a left-moving Virasoro and
a $\GU(1)_L$ current algebras in the $\bQb$-cohomology of the UV theory \cite{Silverstein:1994ih}.
The operators
\begin{align}
\label{eq:leftmovopers}
\cJ_{L} &= Q_A \cX^A \cXb_A - q_\alpha Y^\alpha P_\alpha~,\nonumber\\
\cT_0 &= -g_{\alpha\betab} \p Y^\alpha \p \Yb^{\betab} - \cX^A D^H \cXb_A 
=-\p Y^\alpha P_\alpha - \cX^A \p \cXb_A
\end{align}
have weights (1,0) and (2,0) respectively. By using the equations of motion \eqref{eq:eomsuperspace} and the chirality of the fields it is easy to show that $\cDb \cT_0=0$, 
while $\cDb \cJ_{L}=0$ holds only when the superpotential satisfies the quasi-homogeneity condition \eqref{eq:quasihom}.
In particular, with our assumptions, $\cDb \p \cJ_L =0$ holds as well, and it is natural to define
\begin{align}
\cT \equiv \cT_0 - \half \p \cJ_L~.
\end{align}
This particular linear combination is fixed by requiring that the $\GU(1)_L$ current in the IR determined by $\cJ_L$ is non-anomalous, 
as in (0,2) LG theories \cite{Melnikov:2009nh}.
Let us evaluate the lowest components\footnote{Since $\cDb$ and $\bQb$ are conjugate operators, the lowest components
of these operators are $\bQb$-closed.} of the operators we just defined. We have
\begin{align}
\label{eq:leftcurrents}
J_L&\equiv \cJ_{L}|_{\theta=0} = Q_A \chi^A  \chib_A - q_\alpha y^\alpha \rho_\alpha~, \nonumber\\
T&\equiv \cT|_{\theta=0} = -\p y^\alpha\rho_\alpha - \chi^A \p \chib_A -\half \p\left[ Q_A \chi^A  \chib_A - q_\alpha y^\alpha \rho_\alpha \right]~.
\end{align}
When the theory possesses (2,2) superconformal symmetry, the generators above 
are supplemented by two additional left-moving operators (supercurrents), completing \eqref{eq:leftcurrents}
to a left-moving $N=2$ algebra. 

Having constructed the left-moving algebras above in $\bQb$-cohomology, 
we can use this structure to probe the most basic low-energy properties of our models. 
Using the free fields OPEs
\begin{align}
\label{eq:freefOPEs}
y^\alpha(z) \rho_{\beta} (w) &\sim \frac1{z-w}\delta^\alpha_\beta~,		&\chi^A(z) \chib_B (w) &\sim \frac1{z-w}\delta^A_B~,
\end{align}
which follow from the action \eqref{eq:compaction} we derive the algebra 
\begin{align}
T(z) T(w)&\sim \frac{d+n-R -3\left(\sum_iq_i+\sum_AQ_A\right)+\ff32\left( \sum_i q_i^2-\sum_A Q_A^2\right) }{(z-w)^4} +\frac{2T(w)}{(z-w)^2}+\frac{\p_w T(w)}{z-w}~,\nonumber\\
T(z) J_L(w) &\sim\frac{J_L(w)}{(z-w)^2}+\frac{\p_w J_L(w)}{z-w}~, \nonumber\\
J_L(z) J_L(w) &\sim\frac{\sum_A Q_A^2-\sum_i q_i^2}{(z-w)^2}~.
\end{align}
This is precisely the structure expected from a Kac-Moody (KM) $\mathfrak{u}(1)$ current of level
\begin{align}
r = \sum_A Q_A^2 - \sum_i q_i^2~,
\end{align}
and an energy-momentum tensor with central charge
\begin{align}
\label{eq:cleft}
c=2d+3r+2(n-R)~,
\end{align}
where we have used \eqref{eq:mixanom}.
We can perform some simple consistency checks to make sure that we recover
the expected results when our hybrid reduces to familiar constructions. 
First of all, for a (2,2) theory we have $n=N$, $R_B=d$ and $Q_i=q_i-1$ so that
\begin{align}
c_{(2,2)}&=3d+3\sum_i(1-2q_i)~.
\end{align}
Also, if $d=R_B=0$, we obtain the known formula for a (0,2) LG theory
\begin{align}
c_{\text{LG}}&=3r_{\text{LG}}+2(n-N)~,		&r_{\text{LG}}&=-\sum_I Q_I-\sum_i q_i~.
\end{align}
Now we can conclude the analysis by 
deriving the right-moving central charge of the IR CFT. To do this we use the fact that the gravitational anomaly is invariant under RG. In the UV it is simply
\begin{align}
\cb_{\text{UV}}-c_{\text{UV}}&=n+d-R~,
\end{align}
which determines in the IR
\begin{align}
\label{eq:cbright}
\cb=3d+3r+3(n-R)~.
\end{align}
Finally, in order for our hybrid model to be suitable for heterotic compactifications, we must have the conditions $c=2D+r$, $\cb=3D$, 
which imply 
\begin{align}
\label{eq:stringcond}
D=d+r+n-R~. 
\end{align}

\section{Massless spectrum}
\label{s:spectrum}

In this section we are going to describe the techniques to compute the massless spectrum for a compactification of the $\GE_8\times\GE_8$ heterotic string based
on a (0,2) hybrid CFT satisfying \eqref{eq:stringcond}, generalizing the methods of \cite{Bertolini:2013xga}.

\subsection*{Heterotic hybrids}

We follow \cite{Gepner:1987vz} to complete our hybrid model to a critical heterotic background.
We supplement our internal (0,2) hybrid theory ($c=2D+r$, $\cb=3D$) by adding a set of free left-moving fermions $\lambda$ realizing a level 1 $\so(16-2r)$ current algebra 
($c=8-r$, $\cb=0$), a left-moving level 1 $\e_8$ current algebra for the hidden sector ($c=8$, $\cb=0$) and the degrees of freedom 
of the extended spacetime $\R^{1,9-2D}$ ($c=10-2D$, $\cb=15-3D$).

We perform separate GSO projections on both left- and right-moving fermion numbers. The left-moving GSO projection is onto $e^{i\pi J_0}(-1)^{F_\lambda}=1$, where 
$J_0$ denotes the $\GU(1)_L$ conserved charge, and $F_\lambda$ is the fermion number for the $\so(16-2r)$ left-moving fermions.
This is responsible for modular invariance and for enhancing the linearly realized $\so(16-2r)\oplus\mathfrak{u}(1)_L$ to the full 
spacetime gauge group $G$ \cite{Distler:1995mi}
determined by
\begin{align}
\xymatrix@R=1mm@C=8mm{
r		&1		&2		&3		&4		&5\\
G		&\e_8	&\e_7	&\e_6	&\so(10)	&\su(5)~.
}
\end{align}
The right-moving GSO projection implies spacetime supersymmetry, and in particular it determines an isomorphism between
the right-moving Neveu-Schwarz (NS) and Ramond (R) sectors. The former
corresponds to spacetime bosons and in particular contains
the topological heterotic ring \cite{Adams:2005tc}, which is the (0,2) generalization of the usual A/B topological rings in (2,2) theories.
The latter contains the massless spacetime fermions which will be the subject of our investigation in this section.

Massless string states are in the kernel of the right-moving Hamiltonian $\Lb_0$, and we use the fact \cite{Kachru:1993pg,Bertolini:2013xga}
that this is isomorphic to the cohomology of $\bQb$ to compute the
spectrum by
exploiting
the left-moving algebra structure derived in section \ref{s:IRphys}.

\subsection*{The orbifold}

While the construction of section \ref{s:model} and \ref{s:anomalies} determines a UV completion of a (0,2) CFT it is not in general suitable for a string compactification,
 as it does not possess integral $\GU(1)_L\times\GU(1)_R$ charges. 
The resolution is well known \cite{Gepner:1987vz,Vafa:1989xc} and amounts to perform an orbifold by
quotienting the hybrid theory by the symmetry generated by $e^{2\pi i J_0}$, where $J_0$ is the charge associated to $J_L$.
In particular, let $q_i=a_i/L_i$ for $a_i,L_i\in\mathbb{\Z}_{>0}$ , then we orbifold the theory by $\Z_L$ where
$L=\lcm(L_1,\dots,L_n)$. The fact that the superpotential satisfies the condition $J^{-1}(0)=B$ implies that including the charges $Q_I$ in the discussion is not necessary.
Moreover, we can combine the orbifold with the left-moving $\Z_2$ GSO projection in a single $\Z_{2L}$ orbifold. 
The orbifold theory consists of
a collection of twisted sectors labelled by an integer $k=0,\dots,2L-1$. Sectors with even $k$ are associated to (R,R) sectors and their study 
is sufficient to determine the $G$-charged content of the theory. Odd $k$ sectors yield additional $G$-neutral states, which correspond
to $G$-preserving first-order deformations of the theory.
Moreover, the spectrum must be invariant under CPT, which exchanges the sectors $k$ and $-k\mod 2L$. Thus, in the following
it will suffice to analyze twisted sectors labeled by $k=0,\dots,L$. 

\subsection*{The Hilbert space in the hybrid limit}

We anticipated above that right-moving zero-energy states are selected by $\bQb$-cohomology. 
By construction, $\bQb$ commutes with the left-moving $\GU(1)_L$ charge and Hamiltonian. These are obtained from the currents \eqref{eq:leftcurrents} as follows
\begin{align}
\label{eq:leftcharges}
J_0&=\oint {dz\over 2\pi i} J_L(z)~,		&L_0&=\oint {dz\over 2\pi i} z T(z)~,
\end{align}
and we can use these to assign charges and weights to all the fields in the theory, which we list in table \ref{t:charges}.
\begin{table}[t]
\begin{center}
\begin{tabular}{|c|c|c|c|c|c|c|c|c|}
\hline
~ 	
				&$y^\alpha$			&$\rho_\alpha$			&$\eta^\alpha$			&$\etab^{\alphab}$		&$\chi^A$			&$\chib_A$  \\ \hline
$\bq$			&$q_\alpha$			&$-q_\alpha$			&$q_\alpha$			&$-q_\alpha$			&$Q_A$			&$-Q_A$	\\ \hline
$2h$				&$q_\alpha$			&$2-q_\alpha$			&$q_\alpha$			&$2-q_\alpha$			&$Q_A+2$ 		&$-Q_A$ \\ \hline
$\bqb$ 			&$q_\alpha$			&$-q_\alpha$			&$q_\alpha-1$			&$-q_\alpha+1$		&$Q_A+1$		&$-Q_A-1$ 
\\ \hline
\end{tabular}
\caption{Weights and charges of the fields.}
\label{t:charges}
\end{center}
\end{table}
The right-moving $\GU(1)_R$ R-charge $\Jb_0$ can be obtained in the UV theory as $\Jb_0=J_0+J_R$, 
where $J_R$ assigns charge $-1$ to $\eta^\alpha$ and $+1$ to $\chi^A$.
In particular, the charges \eqref{eq:leftcharges} induce a grading on the Hilbert space $\cH=\oplus_{\bq,h}\cH_{\bq,h}$, 
and in computing $\bQb$-cohomology we can restrict to each of the subspaces $\cH_{\bq,h}$. We denote this grading as the {\it physical grading}. In particular, 
as we are interested in the massless spectrum, we restrict our attention to $\cH_{\bq,h\leq0}$. Hilbert subspaces with strictly negative left-moving energy are 
paired with the appropriate $\so(16-2r)$ left-moving oscillators in such a way that the total left-moving energy 
vanishes.\footnote{The hidden $\e_8$ is always in the NS vacuum and does not contribute.}

A model based on a non-trivial hybrid geometry exhibits the same features as the more familiar setting
of a compact NLSM: worldsheet instantons wrapping non-trivial cycles in $B$ are expected to generate non-perturbative
corrections to RG-invariant quantities, such as massless spectra, Yukawa couplings, etc.
We then perform these computations in the limit where the K\"ahler class of $B$ lies deep in its K\"ahler cone. 
This defines the {\it hybrid limit}, which identifies a (limiting) point in the K\"ahler moduli space around which
a perturbative expansion in worldsheet instantons is well-defined, as the corrections due to topologically non-trivial  
maps are negligible. We stress that this point lies at infinite distance in the K\"ahler moduli space, and the existence
of such limit is tied to our assumption that the R-symmetry does not act on the base coordinates. 
When this fails, it can be shown \cite{Aspinwall:2009qy,Hori:2013gga} that the hybrid limit is at finite distance in the moduli space and 
corresponds to a singular CFT.

\subsection{Twisted sectors and quantum numbers}

Next we determine, given the algebra above, the quantum numbers of the vacuum state in each twisted sector $k$ relevant for the orbifold $\Z_{2L}$. 
The procedure is familiar: we expand the fields in modes
\begin{align}
\label{eq:fieldmodeexp}
y^\alpha&=\sum_{s\in \Z-\nu_\alpha} y^\alpha_s z^{-s-h_\alpha}~,	&\rho_\alpha&=\sum_{s\in \Z+\nu_\alpha} \rho_{\alpha,s} z^{-s+h_\alpha-1}~,\nonumber\\
\chi^A&=\sum_{s\in \Z-\nut_A} \chi^A_s z^{-s-\htt_A}~,				&\chib_A&=\sum_{s\in \Z+\nut_A} \chib_{A,s} z^{-s+\htt_A-1}~,
\end{align}
where
\begin{align}
\nu_\alpha&=\frac{kq_\alpha}2 \mod 1~,		&\nut_A&=\frac{kQ_A}2 \mod 1~,		&h_\alpha&=\frac{q_\alpha}2~,		&\htt_A&=1+\frac{Q_A}2~,
\end{align}
and such that $0\leq\nu_\alpha<1$ and $-1<\nut_A\leq0$.
We define the Fock vacuum $|k\ra$ to be annihilated by all positive modes, and if there are Fermi zero modes $\chi_0^A$ and $\chib_{A,0}$, we additionally
choose $\chi^A_0|k\ra=0$ for all relevant $A$.
In terms of modes the OPEs \eqref{eq:freefOPEs} read
\begin{align}
\CO{y^\alpha_s}{\rho_{\alpha,r}}&=\delta_{r+s,0}~,		&\AC{\chi^A_s}{\chib_{A,r}}=\delta_{r+s,0}~.
\end{align}
With this set-up at hand, one-point functions of $J_0$ and $\Jb_0$  in the twisted vacuum compute the charges
\begin{align}
\bq_{|k\ra}&=-\frac{r}2 -\sum_\alpha \nu_\alpha q_\alpha+\sum_A\nut_A Q_A~,\nonumber\\
\bqb_{|k\ra}&=-\frac{D}2 +\sum_\alpha \nu_\alpha (1-q_\alpha)+\sum_A\nut_A (Q_A+1)~,
\end{align}
and the one-point function $\la k| T|k\ra$ determines the vacuum energy
\begin{align}
\label{eq:energyvacuum}
E_{|k\ra}&=\half \sum_\alpha \nu_\alpha(1-\nu_\alpha)+\half \sum_A\nut_A(\nut_A+1)~,  &\text{for $k$ even}&~,\nonumber\\
E_{|k\ra}&=-1+\frac{r}8 +\half \sum_\alpha \nu_\alpha(1-\nu_\alpha)+\half \sum_A\nut_A(\nut_A+1)~,  &\text{for $k$ odd}&~. 
\end{align}
The final data we need to determine is how the vacuum transforms over the base. It is in fact a section of the holomorphic line bundle
\begin{align}
L_{|k\ra}&=\otimes_i X_i^{-\nu_i}\otimes\otimes_{\Gammat}\cF_{\Gammat}^{\nut_{\Gammat-m}}\otimes\wedge^{R_B}\cE_B^{-\nut_M}~,
\end{align}
where $\nut_{M}=0$ for $k$ even and $\nut_{M}=-1/2$ for $k$ odd.
By using the anomaly cancellation conditions \eqref{eq:anomaly2} and \eqref{eq:anomaly3} we obtain for $k$ even
\begin{align}
\label{eq:bundlevaceven}
L_{|k\ra}&=\otimes_i X_i^{-\nu_i}\otimes\otimes_{\Gammat}\cF_{\Gammat}^{\nut_{\Gammat-m}}~,
\end{align}
while for $k$ odd
\begin{align}
\label{eq:bundlevacodd}
L_{|k\ra}&=\otimes_i X_i^{-\nu_i+\ff12q_i}\otimes\otimes_{\Gammat}\cF_{\Gammat}^{\nut_{\Gammat-m}-\ff12 Q_{\Gammat}}~.
\end{align}

\subsection{$\bQb$-cohomology}
\label{ss:spetrseq}

The technique we use to compute the massless spectrum was first developed in \cite{Kachru:1993pg} in the context of LGOs and extended to (2,2) hybrid
models in \cite{Bertolini:2013xga}. 
We review this briefly here, emphasizing a few technically challenging
issues new to our models.
From the structure of the right-moving $N=2$ algebra
\begin{align}
\bQ^2 &= \bQb^2 =  0~, &\AC{\bQ}{\bQb} &= -2\pb_{\zb}~,
\end{align}
it follows that the kernel of $\pb_{\zb}$ is isomorphic to the cohomology of $\bQb$. 
Working at fixed $\bq,h$, the object of interest is the cohomology of the complex
\begin{align}
\label{eq:totalcohomQ}
\xymatrix@R=0mm@C=8mm{
\cdots \ar[r]   &\cH^{\bqb-1} \ar[r]^-{\bQb} &\cH^{\bqb} \ar[r]^-{\bQb} &\cH^{\bqb+1} \ar[r]^-{\bQb} & \cdots~.
}
\end{align}
Moreover, $\cH^{\bqb}$ admits a second grading $u$ which assigns
charge $+1$ to $\etab$ and $-1$ to $\eta$ . Since we have a
decomposition $\bQb=\bQb_0 + \bQb_J$, where
$\bQb_0\equiv \bQb\big|_{J=0}$ and $\bQb_J$ contains all the
dependence on the superpotential, and since these satisfy
\begin{align}
\bQ_0^2 &= \bQb_J^2=\AC{\bQb_0}{\bQb_J} =0~,
\end{align}
it follows that $\bQb_0$ and $\bQb_J$ act respectively as vertical and horizontal differentials on the double graded complex $\cH^{p,u}$, where $p=\bqb-u$.
The total $\bQb$-cohomology \eqref{eq:totalcohomQ} is then computed by a spectral sequence $E_l^{p,u}$ equipped with differentials
\begin{align}
d_l : E_l^{p,u} \rightarrow E_l^{p+l,u+1-l}~,
\end{align}
such that $E_{l+1}=H_{d_l}(E_l)$ \cite{Bott:1982df}. Convergence of the spectral sequence is ensured by the fact that $u\leq d$ implies $d_l=0$ for $l\geq d+2$.
Recall that the non-trivial action of $\bQb$ on the fields \eqref{eq:fieldcontent} restricted to $\cH_{\bq,h}$ is
\begin{align}
\label{eq:Qbaraction}
\bQb_0\cdot \yb^{\alphab}&=-\etab^{\alphab}~,	&\bQb_0\cdot \eta^{\alpha}&=-\pb y^{\alpha}~,	&\bQb_J\cdot \chib_A&=J_A~,	&\bQb_J\cdot \rho_{\alpha}&=\chi^AJ_{A\alpha}~.
\end{align}
Since the right-moving fields are always restricted to the R sector, we can restrict to zero modes, and we choose the Ramond ground state to be annihilated by the zero modes of $\eta^\alpha$. 
Thus, for all practical purposes we can drop $\eta^\alpha$ henceforth and restrict to zero modes for $\etab^{\alphab}\equiv \etab^{\alphab}_0$. 
The first equation in \eqref{eq:Qbaraction} 
implies that $\bQb_0$-cohomology is equivalent to Dolbeault
cohomology, and since $\bY$ retracts to $B$ this is given by
horizontal forms. In particular, we can represent the 
action of the supercharges as follows
\begin{align}
d_0=\bQb_0 &= -\etab^{\mub} {\p\over \p\yb^{\mub}}~,		&d_1=\bQb_J&=\oint {dz\over 2\pi i} [\chi^AJ_A](z)~.
\end{align}
While conceptually this framework parallels the (2,2) case, there is a substantial  
technical difference, which we mention here and we will tackle later in this section.
As pointed out in \cite{Bertolini:2013xga} the cohomology groups 
$H^\bullet_{\bQb_0}(\bY,\cH_{\bq,h})$ are generically infinite-dimensional due to the 
non-compact geometry of $\bY$. 
The strategy to obtain a well-defined counting problem is the following:
\begin{enumerate}
\item Even at $J=0$ we can still keep track of the weights and charges of the fields, using the quantities \eqref{eq:leftcharges}. 
It turns out that the geometric structure defined in this work admits a {\it coarse grading} $\bR$ -- which generalizes
the fine grading $\br$ assigning the grade vector $(r_1,\dots,r_n)$ to the fiber monomial $\prod_i (\phi^i)^{r_i}$ -- which is a refinement of the physical grading $\bq,h$.
In particular, only a finite number of these will contribute to the total $\bQb$-cohomology for $h\leq0$.
\item The graded cohomology groups $H^\bullet_{\bR}(\bY,\cH_{\bq,h})$ can be evaluated in terms of cohomology groups 
over the base, which are easier to compute. 
\end{enumerate}
We will develop the relevant techniques to evaluate sheaf cohomology over $\bY$ of various bundles constructed from $\cE$ 
(and therefore all relevant elements in $\cH_{\bq,h}$) in section \ref{ss:sheafy}.

\subsection{Geometry of massless states}
\label{ss:geomstates}

In this section we 
determine the geometric structure describing the massless states of a (0,2) hybrid model in a given twisted sector labeled by an integer $k$.
From the discussion above, a candidate for an element in $\cH$, that is, the most general $\bQb_0$-closed combination of fields of the theory,
takes the form
\begin{align}
\label{eq:generalstates}
S(y^\alpha;\yb^\mu)^{B_1\cdots B_v\alpha_1\cdots\alpha_t}_{A_1\cdots A_s;\mub_1\cdots\mub_u} \rho_{\alpha_1}\cdots\rho_{\alpha_t}\chi^{A_1}\cdots\chi^{A_s}\chib_{B_1}\cdots\chib_{B_v}\etab^{\mub_1}\cdots\etab^{\mub_u}|k\ra~.
\end{align}
$S$ is taken to be polynomial, hence holomorphic, in the fiber
coordinates as indicated.
The geometric properties of the states \eqref{eq:generalstates} are derived from how 
the various fields transform across patches. Let $\{\mathfrak{U}\}$ be a cover of $\bY$ and 
let $\mathfrak{U}_{a,b}$ be two patches with local coordinates $y^{\alpha}_{a,b}$ such that $\mathfrak{U}_a\cap\mathfrak{U}_b\neq\emptyset$. Then we have for the left-moving fields
\begin{align}
\label{eq:transfunctfields}
y^\alpha_b&=y^\alpha_b(y_a)~,		&\rho_{b\alpha}&=:(\cT^{-1}_{ba})^\beta_\alpha\left[\rho_{a\beta}-(S_{ba})_{\beta A}^B \chib_{aB}\chi^A_a \right]:~,\nonumber\\
\chi^A_b&=\chi^B_a(G_{ba})^A_B~,		&\chib_{bA}&=(G^{-1}_{ba})^B_A \chib_{aB}~,
\end{align}
where the transition functions $G_{ba}$ for the bundle $\cE$ are defined in \eqref{eq:Etransfuncts}, while 
$\cT_{ba}{}^{\alpha}_\beta \equiv {\p y_b^{\alpha}\over \p y^\beta_a}$
are the transition functions for the tangent bundle $T_{\bY}$
and $(S_{ba})_{\beta A}^B \equiv (G_{ba}^{-1})^B_C(G_{ba})^C_{A,\beta}$.
We recall that the normal ordering in the definition of $\rho_\alpha$
is needed to cure the divergencies in the OPEs \eqref{eq:freefOPEs}. 
More importantly, the Fermi bilinear term in the definition of $\rho_\alpha$ implies
that $\rho_\alpha$ does not transform as a section of a bundle on $\bY$.
It is possible however to build well-defined states, i.e., states that patch as sections of a holomorphic bundle over $\bY$,
by replacing $\rho_{\alpha}\rightarrow \rho_{\alpha} - \Gamma_{\alpha B}^A \chib_A\chi^B$ in \eqref{eq:generalstates}, 
where $\Gamma_{\alpha B}^A$ is the Hermitian connection. 
However, these states are in general not $\bQb_0$-closed, since
\begin{align}
\label{eq:bQb0closdrho}
\bQb_0 \cdot \left( \rho_{\alpha} - \Gamma_{\alpha B}^A \chib_A\chi^B\right) = -\cF_{\Mb \alpha B}^A  \chib_A\chi^B \etab^{\Mb}~.
\end{align}
The geometric interpretation of this fact is the appearance of obstructions to the states being massless. 
In (2,2) models such obstructions are required to vanish in order for
CPT invariance to hold as a symmetry in $\bQb_0$-cohomology
\cite{Bertolini:2013xga}. 
In (0,2) models there can be obstructions compatible with CPT
invariance, although we do not have an example in which these are nontrivial. We describe these in detail in 
section \ref{ss:obstructions}. We now turn to a description of the geometry in various twisted sectors.

In the untwisted (R,R) sector ($k=0$), the vacuum $|0\ra$ is characterized by
\begin{align}
E_{|0\ra}&=0~,		&(\bq_{|0\ra},\bqb_{|0\ra})&=(-r/2,-D/2)~,		&L_{|0\ra}&=\cO_B~.
\end{align}
Thus, we can simply restrict to zero modes for all the fields. With our conventions, this implies that $\rho_\alpha$ and $\chi^A$
drop out of the Hilbert space and the relevant states in $\bQb_0$-cohomology are 
associated to the wave-functions
\begin{align}
\label{eq:generalstatesk0}
S(y^\alpha;\yb^\mu)^{B_1\cdots B_v}_{\mub_1\cdots\mub_u}\in H^u_{\pb}\left(\bY,\wedge^v \cE\right)~.
\end{align}

In the untwisted (NS,R) sector ($k=1$), 
anomaly cancellation yields
\begin{align}
E_{|1\ra}&=-1~,		&(\bq_{|1\ra},\bqb_{|1\ra})&=(0,-D/2)~,		&L_{|1\ra}&=\cO_B~.
\end{align}
Here, the fields transform as prescribed in \eqref{eq:transfunctfields} 
and the relevant geometry for describing the states \eqref{eq:generalstates}
is simply given by $\bY$ and appropriate powers (and duals)
of the bundles $\cE\rightarrow \bY$ and $T_{\bY}\rightarrow \bY$.
In particular, assuming the obstructions \eqref{eq:bQb0closdrho} to vanish we have that upon taking 
$\bQb_0$-cohomology on the states \eqref{eq:generalstates} we obtain
\begin{align}
S(y^\alpha;\yb^\mu)^{B_1\cdots B_v\alpha_1\cdots\alpha_t}_{A_1\cdots A_s;\mub_1\cdots\mub_u} \in H^u_{\pb}\left(\bY,\wedge^v \cE \wedge^s \cE^\ast \otimes \Sym^t(T_{\bY})\right)~.
\end{align}
In general obstructions might not vanish, and the action of $\bQb_0$ must be corrected appropriately, as we discuss in detail below. 

In twisted sectors ($k\geq2$), two new features arise in our (0,2) models. 
First, as it can be observed from \eqref{eq:energyvacuum}, sectors for $k\in2\Z$ can have $E_{|k\ra}<0$ and it does not
suffice to reduce to zero modes. Second, the splitting of the coordinates into {\it light} modes -- which organize themselves in a non-trivial geometry -- 
and {\it heavy} modes -- which instead transform simply as pullbacks -- must be generalized for (0,2) models.
A great simplification comes from the fact that we always have $E_{|k\ra}>-1$. 
Thus the massless states are contained in 
the subspace of the Hilbert space
obtained by exciting only the lowest energy modes of the left-moving fields.
In particular, we truncate the expansions \eqref{eq:fieldmodeexp} to $s\in (-1,1)$, that is, for the bosonic fields
\begin{align}
y^\mu&\equiv y^\mu_{0}~,		&\phi^i&\equiv\phi^i_{-\nu_i}~,				&\rho_i^\dagger&\equiv \phi^i_{1-\nu_i}~,		
&y^\mu{}^\dagger&\equiv\rho_{\mu,0}~,	&\phi^i{}^\dagger&\equiv\rho_{i,\nu_i}~,		&\rho_i&\equiv\rho_{i,\nu_i-1}~,
\end{align}
while for the left-moving fermions we distinguish between $k\in2\Z$ (where $\nut_M=0$)
\begin{align}
\chib_M^\dagger&\equiv\chi^M_{0}~,			&\chib_I^\dagger&\equiv\chi^I_{-\nut_I}~,		& \chi^I&\equiv\chi^I_{-1-\nut_I}~,		&\chib_M&\equiv\chib_{M,0}~,	
&\chib_I&\equiv\chib_{I,\nut_I}~,			&\chi^I{}^\dagger&\equiv\chib_{I,1+\nut_I}~,
\end{align}
and $k\in2\Z+1$ (where $\nut_M=-1/2$)
\begin{align}
\chib_A^\dagger&\equiv\chi^A_{-\nut_A}~,		&\chi^A&\equiv\chi^A_{-1-\nut_A}~,		&\chib_A&\equiv\chib_{A,\nut_A}~,		
&\chi^A{}^\dagger&\equiv\chib_{A,1+\nut_A}~.	
\end{align}
Next, we expand $G^A_B$ in \eqref{eq:transfunctfields} in terms of these modes, and we obtain
\begin{align}
\chi^I_b=C^I_{ab}\chi^M_a + g^{I}_{ab}\chi^I_{a} = \sum_{\br\in\Delta_{I+m}} \mathsf{M}_{\br} f_{ab,\br}^I\chi^M_a + g^{I}_{ab}\chi^I_{a}~.
\end{align}
The term defined by each monomial $\mathsf{M}_{\br}=\prod_i (\phi^i)^{r_i}$ 
contributes in the twisted sector, i.e., it is not projected out by the truncation to lowest-energy modes, 
if and only if the energy contributions of the various terms match, that is,
\begin{align}
\label{eq:condtwistkev}
1+\nut_I&= 1+\nut_M +\sum_i r_i\nu_i~. 
\end{align}
In other words, even if all the modes $\phi^i$ in $\mathsf{M}_{\br}$ survive the projection, the monomial itself
might not contribute to the transition functions for the bundle $\cE$, and it does so if and only if \eqref{eq:condtwistkev} is satisfied.

For $k\in2\Z$, in particular, $\nut_M=0$, thus we have $\nut_I= \sum_i r_i\nu_i$.
However, $\nut_I \leq0$ while $ \sum_i r_i\nu_i$ is a sum of non-negative quantities, thus \eqref{eq:condtwistkev}
is satisfied only if $\nut_I= r_i\nu_i=0$.
This implies that we can distinguish 
the fields as (base,zero modes,heavy) coordinates $y^\alpha$ 
according to $\alpha=(\mu,i',\iota)$ such that $\nu_{i'}=0$ and $\nu_\iota>0$.
Thus, base and fiber zero modes are coordinates on a sub-bundle $\bY_k\subseteq \bY$ defined as
\begin{align}
\bY_k:\tot\left( \oplus_{i'}X_{i'} \overset{\pi_k}{\longrightarrow} B\right)~,
\end{align}
while the heavy coordinates transform as pullbacks $\pi_k^\ast (X_\iota)$.
Note that the modes $\rho_{i'}$ cannot contribute to the massless spectrum in these sectors as they carry weight $h=1$, while
 $\rho_{\iota}$ transform as sections of the pullback bundle $\pi_k^\ast (X_\iota^\ast)$ over $\bY_k$, 
 thus not needing any Fermi bilinear correction.
This shows already that obstructions of the form \eqref{eq:bQb0closdrho} are absent in sectors $k\in2\Z$.
For the Fermi modes, we have a similar splitting into $A=(M,I',\Lambda)$, where $\nut_{I'}=0$ and $\nut_{\Lambda}<0$.
Indeed, \eqref{eq:condtwistkev} can only be satisfied for $A=I'$ and for $\mathsf{M}_{\br}=\prod_{i'} (\phi^{i'})^{r_{i'}}$, that is, all the
monomials $\mathsf{M}_{\br}$ are restricted to zero modes.
Thus, the zero modes $\chi^{M,I'}$ organize themselves as a sub-bundle $\cE_k\rightarrow \bY_k$ of $\cE$ defined by the SES
\begin{align}
\label{eq:EkSESdef}
\xymatrix@R=0mm@C=8mm{
0\ar[r] &\oplus_{I'} \pi_k^\ast \cF_{I'+m}    \ar[r] &\cE_k   \ar[r] & \pi_k^\ast \cE_B \ar[r] &0~,
}
\end{align}
with transition functions defined by the projection onto zero modes of \eqref{eq:Etransfuncts}.
The heavy modes $\chi^\Lambda$ transform instead as sections of the pullback bundles $\pi_k^\ast (\cF_\Lambda)$. 
We summarize these in the following table:
\begin{align}
\xymatrix@R=0mm@C=5mm{
\text{modes}		&y^{\mu,i'}		&y^\iota			&\rho_\iota				&\chi^{M,I'}		&\chi^\Lambda				&\chib_{M,I'}		&\chib_A\\
\text{bundle}		&\bY_k		&\pi_k^\ast X_\iota	&\pi_k^\ast(X_\iota^\ast)		&\cE_k			&\pi^\ast_k\cF_{\Lambda+m}	&\cE^\ast_k		&\pi^\ast_k\cF^\ast_{\Lambda+m}
}
\end{align}

In $k>1$ odd sectors, $\nut_M=-1/2$, and in this case \eqref{eq:condtwistkev} reads
\begin{align}
\label{eq:condtwistkodd}
\nut_I- \sum_i r_i\nu_i=-\frac12~.
\end{align}
Thus, we split the fields as (base,light,heavy) modes,
generalizing the same notation we adopted for the even $k$ case, such that $\nu_{i'} \leq 1/2$ and $\nu_{\iota} > 1/2$. In fact,
any monomial $\mathsf{M}_{\br}$ with a non-trivial dependence on
any of the heavy coordinates $\phi^\iota$ cannot satisfy \eqref{eq:condtwistkodd}.
Note that in $k>1$ sectors of sensible models 
$E_{|k\ra}>-1$ and we can again suppress the 
base oscillator $\rho_\mu$ dependence. 
Moreover, $\rho_\iota$ will transform again as sections of the pullback bundles $\pi_k^\ast (X_\iota^\ast)$.
For the Fermi modes, 
we have a similar splitting into $A=(M,I',\Lambda)$ such that \eqref{eq:condtwistkodd} is satisfied 
for some $\br\in\Delta_{I'+m}$ and is never satisfied for $\br\in\Delta_{\Lambda+m}$. 
The light fields $\chi^{M,I'}$ then organize themselves as a sub-bundle $\cE_k\rightarrow \bY_k$ of $\cE$ defined by the SES \eqref{eq:EkSESdef},
whose transition functions are obtained from \eqref{eq:Etransfuncts} 
by projecting onto $\mathsf{M}_{\br}$ such that \eqref{eq:condtwistkodd} holds.
As before, the heavy fields instead transform as pullbacks $\pi_k^\ast (\cF_{\Lambda+m})$
and the modes $\chib_A$ transform in the dual of the corresponding bundles.
Finally, we need to pair the light modes $\rho_{i'}$ to the Fermi bi-linear using a connection on $\cE_k$, and the well-defined 
combination will transform as a section of $T_{\bY_k}$. We summarize again all the relevant modes and their transformation properties in a 
generic twisted sector $k\in 2\Z_{\geq0}+3$:
\begin{align}
\xymatrix@R=0mm@C=1mm{
\text{modes}		&y^{\alpha,i'}		&y^\iota			&\rho_{i'}-\cA_{i'(M,I')}^{(N,J')}\chib_{N,J'}\chi^{M,I'}		&\rho_\iota				&\chi^{M,I'}		&\chi^\Lambda			&\chib_{M,I'}		&\chib_\Lambda\\
\text{bundle}				&\bY_k			&\pi_k^\ast X_\iota	&T_{\bY_k}									&\pi_k^\ast(X_\iota^\ast)		&\cE_k			&\pi^\ast_k\cF_{\Lambda+m}	&\cE^\ast_k		&\pi^\ast_k\cF^\ast_{\Lambda+m}~.
}
\end{align}
Thus, from the discussion above, we conclude that in a given twisted sector the states \eqref{eq:generalstates} are organized as $(0,u)$-horizontal forms valued in the bundle
\begin{align}
B^k_{r,s,t}&\equiv \wedge^r \cE_k\wedge^s \cE^\ast_k \otimes \Sym^t(T_{\bY_k})\otimes L_{|k\ra}^\ast~,
\end{align} 
tensored with an appropriate pullback bundle $\pi_k^\ast(\cF_H)$ which takes into account the heavy fields. 
The goal of section \ref{ss:sheafy} is to show how to compute the 
cohomology groups $H^\bullet(\bY_k,B^k_{r,s,t}\otimes \pi_k^\ast(\cF_H))$.

\subsection{(0,2) sheaf cohomology}
\label{ss:sheafy}

We now turn to the description of the techniques to compute the graded cohomology groups relevant
for the massless spectrum analysis. In particular, we will focus on the cohomology groups $H^\bullet(\bY,\cE)$.
The analysis can be straightforwardly generalized to the dual bundle $\cE^\ast$ and more general 
tensor (and wedge) products, thus covering all the relevant elements in $\cH_{\bq,h}$.

Our starting point is the fine grading $\br\in\Z^{\oplus n}$, which we recall assigns to each fiber monomial $\prod_i (\phi^i)^{r_i}$ the vector $\br=(r_1,\dots,r_n)$. 
In the case $\cE=T_{\bY}$ it has been shown in \cite{Bertolini:2013xga} that the SES defining $T_{\bY}$ can be graded with respect to $\br$
as follows
\begin{align}
\label{eq:22TYcase}
\xymatrix@R=0mm@C=8mm{
0\ar[r] &\oplus_{i} (\pi^\ast X_i)_{\br+\bx_i}   \ar[r] & (T_{\bY})_{\br}   \ar[r] & (\pi^\ast T_B)_{\br} \ar[r] &0~,
}
\end{align}
where $(\bx_i)_j=\delta_{ij}$. From this form, by passing to the induced LES,
it is possible to reduce the computation to cohomology groups over the base.
Generically, there is an obstruction in doing so when $\cE\neq T_{\bY}$. 
The sections $E^{\Gammat}$ in \eqref{eq:Eexplexpr} do not respect in general the fine grading $\br$,
as different monomials consistent with \eqref{eq:VactionEs} are
associated to different fine grades.
We then start by defining
\begin{align}
\label{eq:DeltaI2def}
\Delta'_{\Gammat} = \Big\{ \br\in \Delta_{\Gammat} \Big| S^{\br}_{\Gammat}\neq0 \Big\}~.
\end{align}
This selects the monomials that do appear (with non-zero coefficient)
in the maps \eqref{eq:Eexplexpr}.
It is also convenient to define
\begin{align}
\delta_{\Gammat}=\Big\{ \br_1-\br_2, \forall \br_{1,2}\in\Delta'_{\Gammat} \Big\}~,
\end{align}
and let $\Pi_{\Gammat}$ be the (codimension at least one) 
sublattice of $\Z^{\oplus n}$ generated by the elements of $\delta_{\Gammat}$.
More generally, let $\Pi$ be the sublattice of $\Z^{\oplus n}$
generated by the elements of $\delta_{\Gammat}$,
$\Gammat=P+1,\dots,R+m$. 
While the fine grading by $\br$ is not preserved by the transition
functions, it is clear by construction that the coarser grading by 
\begin{align}
\label{eq:coarsedef}
\bR\in\frac{\Z^{\oplus n}}{\Pi}~,
\end{align}
is preserved; we denote this the {\it coarse grading}.

Now, let $H_{\bq}\subset \R^n$ be the ``charge $\bq$" hyperplane defined by the relation $\sum_i q_i x_i=\bq$. Then we have that $\Pi\subseteq H_0\cap\Z^{\oplus n}$.
We see therefore the relation between the various gradings introduced so far:
\begin{itemize}
\item If $\Pi=\emptyset$, the bundle $\cE$ splits as a sum of line bundles. More generally, 
if $\Delta_{\Gammat}$ is empty, $\delta_{\Gammat}$ is empty as well and we can write $\cE=\cE'\oplus\pi^\ast\cF_{\Gammat}$ for some $\cE'$. 
\item If $\Pi=\{\mathbf{0}\}$, then $\bR$ reduces simply to the fine grading $\br$. This occurs when $\Delta_{\Gammat}=\{ \widetilde\br_{\Gammat} \}$ is 
one-dimensional for all $\Gammat$. 
\item If $\Pi=H_0\cap\Z^{\oplus n}$, then $\bR=H_{\bq(\bR)}\cap\Z^{\oplus n}$ is the physical grading, where $\bq(\bR)=\bq(\br)$ for $\bR=[\br]$. 
In this case, the energy grading simply follows from the charge grading
since $q_i=2h_i$ for $\phi^i$.
\item If $\Pi\subset H_0\cap\Z^{\oplus n}$, and if $\Pi$ is non-trivial (in the sense that it is not empty and it does not consist only of the zero element), 
then $\bR\subset H_{\bq(\bR)}\cap\Z^{\oplus n}$, and it is apparent that this is a refinement of the physical grading. 
This occurs whenever $\Delta_{\Gammat}$, for some $\Gammat$, contains at least two elements.
This is precisely the case we are interested in solving.
\end{itemize}

Note that in general each conjugacy class in \eqref{eq:coarsedef} is
infinite as a subset of $\Z^{\oplus n}$.  Our interest is in sections with local polynomial
dependence on the fiber coordinates,  so we 
define the {\it truncated coarse grading class} as the intersection of the class $\bR$ with the positive orthant,
$\bR_+\equiv \bR \cap \Z^{\oplus n}_+$.  This will contain a finite
number of points (monomials).

Finally, we show how to compute such graded cohomology groups.
From the form of \eqref{eq:Etransfuncts} it follows that we can grade \eqref{eq:EasExt} as
\begin{align}
\label{eq:EasExtfinegrad}
\xymatrix@R=0mm@C=8mm{
0\ar[r] &\oplus_{\Gammat} \oplus_{\br\in(\bR+\Delta'_{\Gammat})_+} (\pi^\ast \cF_{\Gammat} )_{\br}   \ar[r] & (\cE)_{\bR}   \ar[r] & \oplus_{\br\in\bR_+}(\pi^\ast \cE_B)_{\br} \ar[r] &0~,
}
\end{align}
where
\begin{align}
\bR+\Delta'_{\Gammat} = \Big\{ \br_1+\br_2 \big| \br_1\in\bR, \br_2\in\Delta'_{\Gammat} \Big\}\subseteq H_{\bq(\bR)+Q_{\Gammat}+1}\cap \Z^{\oplus n}~.
\end{align}

Finally, the graded cohomology groups $H^\bullet_\bR(\bY,\cE)$ can be promptly computed by passing to the LES associated to \eqref{eq:EasExtfinegrad},
and using the fact that for a pullback bundle $H_{\br}^\bullet(\bY,\pi^\ast \cF)=H^\bullet(B,\cF\otimes \mathbb{L}_{\br})$ where
$\mathbb{L}_{\br}\equiv \otimes_i (X_i^\ast)^{r_i}$.\footnote{For a proof of this fact we refer to appendix C of \cite{Bertolini:2013xga}.}

An example will illustrate this best. 

\subsection*{The coarse grading: an example}

Let us take $\bY=\tot\left(\cO(-1)^{\oplus2}\rightarrow \P^1 \right)$ and 
\begin{align}
\label{eq:tandisg}
E_1 &= a\phi^1 + b\phi^2~,		&E_2&=c\phi^1+d\phi^2~.
\end{align}
Here $\phi^i$, $i=1,2$ are the fiber coordinates of $\bY$ and $a,b,c,d\in\C$. This particular choice for the map $E$ forces $\cF_I = \cO(-1)$ for $I=i=1,2$, and we
choose $\cE_B=T_B=\cO(2)$. Since we are not going to build a full hybrid model out of this geometric set-up, only for this example we do not follow our
conventions for the indices as in the rest of the work.
Note that upon performing a $\GL(2,\C)$ field redefinition of the $\phi$'s, it is possible to diagonalize the maps \eqref{eq:tandisg}, 
recovering the (2,2) form $E_i=\phi^i$. Hence, the bundle $\cE$
defined by \eqref{eq:tandisg} is isomorphic to the tangent sheaf $T_{\bY}$. 
We are going to compute the sheaf cohomology of this (0,2) disguised tangent bundle
and check our answer with the known result.
If we assume that $a,b,c,d\neq0$ then \eqref{eq:tandisg} implies $\Delta'_i =\{ (1,0),(0,1)\}$ and $\delta_i=\{ (0,0),\pm(1,-1) \}$.
Thus, $\Pi=\{(r_1,r_2)\in \Z^{\oplus2} | r_2=-r_1 \}$
and \eqref{eq:coarsedef} determines the coarse grading $\bR$ by an integer $L$ as follows
\begin{align}
\bR_L &=\big\{ (r_1,r_2)\in \Z^{\oplus2} | r_1+r_2 = L \big\}~.
\end{align}
Now, for a fixed grading $\bR_L$ we then have
\begin{align}
\bR_L + \Delta'_i = \Big\{ (r_1,r_2) \big| r_1+r_2=L+1 \Big\}=\bR_{L+1}~.
\end{align}
Thus, \eqref{eq:EasExtfinegrad} reads
\begin{align}
\xymatrix@R=0mm@C=8mm{
0\ar[r] & (\pi^\ast \cO(-1) )^{\oplus2}_{\br\in(\bR_{L+1})_+}   \ar[r] &(\cE)_{\bR_L}   \ar[r] &(\pi^\ast \cO(2))_{\br\in(\bR_L)_+} \ar[r] &0~.
}
\end{align}
In this case the grading bundle is $\mathbb{L}_{(r_1,r_2)}=\cO(1)^{ r_1}\otimes\cO(1)^{ r_2}=\cO(r_1+r_2)$. We can pass to the induced LES
\begin{align}
\xy {\ar(0.05,-10)*{};(0.05,-11)*{}}; 
\xymatrix@R=8mm@C=0mm{ 
0 \ar[r] &H^0_{(\bR_{L+1})_+}(B,\cO(L))^{\oplus2} \ar[r] & H_{\bR_L}^0 (\bY,\cE) \ar[r] &H_{(\bR_L)_+}^0(B,\cO(L+2)) 
\ar@{-} `d[l]`[llld]  \\
H^1_{(\bR_{L+1})_+}(B,\cO(L))^{\oplus2} \ar[r] & H^1_{\bR_L} (\bY,\cE) \ar[r] & H^1_{(\bR_L)_+}(B,\cO(L+2)) \ar[r] & 0 
}
\endxy
\end{align}
where 
\begin{align}
\label{eq:Hdefsumsub}
H^\bullet_{(\bR_{L'})_+}(B,\bullet)\equiv\oplus_{\br\in(\bR_{L'})_+} H^\bullet (B,\bullet)=H^\bullet (B,\bullet)^{\oplus  |(\bR_{L'})_+|}~. 
\end{align}
If $L\geq0$ we then have
\begin{align}
\label{eq:disgtangresult}
H^0_{\bR_L} (\bY,\cE)&=\C^{3L^2+10L+7}~,			&H^1_{\bR_L} (\bY,\cE)&=0~,
\end{align}
and zero otherwise.
We recall that for the diagonal form of the $E$'s we have
\begin{align}
H^0_{(r_1,r_2)} (\bY,T_{\bY})&=\C^{3(r_1+r_2)+5}~,		&H^1_{(r_1,r_2)} (\bY,T_{\bY})&=0~,
\end{align}
while if $r_1=-1$
\begin{align}
H^0_{(-1,r_2)} (\bY,T_{\bY})&=\C^{r_2}~,		&H^1_{(-1,r_2)} (\bY,T_{\bY})&=0~,
\end{align}
and similarly for $r_2=-1$. All other graded cohomology group vanish.
Considering the appropriate sum of these cohomology groups we obtain
\begin{align}
 \oplus_{r_1+r_2=L} H^0_{(r_1,r_2)} (\bY,T_{\bY})&=\C^{3L^2+10L+7}~,			& \oplus_{r_1+r_2=L} H^1_{(r_1,r_2)} (\bY,T_{\bY})&=0~,
\end{align}
which agrees as expected with \eqref{eq:disgtangresult}.

\subsection{Obstructions and CPT}
\label{ss:obstructions}

Consistent physical theories must exhibit CPT as a symmetry. 
In our (0,2) models, it follows from energy and charge considerations that 
a state in the $k$-th sector with charges $(\bq,\bqb)$ is paired with a state in the $2L-k$ sector with charges $(\bq,-d-\bqb)$ 
by replacing the fields/modes 
\begin{align}
\label{eq:CPTpairing}
y^i &\leftrightarrow \rho_i~,		&\chi^A&\leftrightarrow \chib_A~,
\end{align}
and recalling that the quantum numbers for the conjugate vacua satisfy
\begin{align}
(\bq_{|2L-k\ra},\bqb_{|2L-k\ra})&=(-\bq_{|k\ra},-\bqb_{|k\ra}-d)~,		&L_{|k\ra}\otimes L_{|2L-k\ra}&=K^\ast_B~.
\end{align}
Whenever $\bQb_0=\pb$, Serre duality on the base ensures that this pairing descends to $\bQb_0$-cohomology.
As shown in \eqref{eq:bQb0closdrho}, this is not always the case, signaling the possibility of
non-trivial obstructions. In (2,2) models, it has been argued that such obstructions would lead to 
a violation of CPT, therefore these had to be required to vanish in a physically relevant model. This is not
necessarily the case outside the realm of (2,2) hybrids, and we 
now turn to the description of a class of obstructions that potentially arise in our (0,2) models.

The case where the hybrid formalism is employed to describe the massless spectrum
of a (0,2) NLSM with a CY${}_3$ manifold target space equipped with a left-moving $\SU(n)$ stable bundle $\cV$ has been already 
worked out in detail in \cite{Bertolini:2013xga}. Thus, we consider here hybrid models where the
orbifold action $\Z_{2L}$ is non-trivial. This implies, in particular, that the CPT conjugate twisted sectors 
$k=1$ and $k=-1$ do not coincide, and the class of obstructions we are going to discuss
will involve gauge singlets arising in these two sectors.
For ease of exposition, only in this section we make a few simplifying assumptions: we assume
$\cb=9$ and that the charges of the fields satisfy the bounds $q_i<1/2$ and $Q_A<-1/2$.  
The motivation behind the first assumption is that it allows us to be more explicit in our
presentation and it corresponds to the phenomenologically more interesting class of models. 
Second, fields which do not satisfy the bounds on the charges above give rise to extra massless states, 
which however do not take part in the class of obstructions we are going to describe. 
Moreover, a model with $d=3$ is necessarily equipped with a $\cb=0$ fiber LG theory, in
which all the fiber fields are massive in the sense that the superpotential is linear in the fiber fields and the above bounds
on the charges cannot be satisfied. This means that, in our current analysis,
we can restrict our attention to the case $d\leq2$.

From the general discussion above, all the fields in the $k=1$ sector are light, and well-defined states 
transform as sections of the appropriate bundles over $\bY$. In the $k=-1$ sector instead, all the fields are
heavy, and the geometry just reduces to the base. The vacua have the following universal quantum numbers
\begin{align}
E_{|1\ra}&=E_{|-1\ra}=-1~,	&\bq_{|1\ra}&=\bq_{|-1\ra}=0~,	&\bqb_{|1\ra}&=-\bqb_{|-1\ra}-d=-\frac32~.	
\end{align}
Moreover, while $|1\ra$ transforms as a section of the trivial bundle over the base, 
the conjugate vacuum transforms non-trivially, $|-1\ra\in\Gamma(K^\ast_B)$.

With this set-up, we can list all the states that can give rise to singlets in both sectors. Starting with $k=1$, we have 
\begin{align}
\xymatrix@R=0mm@C=5mm{
	&\Psi^1_\alpha \p_z y^\alpha |1\ra	&\Psi^{2 \alpha} (\rho_\alpha - \cA_{\alpha B}^A \chib_A\chi^B)|1\ra	&\Psi^{3A}_B\chib_A\chi^B|1\ra		&\Psi^4_A\chi^A|1\ra\\
\bq	&0							&q_\Psi-q_\alpha										&q_\Psi+Q_B-Q_A				&q_\Psi+Q_A			\\
\bqb	&-3/2						&q_\Psi-q_\alpha-3/2										&q_\Psi+Q_B-Q_A-3/2			&q_\Psi+Q_A-1/2
}
\end{align}
The requirement that these are singlets, $\bq=0$, determines the restriction to the appropriate physical grades of the various sections $\Psi$,
which are sections over $\bY$ of the following bundles
\begin{align}
\left(\Psi^1_\alpha\right)_{0} \in \Gamma(\pi^\ast T^\ast_B)~,	&&\left(\Psi^{2 \alpha}\right)_{\bq_\alpha} \in\Gamma(T_{\bY})~,	
&&\left(\Psi^{3A}_B\right)_{\bQ_A-\bQ_B}\in\Gamma(\End\cE)~,	&&\left(\Psi^4_A\right)_{-\bQ_A}\in\Gamma(\cE^\ast)~,
\end{align}
where $J_0\cdot \left(\Psi \right)_{\bq_\Psi}=\bq_\Psi \Psi$.
These states fit into a complex
\begin{align}
\xymatrix@R=0mm@C=5mm{
\Psi^1_\alpha \p y^\alpha	\ar[r]	&\Psi^1_{\alpha\mub} \p y^\alpha\etab^{\mub} \ar[r]	&\Psi^1_{\alpha\mub\nub} \p y^\alpha \etab^{\mub}\etab^{\nub}\\
{\begin{matrix}
\Psi^{2 \alpha} (\rho_\alpha - \cA_{\alpha B}^A \chib_A\chi^B)\\\oplus\\
\Psi^{3A}_B\chib_A\chi^B
\end{matrix}} \ar[r] 
&{\begin{matrix}
\Psi^{2\alpha}_{\mub} (\rho_\alpha - \cA_{\alpha B}^A \chib_A\chi^B)\etab^{\mub}\\\oplus\\
\Psi^{3A}_{B\mub}\chib_A\chi^B\etab^{\mub}
\end{matrix}} \ar[r] 
&{\begin{matrix}
\Psi^{2\alpha}_{\mub\nub} (\rho_\alpha - \cA_{\alpha B}^A \chib_A\chi^B)\etab^{\mub}\etab^{\nub}\\\oplus\\
\Psi^{3A}_{B\mub\nub}\chib_A\chi^B\etab^{\mub}\etab^{\nub}
\end{matrix}} \\
&\Psi^4_A\chi^A	\ar[r]	&\Psi^4_{A\mub}\chi^A\etab^{\mub}	\ar[r]	&\Psi^4_{A\mub\nub}\chi^A\etab^{\mub}\etab^{\nub}~\\
-3/2	&-1/2&1/2&3/2
}
\end{align}
where we omitted the vacuum $|1\ra$ and the appropriate grading on the various sections. Each column corresponds to the indicated eigenvalue
of the right-moving charge $\bqb$.
Taking $\bQb_0$-cohomology we obtain
\begin{align}
\label{eq:k1Q0cohmsinglt}
\xymatrix@R=0mm@C=8mm{
H^0(T^\ast_B)	&H^1(T^\ast_B) 	&H^2(T^\ast_B) \\ 
{\begin{matrix}
H^0(T_{\bY})\\\oplus\\
H^0(\End \cE)
\end{matrix}} \ar[r]^-{\obs_0}
&{\begin{matrix}
H^1(T_{\bY})\\\oplus\\
H^1(\End \cE)
\end{matrix}} \ar[r]^-{\obs_1}
&{\begin{matrix}
H^2(T_{\bY})\\\oplus\\
H^2(\End \cE)
\end{matrix}} \\
&H^0(\cE^\ast)	&H^1(\cE^\ast)			&H^2(\cE^\ast)\\ 
-3/2	&-1/2	&1/2	&3/2	
}
\end{align}
where the map $\obs_k:H^k(\bY,T_{\bY})\rightarrow H^{k+1}(\bY,\End\cE)$ acts by contraction with the curvature
\begin{align}
\obs_k(\Psi^{2\alpha}_{\mub_1\cdots\mub_k}) = - \Psi^{2\alpha}_{\mub_1\cdots\mub_k}\cF^A_{\mub_{k+1}\alpha B}~.
\end{align}

We wish now to perform the same analysis for the singlets arising in the CPT conjugate $k=-1$ sector. 
We start again by listing all the 
operators that can give rise to $h=1$ and $\bq=0$ states
\begin{align}
\label{eq:kinvsingl}
\xymatrix@R=0mm@C=2mm{
	&\Xi^1_\alpha y^\alpha |-1\ra		&\Xi^{2 \mu} (\rho_\mu - \cA_{\mu B}^A \chib_A\chi^B)|-1\ra	&\Xi^{3A}_B\chib_A\chi^B|-1\ra		&\Xi^{4A}\chib_A|-1\ra\\
\bq	&q_\Xi+q_\alpha				&0												&q_\Xi+Q_B-Q_A				&q_\Xi-Q_A			\\
\bqb	&3/2-d						&q_\Xi-q_\alpha+3/2-d								&q_\Xi+Q_B-Q_A+3/2-d			&q_\Xi-Q_A+1/2-d
}
\end{align}
where $y^\alpha=(\p y^\mu,\phi^\iota)$. In this notation, $\Xi\equiv\Xi(y^\mu;\yb^{\mub};\rho_\iota;\etab^\mu)$, and the requirement that the above states satisfy $\bq=0$
determines the $\rho$ dependence.  That is, we can expand the wavefunctions as
\begin{align}
\Xi=\sum_{\br}\sum_k\left(\Xi\right)_0(y^\mu,\yb^{\mub})_{\mub_1\cdots\mub_k}\prod_\iota \rho_\iota^{r_\iota} \etab^{\mub_1}\cdots\etab^{\mub_k}~,
\end{align}
and $J_0\cdot \left(\Xi\right)_{\bq_\Xi} = \bq_\Xi \Xi$ implies $-q_\iota r_\iota = \bq_\Xi$. We can then characterize the various bundles over $B$ 
which determine the transformation properties of the relevant states in \eqref{eq:kinvsingl}:
\begin{align}
\label{eq:Xiwavefunctgrad}
&\left(\Xi^1_\mu\right)_{0} \in \Gamma(K_B\otimes T^\ast_B)~,	
&&\left(\Xi^1_i\right)_{\bq_i} \in \Gamma\left(K_B\otimes X^\ast_i\otimes \left( \oplus_{\br|-r_\iota q_\iota=\bq_\iota} \otimes_\iota X_\iota^{r_\iota}\right)\right)~,	\nonumber\\
&\left(\Xi^{2\mu}\right)_0\in\Gamma(K_B\otimes T_B)~,	
&&\left(\Xi^{3I}_J\right)_{\bQ_A-\bQ_B}\in\Gamma\left(K_B\otimes\End\cE|_B\otimes \left( \oplus_{\br|-r_\iota q_\iota=\bQ_A-\bQ_B} \otimes_\iota X_\iota^{r_\iota}\right)\right)~,	\nonumber\\
&&&\left(\Xi^{4A}\right)_{\bQ_A}\in\Gamma\left(K_B\otimes\cE|_B\otimes\left( \oplus_{\br|-r_\iota q_\iota=\bQ_A} \otimes_\iota X_\iota^{r_\iota}\right)\right)~.
\end{align}
Then, we can construct the complex (again, we drop the vacuum $|-1\ra$, but we need to remember it transforms non-trivially)
\begin{align}
\xymatrix@R=0mm@C=7mm{
&\Xi^1_\alpha y^\alpha	\ar[r]	&\Xi^1_{\alpha\mub} y^\alpha\etab^{\mub} \ar[r]	&\Xi^1_{\alpha\mub\nub}  y^\alpha \etab^{\mub}\etab^{\nub} \\
&{\begin{matrix}
\Xi^{2\mu} (\rho_\mu - \cA_{\mu B}^A \chib_A\chi^B)\\\oplus\\
\Xi^{3A}_B\chib_A\chi^B
\end{matrix}} \ar[r] 
&\ar[r]
{\begin{matrix}
\Xi^{2\mu}_{\mub} (\rho_\mu - \cA_{\alpha B}^A \chib_A\chi^B)\etab^{\mub}\\\oplus\\
\Xi^{3A}_{B\mub}\chib_A\chi^B\etab^{\mub}
\end{matrix}} \ar[r] 
&{\begin{matrix}
\Xi^{2\mu}_{\mub\nub} (\rho_\mu - \cA_{\mu B}^A \chib_A\chi^B)\etab^{\mub}\etab^{\nub}\\\oplus\\
\Xi^{3A}_{B\mub\nub}\chib_A\chi^B\etab^{\mub}\etab^{\nub}
\end{matrix}} \\
\Xi^4_A\chi^A	\ar[r]	&\Xi^4_{A\mub}\chi^A\etab^{\mub}	\ar[r]	&\Xi^4_{A\mub\nub}\chi^A\etab^{\mub}\etab^{\nub}\\
1/2-d&3/2-d&5/2-d&7/2-d
}
\end{align}
and by taking $\bQb_0$-cohomology we obtain\footnote{Here and in what follows we leave the grading of the wavefunctions \eqref{eq:Xiwavefunctgrad} implicit to simplify notation.}
\begin{align}
\label{eq:kinvQ0cohmsinglt}
\xymatrix@R=0mm@C=10mm{
&H^0(K_B\otimes T^\ast_{\bY}|_B)	&H^1(K_B\otimes T^\ast_{\bY}|_B) 	&H^2(K_B\otimes T^\ast_{\bY}|_B) \\
&{\begin{matrix}
H^0(K_B\otimes T_B)\\\oplus\\
H^0(K_B\otimes \End \cE|_B)
\end{matrix}} \ar[r]^-{\widetilde{\obs}_0}
&{\begin{matrix}
H^1(K_B\otimes T_B)\\\oplus\\
H^1(K_B\otimes \End \cE|_B)
\end{matrix}} \ar[r]^-{\widetilde{\obs}_1}
&{\begin{matrix}
H^2(K_B\otimes T_B)\\\oplus\\
H^2(K_B\otimes \End \cE|_B)
\end{matrix}}
\\
H^0(K_B\otimes\cE|_B)	&H^1(K_B\otimes\cE|_B)			&H^2(K_B\otimes\cE|_B)\\
1/2-d&3/2-d&5/2-d&7/2-d
}
\end{align}
The obstruction maps $\obst_k : H^k(B,K_B\otimes T_B)\rightarrow H^{k+1}(B,K_B\otimes\End\cE|_B)$ act again as
the contraction with the curvature
\begin{align}
\widetilde{\obs}_k(\Xi^{2\mu}_{\mub_1\cdots\mub_k})=-\Xi^{2\mu}_{\mub_1\cdots\mub_k}\cF_{\mub_{k+1}\mu B}^A~.
\end{align}
It is not hard to show that the cohomology groups in \eqref{eq:k1Q0cohmsinglt} and \eqref{eq:kinvQ0cohmsinglt} are pairwise CPT dual. 
Therefore, in order for CPT to hold as a symmetry in $\bQb_0$-cohomology, the ranks of the obstruction maps
in the dual descriptions must match. In order to be more explicit, and to give a geometric interpretation to these obstructions, let us consider the 
two cases $d=1,2$ separately.

Suppose $d=1$, then all second degree cohomology groups on the base vanish, as well as $\obs_1=\obst_1=0$. 
Then, CPT holds as a symmetry in $\bQb_0$-cohomology iff 
\begin{align}
\label{eq:Kahlerobs}
\rank\obs_0 = \rank\obst_0~.
\end{align}
When this holds and these maps are non-trivial, the dual maps lift pairs of singlets, of which one is chiral and the other is anti-chiral, 
together with the respective chiral/anti-chiral currents.
However, note that the singlets are not CPT conjugate (in the sense of Serre duality) to each other: the chiral singlet represented by a class
in $H^1(\bY,\End \cE)$ in the image of $\obs_0$
is to be interpreted as a bundle deformation, 
while the anti-chiral singlet which maps non-trivially under $\obst_0$
is Serre dual to a chiral singlet corresponding to a K\"ahler deformation,
as $H^0(B,K_B\otimes T_B)=\left[ H^1(B,T_B^\ast)\right]^\ast$.
In other words, what in the chiral sector looks like an obstruction to a bundle deformation,
in the anti-chiral sector is to be interpreted as a K\"ahler deformation obstruction.
Hence, we interpret \eqref{eq:Kahlerobs} as an obstruction to extending an infinitesimal
K\"ahler deformation of the base to an infinitesimal deformation of the bundle $\cE\rightarrow\bY$.
This is a feature proper of a non-trivial hybrid model, and can be perhaps interpreted
as the worldsheet version of the K\"ahler sub-structure \cite{Sharpe:1998zu}. 
We notice that in all the examples we have considered, such obstructions always vanish.
It would be interesting to determine whether this is always the case, but we have not
being able to do so.

Let us now consider the case $d=2$. Here, CPT will hold as
a symmetry in $\bQb_0$-cohomology iff 
\begin{align}
\label{eq:d2obstrsymm}
\rank\obs_0 &= \rank\obst_1~,		&\rank\obs_1&= \rank\obst_0~.
\end{align}
For the first equation in \eqref{eq:d2obstrsymm} the same discussion from the $d=1$ map \eqref{eq:Kahlerobs}
applies, thus in this case we interpret $\obs_0/\obst_1$ again as a K\"ahler/bundle deformations compatibility obstruction. 
The novelty here is represented by the maps $\obs_1$ and $\obst_0$, which 
we require to be dual in a well-defined model. When non-trivial, these maps
lift each pairs of singlets, of which two chiral and two anti-chiral. The chiral singlets lifted by $\obs_1$ 
are represented by cohomology classes in $H^1(\bY,T_{\bY})$,
which are naturally interpreted as infinitesimal complex structure
deformations of the target space $\bY$, 
while the Serre dual of the anti-chiral singlets in the image of $\obst_0$ 
are classes in $\left[H^1(B,K_B\otimes \End\cE|_B)\right]^\ast=H^0(\bY,\End \cE)$,
which are interpreted as infinitesimal complex structure deformations of the bundle $\cE\rightarrow\bY$. 
Hence, it is natural to interpret these maps as the hybrid version of the Atiyah class \cite{Atiyah:1957xx}, 
which captures an obstruction to the compatibility between a complex structure deformation of $\bY$
and a bundle deformation \cite{Anderson:2011ty}. Again, we remark that we do not know of an
example in which nontrivial obstructions arise. 

To conclude this section, we briefly comment on obstructions in twisted sectors for $k>1$.
We have seen that in our (0,2) models, obstructions are compatible 
with CPT invariance precisely when \eqref{eq:Kahlerobs} or \eqref{eq:d2obstrsymm} holds.
A crucial fact for this to be allowed is the presence in both CPT conjugate sectors $k=\pm1$ of
the base fields $\rho_\mu$. In a generic twisted sector $k>1$, these modes do not
appear, as we discussed at the beginning of this section. Thus, the argument of \cite{Bertolini:2013xga} extends to our (0,2) models:
CPT invariance is violated when non-trivial obstructions appear in twisted sectors $k>1$,
as these cannot arise in the dual states in the dual sectors $2L-k$.

Finally, simple energy considerations show that each monomial in $\bQb_J$ in the $k$ sector generates the same action 
as the dual monomial in the $2L-k$ sector obtained by implementing \eqref{eq:CPTpairing}. Thus, provided CPT is a symmetry in $\bQb_0$-cohomology,
it is also maintained as a symmetry by further taking $\bQb_J$-cohomology, and therefore of the full $\bQb$ cohomology.

\section{Examples}
\label{s:examples}

We now turn to the analysis of two examples of (0,2) hybrid models. The first example will illustrate several technical points, as we will require 
 the full power of the techniques developed in the previous section to explicitly solve the model. The second example, while technically less challenging as
 the left-moving bundle simply pullbacks from the base, displays a rather surprising conceptual feature, concerning accidental symmetries in (0,2) models.
 For both examples, we will limit our discussion in this section to the relevant features of the massless spectrum, and we defer 
 the explicit computations to appendix \ref{app:examplescomp}.

\subsection{A $\so(10)$ model}
\label{ss:so10example}

Let us take the target space to be
\begin{align}
\label{eq:so10Y}
\bY&=\tot\left( \cO(-1)^{\oplus2}\oplus\cO\oplus\cO\rightarrow \P^1\right)~,\nonumber\\
&\bq:					\qquad\quad	\ff15 \qquad\quad\ \ \ff15 \quad\  \ff25
\end{align}
where we denote $\phi_{a}$, $a=1,2$, as the coordinates along the fibers $\cO(-1)^{\oplus2}$ and $\phi_{3,4}$ along the trivial fibers. 
Next, we choose the following data for the $\cE\rightarrow \bY$
\begin{align}
\label{eq:so10E}
\xymatrix@R=0mm@C=8mm{
0	\ar[r]	&\cO_{\bY} \ar[r]^-{E} &\pi^\ast \cO(1)^{\oplus3}\oplus\pi^\ast \cO\oplus\pi^\ast \cO(-1)\oplus\cO(-3)\oplus\pi^\ast \cO(1) \ar[r]  &\cE \ar[r] &0~,\\
&\bQ:&						-1 \qquad\ -\frac45 \qquad\ -\frac45 \qquad\ \ -\frac45 \qquad\ \ -\frac35
}
\end{align}
that is, $m=1$, $P=3$ and $R=6$.
In both \eqref{eq:so10Y} and \eqref{eq:so10E} we have indicated the $\GU(1)_L$ action, which determines
the orbifold group $\Gamma=\Z_{10}$. This data yields $c=10$, $\cb=9$ and $r=4$,
thus we expect a $\so(10)$ theory in four dimensions. 
In particular, we choose in \eqref{eq:so10E} the following injective map
\begin{align}
\label{eq:so10Echio}
E^1&=z_1~,		&E^2&=z_2~,		&E^3&=E^4=E^5=E^6=0~,		&E^7&=-z_1^3\phi_1^2-z_2^3\phi_2^2~,
\end{align}
where $z_{1,2}$ are homogeneous coordinates on $B=\P^1$. For local coordinates we will use $u=z_1/z_2=v^{-1}$. 
It is clear from this data that $\cE_B=T_{\P^1}\oplus \cO(1)$, and more in general the bundle is
isomorphic to $\cE=\pi^\ast \cO(-1)\oplus\pi^\ast \cO(-3)\oplus\pi^\ast \cO\oplus\pi^\ast\cO(1)\oplus\cE'$ where $\cE'$ is a rank-2 bundle
defined by the transformations
\begin{align}
\label{eq:transforEpr}
G_{uv}&=\begin{pmatrix}
-v^{-2}		&v^{-2}\phi_1^2+v\phi_2^2\\
0			&v^{-1}
\end{pmatrix}~.
\end{align}
A choice of (0,2) superpotential $J\in\Gamma(\cE^\ast)$ is given by (recall that $A=(M,I)$ where $M=1,2$ and $I=3,\dots,6$)
\begin{align}
\label{eq:so10supchioc}
J_1^u&=(u^4+\alpha u^2-\gamma)\phi_1^5 + \betah u^3 \phi_2^5 +\beta u^3\phi_1^3\phi_2^2		&J_1^v&=\gamma v^3\phi_1^5 + (\alphah v^4+v^2-\betah)\phi_2^5\nonumber\\
&\quad+(u^4+\alphah u^2 -\gammah) \phi_1^2\phi_2^3~,		&&\quad+ (\alpha v^4+v^2-\beta)\phi_1^3\phi_2^2 + \gammah v^3 \phi_1^2\phi_2^3~, \nonumber\\
J_2^u&=(u^4+1)\phi_2^5~,		&J_2^v&=(1+v^4)\phi_2^5~,\nonumber\\
J_3^u&= \phi_4^2~,				&J_3^v&= \phi_4^2~,\nonumber\\
J_4^u&=(u+1)\phi_3^4 ~,			&J_4^v&=(1+v)\phi_3^4 ~,\nonumber\\	
J_5^u&=(u^3+2)\phi_3^4~,		&J_5^v&=(1+2v^3)\phi_1^4~,\nonumber\\		
J_6^u&=(u^2+\alpha)\phi_1^3+(u^2+\alphah)\phi_2^3~,		&J_6^v&=(1+\alpha v^2)\phi_1^3+(1+\alphah v^2)\phi_2^3~.
\end{align}
It is easy to verify that for general values of the parameters $\alpha, \beta, \gamma, \alphah, \betah, \gammah$, \eqref{eq:so10supchioc} defines a non-singular superpotential.

\subsubsection*{GSO projection for $\SO(10)$ bundles}

Let us recall some facts about the GSO projection for $r=4$ theories. 
In this case the group $\SO(16-2r)=\SO(8)$ is linearly realized by eight free fermions $\lambda$ 
and the GSO projection is enforced on the ground states as 
\begin{align}
(-1)^{F_\lambda} e^{-i\pi J_L} =1~,
\end{align}
where $F_\lambda$ is the fermion number for the free $\lambda$ system. 
The eight $\lambda$ combine into four Weyl fermions, whose action is 
 $F_\lambda=0$ on the NS and $F_\lambda=\half$ on the R ground states.
Hence, $(-1)^{F_\lambda}=1$ on the both NS and R ground states. Thus, in the R sectors, to a state
with $J_L$ odd we associate the representation $\rep{8^s}$ of $\so(8)$, while if $J_L$ is even we associate a $\rep{8^c}$.
In the NS sectors, if $J_L$ is odd, the state must have weight $-1/2$ and it is associated to a $\rep{8^v}$,
while if $J_L$ is even, we associate the representation $\rep{28}$ if the state has weight $-1$ or a $\rep{1}$ if it has weight zero.

\subsubsection*{Summary}

We organize the result for the massless spectrum of this model, derived in appendix \ref{app:examplescomp}, in table \ref{t:summaryso10}. 
We verify that the states assemble themselves as expected into representations of $\so(10)$: 
we find the gauginos from the adjoint representation of $\so(10)$, while
the chiral matter is organized in $\rep{16}$'s, $\rep{\overline{16}}$'s and $\rep{10}$'s of $\so(10)$.
In particular, we find a 443-dimensional space of $\so(10)$-preserving first order deformations.
Moreover, the spectrum is supersymmetric,
thus the model does not suffer from the kind of pathology along the lines of \cite{Distler:1993mk}. 
As will we show below, it is not clear how to embed this model into a GLSM. Hence, it is natural to conjecture
that our hybrid model flows in the IR to a previously unexplored CFT.

\begin{table}
\begin{center}
\begin{tabular}{|c|c|c|c|c|c|c|c|c|c|c|c|c|c|c|}
\hline
\multicolumn{1}{|c|}{$\so(10)$}&\multicolumn{4}{c|}{$\rep{45}$}& \multicolumn{2}{c|}{ $\rep{16}$}& \multicolumn{2}{c|}{ $\rep{\overline{16}}$}& \multicolumn{3}{c|}{ $\rep{10}$}&$\rep{1}$\\\hline
\multicolumn{1}{|c|}{$\so(8)\oplus \mathfrak{u}(1)$}	&$\rep{8^c}_{-2}$&$\rep{28}_{0}$&$\rep{1}_{0}$&$\rep{8^c}_{2}$ &$\rep{8^s}_{-1}$&$\rep{8^v}_{1}$&$\rep{8^v}_{-1}$&$\rep{8^s}_{1}$&$\rep{1}_{-2}$	&$\rep{8^c}_{0}$&$\rep{1}_{2}$&$\rep{1}_0$\\\hline
$k=0$		&1		&0		&0			&0			&96			&0			&0			&0			&0			&54			&0		&0\\\hline
$k=1$		&0		&1		&1			&0			&0			&96			&1			&0			&2			&0			&54		&393\\\hline
$k=2$		&0		&0		&0			&1			&0			&0			&0			&1			&0			&2			&0		&0\\\hline
$k=3$		&0		&0		&0			&0			&0			&0			&1			&0			&0			&0			&2		&31\\\hline
$k=4$		&0		&0		&0			&0			&0			&0			&0			&1			&0			&0			&0 		&0\\\hline
$k=5$		&0		&0		&0			&0			&0			&0			&2			&0			&1			&0			&0		&16\\\hline
$k=6$		&0		&0		&0			&0			&0			&0			&0			&2			&0			&1			&0 		&0\\\hline
$k=7$		&0		&0		&0			&0			&0			&0			&0			&0			&0			&0			&1		&3\\\hline
$k=8$		&0		&0		&0			&0			&0			&0			&0			&0			&0			&0			&0		&0\\\hline
$k=9$		&0		&0		&0			&0			&0			&0			&0			&0			&54			&0			&0		&0\\\hline
tot \#			&\multicolumn{4}{c|}{1}						&\multicolumn{2}{c|}{96}		&\multicolumn{2}{c|}{4}		&\multicolumn{3}{c|}{57}				&443\\\hline
\end{tabular}
\end{center}
\caption{Summary of massless spectrum for the $\so(10)$ model.}
\label{t:summaryso10}
\end{table}

\subsection{An $\GE_7$ model}
\label{ss:disaccident}

Our second example is going to reveal an interesting feature of our models. 
One of the assumptions for the correspondence between our UV model and the IR SCFT which we analyzed in section \ref{s:IRphys}
consists in the absence of IR accidental symmetries \cite{Melnikov:2016dnx}. In fact, if these occur, they can mix with the na\"ive $\GU(1)_R^{\text{UV}}$ R-symmetry and
$c$-extremization \cite{Benini:2013cda} might select a different $\GU(1)_R^{\text{IR}}$. 
This can have drastic consequences on the structure of the conformal manifold,
which in the most extreme case might be even empty, that is $c=c^{\text{IR}},\cb=\cb^{\text{IR}}$ is not satisfied for 
any choice of the parameters of the UV model.
As shown in \cite{Bertolini:2014ela} in the context of LG theories this phenomenon does in general occur in (0,2) models, and it is therefore
natural asking how these accidents manifest themselves in more general (0,2) theories. A hybrid theory, which has the structure of a LG fibration, 
provides the perfect setting to begin such a task. In particular, it is reasonable to conjecture that LG accidents carry over to the hybrid structure, which 
would then suffer from the same pathology. Indeed, we can write \eqref{eq:cleft} and \eqref{eq:cbright} in terms of the data of the fiber LG theory as
\begin{align}
c&=2d+R_B+c_{\text{LG}}~,		&\cb&=3d+\cb_{\text{LG}}~,
\end{align}
and one might conclude that the hybrid IR would inherit the pathology of the LG fiber, prompting the model builder to reject such a hybrid theory.  

As this example will show, the situation is not as dramatic as it might seem at this point. We are going to argue
that in some cases the hybrid can resolve a fiber accident, even when this is fairly severe. 
In particular, we study an example obtained as a fibration of the LG model carefully examined in \cite{Bertolini:2014ela},
which we now review.  

The UV LG model has $n=2$, $N=3$ and charges
\begin{align}
\label{eq:devilcharges}
\xymatrix@C=8mm@R=0mm{
			&\Phi^1		&\Phi^2		&\cX^{1,2,3}\\
\bq			&\frac17		&\frac37		&-\frac67\\
\bqb			&\frac17		&\frac37		&\frac17
}
\end{align}
which lead to a putative $r=2,\cb=3$ model. The generic superpotential depends on 9 parameters, and it can be shown
that there exist only three orbits of field redefinitions compatible with \eqref{eq:devilcharges}. For each of these, the corresponding parameter
space reduces to just a point exhibiting enhanced symmetries which lead through $c$-extremization to $\cb^{\text{IR}}>3$. In particular, there is no choice 
of UV parameters that realizes a IR CFT with $r=2,\cb=3$.

We now turn to our hybrid example constructed by fibering the above LG theory over $B=\P^1$ and with the following
geometric set-up
\begin{align}
\bY&=\text{tot}\left( \cO\oplus\cO(-4) \rightarrow \P^1 \right)~,		&\cE=\pi^\ast \cO(2)\oplus \pi^\ast \cO^{\oplus2}&\rightarrow \bY~.
\end{align}
This data satisfies the anomaly conditions \eqref{eq:anomaly1} and \eqref{eq:anomaly2}, \eqref{eq:mixanom} is automatically satisfied by the non-anomalous fiber LG theory,
while \eqref{eq:hetanom} does not impose any constraints in this case since $\dim B=1$.
The most general superpotential compatible with this data is given by
\begin{align}
\label{eq:devfibsup}
J_1&=S_{[2]}\phi_1^3\phi_2+S_{[6]}\phi_2^2~,		&J_{2,3}&=S_{[0]}\phi_1^6+S_{[4]}\phi_1^3\phi_2+S_{[8]}\phi_2^2~,	
\end{align}
where $S_{[d]}\in H^0(\P^1,\cO(d))$. In particular, by considering the following specific choice for \eqref{eq:devfibsup}
\begin{align}
\label{eq:02suppot}
J_1&=u^6\phi_2^2~,		&J_2&=\phi_1^6~,		&J_3&=(u^8+1)\phi_2^2~,
\end{align}
where $u$ is a local coordinate on $\P^1$ in an appropriate patch,
it is obvious to realize that it leads to a non-singular model, i.e., $J^{-1}(0)=B$. 

This data seems to define a theory with $c=\cb=6$ and $r=2$,
which should correspond to a point in the moduli space of K3 compactifications. 

Instead of attempting a rather involved study of orbits of fields redefinitions for \eqref{eq:devfibsup}, we will compute the massless
spectrum of the associated six-dimensional compactification. We will then show that we recover all the expected features
for a K3 compactification, prompting us to conjecture that 
there is a range of UV parameters for which the theory does flow to the 
expected IR fixed point. 

\subsection*{GSO projections for $\su(2)$ bundles in six dimensions}

In this example we have $D=2$ and $r=2$. The unbroken gauge group in spacetime is $\e_7\oplus\e_8$, and in what follows we will restrict
our attention to the degrees of freedom of the $\e_7$ component. 
In a standard $r=2$ NLSM setting the linearly realized algebra is $\so(12)\oplus\su(2)_L\subset\e_7$.
We then decompose the representation of $\e_7$ as follows 
\begin{align}
\e_7 &\mapsto \so(12)\oplus\su(2)_L \nonumber\\
\rep{133}&=(\rep{66},\rep1)\oplus(\rep{32},\rep2)\oplus(\rep1,\rep3)~,\nonumber\\	
\rep{56}&=(\rep{12},\rep2)\oplus(\rep{32'},\rep1)~.
\end{align}
However, in our hybrid application the states are classified according to 
$\so(12)\oplus\mathfrak{u}(1)_L\oplus\mathfrak{u}(1)_R$. In other words, we do not have at our disposal a fully linearly realized $\su(2)_{L}$, but only the $\mathfrak{u}(1)_{L}$ charges 
$\bq$.
However, these values will be enough to reconstruct the full representations. In particular, with our normalization, $\bq=(1,-1)$ will correspond to a doublet under $\su(2)_L$ and 
$\bq=(2,0,-2)$ to the triplet $\rep3\in\su(2)_L$.
Moreover, the left-moving GSO projection is always onto even values of $\bq$ in every sector, due to the fact that
the number of free fermions is even (six to be precise, as they form a $\so(16-2r)$ level 1 algebra).
The only difference is that in R sectors ($k$ even) states with $\bq$ odd correspond to $\rep{32}\in\so(12)$ while states with $\bq$ even to $\rep{32'}\in\so(12)$.

An analogous story holds for the right-moving sector. Here, we have a $\su(2)_R$ R-symmetry but again in our hybrid setting we classify 
states according to the charges $\bqb$. 
The right-moving GSO projection identifies doublets of $\su(2)_R$ with spacetimes vectors and $\su(2)_R$-singlets with $\half$-hypers.

\subsection*{Summary}

We summarize our results from appendix \ref{app:examplescomp}, taking into consideration the contribution from the CPT conjugate sectors. 
For the most part the spectrum is universal, that is, it is independent of the specific
form of the superpotential \eqref{eq:devfibsup}. 
We list in table \ref{t:summarydev} the contribution from each sector, restricting ourselves to $\bqb\leq0$, as this
completely determines the spectrum structure. 
Our notation is such that each entry in the table is of the form $\#_{\bq}$, where $\#$ is the number of states
in the appropriate $\so(12)$ representation, and $\bq$ is the corresponding eigenvalue under $\mathfrak{u}(1)_L\subset \su(2)_L$.
In particular, we find the gauge vectors necessary for a $\GE_7$ theory, as well as 20 $\ff{1}{2}$-hypers corresponding to $\GE_7$-charged matter
and 65 $\GE_7$-neutral hypers.
This is the expected spectrum for a K3 compactification, and it satisfies anomaly cancellation in six dimensions. 
Additional not universal hypers arise for special forms of the superpotential, but only in the combination of vector/hyper, as expected from anomaly cancellation. 
\begin{table}
\begin{center}
\begin{tabular}{|c|c|c|c|c|c|c|c|c|c|c|c|c|c|c|}
\hline
\multicolumn{1}{|c|}{$\GE_7$}&\multicolumn{3}{c|}{$\rep{133}$}& \multicolumn{2}{c|}{ $\rep{56}$}&$\rep{1}$\\\hline
\multicolumn{1}{|c|}{$\so(12)\oplus\su(2)_L$}	&$\rep{66},\rep{1}$&$\rep{1},\rep{3}$&$\rep{32},\rep{2}$&$\rep{12},\rep{2}$ &$\rep{32'},\rep{1}$&$\rep{1},\rep{1}$\\\hline
$k=0$		&0		&0		&$1_{-1}$		&0			&$10_0$		&0			\\\hline
$k=1$		&$1_0$	&$1_0$	&0			&$10_1$		&0			&$30_0$		\\\hline
$k=2$		&0		&0		&$1_1$		&0			&0			&0			\\\hline
$k=3$		&0		&$1_2$	&0			&$5_{-1}$		&0			&$15_0$		\\\hline
$k=4$		&0		&0		&0			&0			&$5_0$		&0			\\\hline
$k=5$		&0		&0		&0			&$5_1$		&0			&$9_0$		\\\hline
$k=6$		&0		&0		&0			&0			&0			&0			\\\hline
$k=7$		&0		&0		&0			&0			&0			&$22_0$		\\\hline
$k=8$		&0		&0		&0			&0			&0			&0			\\\hline
$k=9$		&0		&0		&0			&$5_{-1}$		&0			&$9_0$		\\\hline
$k=10$		&0		&0		&0			&0			&$5_0$		&0			\\\hline
$k=11$		&0		&0		&0			&$5_1$		&0			&$15_0$		\\\hline
$k=12$		&0		&0		&0			&0			&0			&0			\\\hline
$k=13$		&0		&$1_{-2}$	&0			&$10_{-1}$	&0			&$30$		\\\hline
tot\#		&\multicolumn{3}{c|}{1}			&\multicolumn{2}{c|}{20}		&\multicolumn{1}{c|}{130}\\\hline 
\end{tabular}
\end{center}
\caption{Summary of massless spectrum for the $\GE_7$ model.}
\label{t:summarydev}
\end{table}

\subsubsection*{$c$-extremization and field redefinitions}

The above result is quite surprising form the point of view of \cite{Bertolini:2014ela}, which we summarized above. In fact, the hybrid superpotential
\eqref{eq:02suppot} exhibits a $\GU(1)^{\oplus2}$ enhanced symmetry that acts independently on $\phi^1$ and $\phi^2$ and it leaves the base invariant. 
Applying $c$-extremization \cite{Benini:2012cz,Benini:2013cda} leads to $\cb=6$ and the R-charge assignment manifest for the generic superpotential \eqref{eq:devfibsup}. 

The LG accident occurs fiberwise as it is always possible to perform a rotation in the $\Gamma^{1,3}$ plane and reduce the superpotential to the following
\begin{align}
\cW_{\text{fiber}}&=\cX^1 \phi_2^2+\cX^2\phi_1^6~,	
\end{align}
which correspond to a direct sum of (2,2) minimal models and a free Fermi multiplet $\cX^3$. 
In the full hybrid theory, the situation is more involved because field redefinitions have a geometric interpretation. 
More precisely, a hybrid field redefinitions is defined by invertible sections $F\in \Gamma(T_{\bY}\otimes T^\ast_{\bY})$ and $G\in \Gamma(\cE\otimes\cE^\ast)$, which 
rotate the fields $Y^\alpha$ and $\cX^A$, respectively.

In general, a fiberwise field redefinition is not compatible with this geometric structure.
This is precisely what happens in our example. The fiberwise field redefinition
\begin{align}
J_3' &= F_{[2]} J_1 + a J_3=(F_{[2]}u^6+a(u^8+1))(\phi^2)^2=0~,
\end{align}
where $F_{[2]}\in H^0(\P^1,\cO(2))$, is not allowed geometrically since $u^6$ does not divide $u^8+1$. 
Explicitly, trying to extend this globally, we construct 
\begin{align}
\label{eq:globsingfr}
G=\begin{pmatrix}
1	&0		&0\\
0	&1		&0\\
-\frac{u^8+1}{u^6}	&0	&1
\end{pmatrix} \in \Gamma(\cE\otimes\cE^\ast)~,
\end{align}
which indeed yields
\begin{align}
G \begin{pmatrix} J_1 \\ J_2 \\ J_3 \end{pmatrix} = \begin{pmatrix} J_1 \\ J_2 \\ 0 \end{pmatrix} ~.
\end{align}
However, \eqref{eq:globsingfr} is singular at $u^6=0$ and therefore not an allowed field redefinition.

It would be an interesting exercise, but we do not attempt it here, to determine whether there are full hybrid orbits of 
field redefinitions which do suffer from accidents in this model. Certainly, hybrid models do suffer from accidents, 
and we present an example of this in appendix \ref{app:acchybr}.

\section{The minimal GLSM embedding}
\label{s:GLSMembed}

An interesting issue regarding the structure of the (0,2) moduli space 
is whether every CFT realization is a point in the moduli space
of a Calabi-Yau. It is then natural to ask: what about our (0,2) hybrid CFTs?
A particularly useful tool in exploring the relation between geometric and non-geometric (CY-LG correspondence)
realizations of a SCFT is the GLSM \cite{Witten:1993yc}. In this section, we are going to derive the conditions that a (0,2) hybrid
needs to satisfy in order to arise as a phase of a GLSM.
This is in general a challenging question, which is unsolved even in the context of (0,2) LG models.
A simpler goal would be to study hybrid embeddings into linear models which exhibit a geometric phase 
with target space a complete intersection Calabi-Yau (equipped with a stable holomorphic bundle).
This appears to be still quite a formidable task, as the combinatorics becomes rapidly unwieldy, and it is
beyond the scope of this work.
Here, we restrict our attention to a particular subclass of Abelian GLSMs, 
which admit a geometric phase described by a (possibly singular) CY hypersurface in a toric variety
equipped with a (possibly singular) holomorphic bundle.
Our conventions for (0,2) GLSMs follow \cite{Bertolini:2014dna}.

As it is natural in the context of GLSMs, we restrict our attention to $B$ being a compact toric variety. Let
\begin{align}
B=\frac{\C^M - Z(F)}{(\C^\ast)^{M-d}}~,
\end{align}
where $F$ is the irrelevant ideal in the homogeneous coordinate ring $\C[x_1,\dots,x_M]$, and $Z(F)\subset \C^M$
is the associated subvariety. The $(\C^\ast)^{M-d}$ action on the (0,2) bosonic chiral superfields $X^{\mu'}$, $\mu'=1,\dots,M$, 
whose lowest components are the homogeneous coordinates $x^{\mu'}$, is given by 
\begin{align}
X^{\mu'} & \rightarrow \prod_a (\lambda_a)^{m_{\mu'}^a}X^{\mu'}~,		&\lambda_a& \in (\C^\ast)^{M-d}~.
\end{align}
This allows us to construct the $V$-model with target space $\bY$ equipped with a bundle $\cE\rightarrow\bY$.
This model has gauge group $\GU(1)^{p_B}$, where $p_B=\dim \Pic B$.
We introduce $m$ neutral chiral supermultiplets $\Sigma_{\varpi} =(\sigma_\varpi,\lambda_{\varpi,+})$, $\varpi=1,\dots,m$,
which we couple to 
\begin{align}
\label{eq:VmodelY}
\xymatrix@R=0mm@C=8mm{
\text{fields}	&X^{\mu'}			&\Phi^i		&\cX^{\Gamma}				&\text{F.I.}\\
\GU(1)_a		&m^a_{\mu'}		&l_i^a		&L_{\Gamma}^a				&r_a
}
\end{align}
for $a=1,\dots,p_B$. When the F.I.~parameters $r^a$ lie in the K\"ahler cone of $B$, which we denote $\cK_B$, the classical vacua of the theory will be $\bY$.
The $R+m$ Fermi multiplets satisfy the chirality condition
\begin{align}
\cDb \cX^{\Gamma}&=\sqrt2 \Sigma_\varpi E^{\Gamma,\varpi}(X^{\mu'})~,		
\end{align}
and whose lowest components take values in the bundle $\cE\rightarrow\bY$ determined by the SES
\begin{align}
\label{eq:Vmodelbundle}
\xymatrix@R=0mm@C=10mm{
0\ar[r]	& \cO^{\oplus m}	\ar[r]^-{E^\Gamma}	&\oplus_{\Gamma=1}^{R+m} \cO(L_{\Gamma}^a)	\ar[r]	&\cE\ar[r]	&0~,
}
\end{align}
which corresponds to \eqref{eq:Ebundle} with $\cF_{\Gamma}=\cO(L_\Gamma^a)$. 
In order to introduce superpotential interactions, we augment the above construction 
by the bosonic field $P$ and Fermi field $\Lambda$, which we take to be neutral under \eqref{eq:VmodelY},
and we introduce an additional $\GU(1)$ factor to the gauge group, under which the fields have charges
\begin{align}
\label{eq:VpmodelY}
\xymatrix@R=0mm@C=8mm{
\text{fields}		&X^{\mu'}			&\Phi^i		&\cX^{\Gamma}		&P		&\Lambda		&\text{F.I.}\\
\GU(1)_{p_B+1}	&0				&q_i			&Q_{\Gamma}+1		&-1		&-1			&r_{p_B+1}~.
}
\end{align}
We also introduce an additional multiplet $\Sigma_{m+1}$, thus the chirality constraints on the Fermi fields now read
\begin{align}
\cDb \cX^{\Gamma}&=\sqrt2 \Sigma_\varpi E^{\Gamma,\varpi}(X^{\mu'})+\sqrt2\Sigma_{m+1}E^{\Gamma,m+1}(X^{\mu'})~,		&\cDb \Lambda &= -\sqrt2 \Sigma_{m+1} P~.
\end{align}
Now we can introduce the superpotential couplings, which we take to be of the form
\begin{align}
\int d\theta^+ \left( \Lambda H(X,\Phi) + \cX^{\Gamma} P J_{\Gamma}(X,\Phi)\right)\big|_{\thetab^+=0} + \text{h.c.}~.
\end{align}
The model will be (0,2) supersymmetric when the couplings satisfy the relations
\begin{align}
\label{eq:02SUSYconstr}
H&=E^{\Gamma,m+1} J_{\Gamma}~,			&E^{\Gamma,\varpi} J_{\Gamma}&=0~,\quad \forall\varpi~.
\end{align}
Now, for $r_a\in\cK_B$ and $r_{p_B}<0$, $p$ has a non-zero vev through the D-term associated to \eqref{eq:VpmodelY}
and the classical vacua of the theory will be parametrized by $B$. In particular, $\lambda_{m+1,+}$ and $\lambda$ (the lowest
component in $\Lambda$) acquire a mass, and the light left-moving fermions take value in the bundle $\cE$ defined by \eqref{eq:Vmodelbundle}. 
By \eqref{eq:02SUSYconstr}, the (0,2) superpotential $J_\Gamma$ determines in this cone a section $J\in\Gamma(\cE^\ast)$, 
which takes the role of the (0,2) superpotential in the hybrid model. Finally, \eqref{eq:VpmodelY} is broken to a finite subgroup
which determines the orbifold.

In the cone $r_a\in\cK_B$ and $r_{p_B}>0$, we expect the space of classical vacua to be the hypersurface $H=0$
in the toric variety determined by the gauge charges \eqref{eq:VmodelY} and \eqref{eq:VpmodelY}. When nonsingular, 
$\Lambda$ again acquires a mass and the light left-moving fermions take values in the bundle determined
by the cohomology of the short sequence
\begin{align}
\label{eq:Mmodelbundle}
\xymatrix@R=0mm@C=10mm{
0\ar[r]	& \cO^{\oplus m}	\ar[r]^-{E^\Gamma}	&\oplus_{\Gamma=1}^{R+m} \cO(L_\Gamma^a)	\ar[r]^-{J_{\Gamma}}		&\cO(-d_P)		\ar[r]	&0~,
}
\end{align}
restricted to $H=0$, where $d_P=(0,\dots,0,-L)$ is the gauge charge of $P$.\footnote{For this purpose it is convenient to rescale the $\GU(1)_{p_B+1}$
charges such that these are all integers.} 

A couple of remarks are in order. First, it is apparent that we have made quite a few assumptions in constructing the above linear model.
In particular, we have introduced only one extra bosonic
multiplet $P$ and Fermi field $\Lambda$. We refer to this as {\it minimal embedding}.
In general, one could consider adding several $\Lambda$ fields, corresponding to a complete intersection at $r_{p_B+1}>0$,
and/or multiple $P$ fields, generalizing the bundle structure in \eqref{eq:Mmodelbundle}. 

Second, even in this minimal construction, the phase structure of the linear model can be quite complicated
and a classification of the various geometric descriptions in each of the phases depends on the details of the model.
It is therefore conceivable that there might be other geometric phases, or that the resulting toric variety for 
$r_{p_B+1}>0$ exhibits singularities which can be resolved \cite{Distler:1996tj}.
In our construction, when realizable, we are guaranteed at least one such geometric phase.

\subsection{Anomalies}

The action for this class of models is invariant under an additional $\GU(1)_L\times\GU(1)_R$ symmetry, which acts which charges
\begin{align}
\label{eq:U1LU1RGLSM}
\xymatrix@R=0mm@C=8mm{
\text{fields}	&X^\mu			&\Phi^i		&\cX^\Gamma		&P		&\Lambda		&\Sigma\\
\GU(1)_L		&0				&0			&-1				&1		&0			&-1	\\
\GU(1)_R		&0				&0			&0				&1		&1			&1	
}
\end{align}
Let us now derive the constraint from gauge anomalies in our linear model.\footnote{Our convention is to assign a minus sign to right-moving and a plus
sign to left-moving fermions.}
First of all, we have for \eqref{eq:U1LU1RGLSM}
\begin{align}
\label{eq:anomalyU1GLSM}
\sum_\Gamma L_\Gamma^a&=\sum_{\mu'} m_{\mu'}^a + \sum_i l_i^a =0~,	&\sum_\Gamma(Q_\Gamma+1) &=\sum_i q_i =1~.
\end{align}
The first condition in \eqref{eq:anomalyU1GLSM} simply implies $c_1(\cE)=c_1(T_{\bY})=0$. 
We recall that this can always be achieved, at the expense of introducing spectator fields.
The second equation instead can be rephrased as $R+m-r=2$. 
The quadratic conditions from \eqref{eq:VmodelY} and \eqref{eq:VpmodelY} instead read
\begin{align}
\label{eq:gaugequadranom}
&\sum_{\mu'} m_{\mu'}^a m_{\mu'}^b + \sum_i l_i^al_i^b = \sum_\Gamma L_\Gamma^aL_\Gamma^b~, &&\forall a,b\nonumber\\
&-\sum_i q_il_i + \sum_\Gamma (Q_\Gamma+1)L_\Gamma^a=0~,&&\forall a\nonumber\\
&-\sum_i q_i^2 + \sum_\Gamma (Q_\Gamma+1)^2=0~.
\end{align}
While the first equation of \eqref{eq:gaugequadranom} implies the anomaly condition \eqref{eq:hetanom}, the converse
does not hold (e.g., for $d=1$), and this can be seen as an additional geometric condition on the structure of the fibration.
The second and third
equations above are automatically satisfied because of the anomaly conditions and \eqref{eq:anomalyU1GLSM}.

\subsection{Examples}

Let us first consider an example which does not admit a geometric description.
Consider the (2,2) model 
\begin{align}
\label{eq:noLR}
\bY&=\tot\left(\cO(-2)\oplus\cO^{\oplus4}\rightarrow \P^1 \right)~,\nonumber\\
&\bq:					\qquad\quad	\ff12 \qquad\ \ \ff14 
\end{align}
and of course $\cE=T_{\bY}$. This model does in fact embed in a GLSM, which was studied in \cite{Aspinwall:1994cf}. There it has been shown that
for this model $h^{1,1}=1$, confirmed by the spectrum computation of \cite{Bertolini:2013xga},
and that the other phase is described by a LGO. 
Now, it is immediate to realize that $\sum_iq_i=3/2>1$, thus \eqref{eq:anomalyU1GLSM} is violated. 
This simple fact is in fact enough to show that we cannot blow up the fiber LG to reach some sort of large radius phase. 

Let us consider also the examples we have encountered in this work. Applying the above procedure to
the model studied in section \ref{ss:so10example} we obtain the following charges
\begin{align}
\label{eq:noLGcharges}
\xymatrix@R=0mm@C=5mm{
\text{fields}	&X^{1,2}			&\Phi^{1,2}	&\Phi^3	&\Phi^4		&\cX^{1,2,3}	&\cX^{4}		&\cX^5	&\cX^6	&\cX^7	&P		&\Lambda			\\
\GU(1)_1		&1				&-1			&0		&0			&1			&-1			&-3		&0		&1		&0		&0				\\
\GU(1)_2		&0				&1			&1		&2			&0			&1			&1		&1		&2		&-5		&-5
}
\end{align}
to which we add two neutral $\Sigma$ fields. It is easy to check that \eqref{eq:anomalyU1GLSM} is satisfied, however, the quadratic anomaly 
\eqref{eq:gaugequadranom} for $a=b=1$ reads $4\neq 14$. Thus, $\GU(1)_1$ does not get enhanced to a gauge symmetry, 
and we cannot construct our GLSM embedding. 

Finally, for the example we studied in section \ref{ss:disaccident}, $c_1(T_{\bY})\neq0$, thus we need to introduce spectator fields $S\in\Gamma(\cO(2))$, 
$\Xi\in\Gamma(\cO(-2))$, such that we obtain the charges
\begin{align}
\label{eq:disaccharges}
\xymatrix@R=0mm@C=7mm{
\text{fields}	&X^{1,2}			&\Phi^1		&\Phi^2		&S			&\cX^1		&\cX^{2,3}		&\Xi		&P		&\Lambda			\\
\GU(1)_1		&1				&0			&-4			&2			&2			&0			&-2		&0		&0				\\
\GU(1)_2		&0				&1			&3			&3			&1			&1			&4		&-7		&-7
}
\end{align}
Again, however, the quadratic anomaly \eqref{eq:gaugequadranom} for $a=b=1$ is violated.

We conclude this section by presenting a (0,2) model which does admit a GLSM embedding.
Let us consider the model
\begin{align}
\label{eq:GLSMY}
\bY&=\tot\left( \cO(-2)\oplus\cO(-1)\oplus\cO\rightarrow \P^2\right)~,\nonumber\\
&\bq:					\qquad\quad	\ff13 \qquad\quad\ \ \ff13 \quad\quad \ \ff13
\end{align}
and
\begin{align}
\label{eq:GLSME}
\cE&=\pi^\ast\cO(2)\oplus\pi^\ast \cO(1)\oplus\pi^\ast \cO(-1)^{\oplus3}~.\nonumber\\
&\bQ:					\quad-1   \qquad\quad -1 \qquad\quad -\ff23
\end{align}
A choice of non singular superpotential is given by
\begin{align}
J_1&=S_{[4]}\phi_1^3~,	&J_2&=T_{[2]}\phi_2^3~,	&J_3&=S_{[5]}\phi_1^2 + S_{[3]}\phi_2^2+S_{[1]}\phi_3^2~,	\nonumber\\
J_4&=T_{[5]}\phi_1^2 + T_{[3]}\phi_2^2+T_{[1]}\phi_3^2~,	&J_5&=V_{[1]}\phi_3^2~.
\end{align}
where $T_{[d]},S_{[d]},V_{[d]}\in H^0(\P^2,\cO(d))$ do not have any common factors. Now, following the procedure above, this model is realized as a phase in the $\GU(1)^2$ GLSM
with the following charges
\begin{align}
\label{eq:02GLSMcharges}
\xymatrix@R=0mm@C=5mm{
\text{fields}	&X^{1,2,3}		&\Phi^1		&\Phi^2		&\Phi^3		&P		&\cX^1		&\cX^2	&\cX^{3,4,5}		&\Lambda		&\text{F.I.}\\
\GU(1)_1		&1				&-2			&-1			&0			&0		&2			&1		&-1				&0			&r_1\\	
\GU(1)_2		&0				&1			&1			&1			&-3		&0			&0		&1				&-3			&r_2	
}
\end{align}
equipped with one multiplet $\Sigma$ which couples to
\begin{align}
\label{eq:Emaps02GLSM}
E_1&=x~,		&E_2&=x^2~,	&E_{3,4,5}&=\phi_1x+\phi_2~.
\end{align}
where we schematically indicated $x\equiv x_{1,2,3}$.
The geometric phase arises in the cone $r_1,r_2>0$, where the irrelevant ideal is $(x_1x_2x_3)(\phi_1\phi_2\phi_3)$,
and it follows that the map \eqref{eq:Emaps02GLSM} fails to be injective at $\phi_1=\phi_2=x=x^2=0$.
Thus, the linear model is generically singular in this phase.

The secondary fan of this model is divided into five phases and all except the one that gives rise to the hybrid model above exhibit some sort of 
singularity (singular CY phases or bad hybrids), confirming the expectation that a generic phase in a (0,2) GLSM is singular or does not have a useful presentation.
Nevertheless, our hybrid theory is well-defined, and it allows to study the properties of the model which 
would be unattainable by other known techniques or constructions. 

\section{Outlook}

In this work, we have constructed a class of models with (0,2) supersymmetry suitable for compactifications of the heterotic string, and described techniques to compute
the massless spectrum in the limit where the K\"ahler class of the base is taken deep inside its K\"ahler cone. We have
described some properties of these models, and most importantly we have provided evidence that supports the fact that 
there exist flows to the expected non-trivial fixed points in the
IR. However, much work lies ahead of us! In this final section, we
wish 
to outline what hybrid models can teach us about the structure of the moduli space of (0,2) SCFTs. 

Let us start with the tamest class of (0,2) hybrids, namely models which admit a (2,2) locus. Here there are two kinds of deformations one might consider.
First, we can have $\cE=T_{\bY}$ and a (0,2) superpotential $J\in\Gamma (\cE^\ast)$ which is not integrable to a (2,2) superpotential $W$.
Second, there are bundle deformations, that is, $\cE$ is a deformation of $T_{\bY}$. While the former set of deformations
is quite well understood,
the latter corresponds, in GLSM terminology, to $E$-deformations, which 
are far more mysterious \cite{McOrist:2007kp,Kreuzer:2010ph}. In particular, it would be extremely important to determine whether, among these, non-linear deformations
are exactly marginal, redundant, or lifted by worldsheet instantons.

Next, one could consider, when possible, decreasing the rank of the map $E$ defining the bundle $\cE$, thereby increasing its rank. It is natural to conjecture that
this procedure corresponds to deforming to a Higgs branch of the
theory \cite{McOrist:2011bn}.

Finally, in the realm of (0,2) theories without a (2,2) locus, the most pressing question is how worldsheet instantons 
modify the classical picture as we move away from the hybrid limit. In theories with a (2,2) locus, such corrections, although there are indications
that they do occur \cite{Aspinwall:2011us,Aspinwall:2014ava},
are particularly elusive \cite{Aspinwall:2010ve}. 
We expect this behavior to be significantly different in more general (0,2) theories \cite{Bertolini:2014dna}.
As we have seen in this work, the space of instantons in hybrid models
is relatively simple since these are maps to the base $B$, and one can
construct nontrivial models with a simple base, as a laboratory for
exploring instanton corrections to the spacetime superpotential.

It would be also interesting to investigate further the connection between hybrids and GLSMs. 
It seems possible that a deeper understanding of the combinatorics we started in this work might lead
to a new class of models along the lines of \cite{Distler:1993mk}.

Finally, the example we studied in section \ref{ss:disaccident} might prompt a revival of the quest of classifying (0,2) LG theories. 
The results of \cite{Bertolini:2014ela} show that if our interest is to classify IR fixed points of (0,2) LG theories, classifying non-singular superpotentials
does not suffice due to the presence of accidents. However, we have seen that even accidental LG theories can be
employed to construct accident-free hybrid models. If we conjecture that any (0,2) LG can be consistently hybridized, 
it would be also meaningful to have a classification of non-singular (0,2) LG superpotentials.

We plan to address some of these questions in future work.

\appendix

\section{Equivalence of left-moving bundles}
\label{app:proofbundles}

In this appendix we are going to prove the isomorphism between the two definitions \eqref{eq:Ebundle} and \eqref{eq:EasExt} of the bundle $\cE$ defined in the main text.
There are obvious maps between the same position entries in the SES \eqref{eq:Ebundle} and the pullback of \eqref{eq:EBbundle}, which we can represent as follows
\begin{align}
\label{eq:snakeforE}
\xymatrix@R=9mm@C=13mm{
&&0 \ar[d]& 0 \ar[d]\\
&&\oplus_{\Gammat=P+1}^{R+m}\pi^\ast(\cF_{\Gammat}) \ar[d]	&\cK \ar[d]\\
0 \ar[r] &(\cO_{\bY})^{\oplus m}\ar[r]^-{E} \ar[d]^-{a} 
& {\begin{matrix} \oplus_{\Gammah=1}^P \pi^\ast(\cF_{\Gammah})\\\oplus\\\oplus_{\Gammat=P+1}^{R+m}\pi^\ast(\cF_{\Gammat}) \end{matrix}}\ar[r]^-{F}  \ar[d]^-{b}& \cE \ar[r]  \ar[d]^-{c}& 0\\
0 \ar[r] & (\pi^\ast \cO_B)^{\oplus m} \ar[r]^-{E_B} & \oplus_{\Gammah=1}^P \pi^\ast(\cF_{\Gammah}) \ar[r]^-{F_B} \ar[d]& \pi^\ast (\cE_B) \ar[r]\ar[d] & 0\\
&&0& \cC\ar[d]\\
&&& 0
}
\end{align}
In particular, the map $a$ is an isomorphism, while $b$ is the projection onto the $\oplus_{\Gammah=1}^P \pi^\ast (\cF_{\Gammah})$ summand. 
Also, we have included in \eqref{eq:snakeforE} the
obvious facts that $\ker b=\oplus_{\Gammat=P+1}^{R+m}\pi^\ast(\cF_{\Gammat})$ and that $\coker\;b=0$.
The map $c$ is defined by the relation $c\circ F = F_B\circ b$,  and we have indicated $\ker c = \cK$ and $\coker\;c = \cC$. We can apply the snake lemma to \eqref{eq:snakeforE}
which determines
\begin{align}
\cK&=\ker b = \oplus_{\Gammat=P+1}^{R+m}\pi^\ast(\cF_{\Gammat})~,		&\cC&=\coker\;b =0~.
\end{align}
Thus, the last column reproduces precisely the SES \eqref{eq:EasExt}, concluding our proof.

\subsection*{One example: two different presentations}

Let us consider $\bY=\tot\left(\cO(-n)\rightarrow \P^1\right)$ and $\oplus_{\Gammah=1}^2\cF_{\Gammah}=\cO(p)\oplus\cO(q),\ \cF_{\Gammat=3}=\cO(-n)$. In our notation 
$u=v^{-1}$ are the local coordinates on the two patches $\mathfrak{U}_{1,2}$ of $B=\P^1$, while $\phi_{u,v}=\phi|_{u,v}$,
are the restrictions to the above patches of a section of the one-dimensional fiber, which satisfy $\phi_u = v^n \phi_v$.
The SES \eqref{eq:Ebundle}  is defined by the map 
\begin{align}
E|_{\mathfrak{U}_1}&=\begin{pmatrix} u^p & 1 & c \phi_u \end{pmatrix}^\top~,		&E|_{\mathfrak{U}_2}&=\begin{pmatrix} 1 & v^q & c \phi_v \end{pmatrix}^\top~,
\end{align}
for some positive integers $p,q$ and $c\in\C$.
Note that for $p=q=1$ and $c=-n$ we just recover the tangent sheaf $T_{\bY}$. Requiring the sequence to be exact determines the map 
$F$ in \eqref{eq:Ebundle} to be of the form
\begin{align}
\label{eq:Fmap}
F|_{\mathfrak{U}_1}&=\begin{pmatrix}
1&-u^p&0\\
0&-c\phi_u&1
\end{pmatrix}~,
&F|_{\mathfrak{U}_2}&=\begin{pmatrix}
v^q&-1&0\\
-c\phi_v&0&1
\end{pmatrix}~.
\end{align}
A section $\chi$ of $\cE$ is thus determined by 
\begin{align}\label{eq:esec}
\chi|_{\mathfrak{U}_1}&=\begin{pmatrix}
r_u-t_u u^p \\ s_u-c\phi_u t_u
\end{pmatrix}~,
&\chi|_{\mathfrak{U}_2}&=\begin{pmatrix}
t_v-r_v v^q \\ s_v-c\phi_v r_v
\end{pmatrix}~,
\end{align}
where $r_{u,v}, t_{u,v}$ and $s_{u,v}$ are restrictions to the two patches of sections of $\cO(p)$, $\cO(q)$ and $\cO(-n)$ respectively. 
In each patch \eqref{eq:Fmap} satisfies $F\cdot E=0$, and it is easy to show that at $c=0$ the map $F|_{c=0} : \cO(p)\oplus\cO(q)\oplus\cO(-n) \longrightarrow \cO(p+q)\oplus\cO(-n)$
is surjective. Turning on $c$ does not affect surjectivity and it parametrizes a family of rank-two bundles described by 
\begin{align}
\xymatrix@R=0mm@C=8mm{
0\ar[r] & \pi^\ast \cO(-n)   \ar[r] &\cE   \ar[r] & \pi^\ast \cO(p+q) \ar[r] &0~.
}
\end{align}
On the overlap $\mathfrak{U}_1\cap\mathfrak{U}_2$ \eqref{eq:esec}  determines the transition functions 
\begin{align}
G_{uv}&=\begin{pmatrix}
-v^{-q-p}	&-cv^{n-q}\phi_v \\
0		&v^n
\end{pmatrix}~,
&G_{vu}&=\begin{pmatrix}
-u^{-q-p}	&-cu^{n-p}\phi_u \\
0		&u^n
\end{pmatrix}
\end{align}
for the bundle $\cE$.
This is indeed what we expect. For fixed $p,q$, at $c=0$ we have the trivial extension $\cE|_{c=0}=\cO(p+q)\oplus\cO(-n)$ while for any non-zero value of $c$
we obtain a family of irreducible rank-two bundles parametrized by $c$. If we set the parameters to their (2,2) values we recover 
the tangent sheaf $T_{\bY}$.

\section{Massless spectrum computations}
\label{app:examplescomp}

In this appendix we collect the details of the massless spectrum computations for the examples in section \ref{s:examples}.

\subsection{A $\so(10)$ example}

\subsubsection*{(0,2) sheaf cohomology for $\cE'$}

We now apply the techniques developed in section \ref{s:spectrum} to compute the relevant graded cohomology groups. 
It suffices for our purpose to restrict our attention to the non trivial part of the bundle, which is defined by \eqref{eq:transforEpr}.
This data yields $\Delta'_7=\left\{(2,0),(0,2)\right\}$ and $\delta_7=\left\{ (0,0),\pm(2,-2)\right\}$. Note that this is an example
where $\Delta' \neq \Delta$, as the element $(1,1)\in \Delta_7$ but not in $\Delta'_7$ 
as we have set to zero the coefficient of the corresponding monomial $\phi_1\phi_2$ in $E^7$.  
Thus
\begin{align}
\Pi=\left\{(2r_1,2r_2)\in\Z^{\oplus2}| r_1=-r_2\right\}~,
\end{align}
and the coarse grading is
determined by an integer $r_1+r_2=L$ as well as whether $r_1$ (or equivalently $r_2$) vanishes$\mod 2$, that is
\begin{align}
\bR^{\text{E}}_L&=\{(r_1,r_2)|r_1+r_2=L, r_1\in 2\Z\}~,		&\bR^{\text{O}}_L&=\{(r_1,r_2)|r_1+r_2=L,r_1\in 2\Z+1\}~,
\end{align}
which yields
\begin{align}
\bR^{\text{E},\text{O}}_L+\Delta'_7&=\bR^{\text{E},\text{O}}_{L+2}~.
\end{align}
Thus, for a fixed grading $\bR^{\text{E},\text{O}}_L$ we have
\begin{align}
\xymatrix@R=0mm@C=8mm{
0\ar[r] & \pi^\ast \cO(1)_{\left(\bR^{\text{E},\text{O}}_{L+2}\right)_+}  \ar[r] &(\cE')_{\bR^{\text{E},\text{O}}_L}   \ar[r] &(\pi^\ast \cO(2))_{\left(\bR^{\text{E},\text{O}}_L\right)_+} \ar[r] &0~.
}
\end{align}
The grading bundle is again $\mathbb{L}_{(r_1,r_2)}=\cO(1)^{r_1}\otimes\cO(1)^{r_2}=\cO(r_1+r_2)$, and from the induced LESes (we have two of these, one for $\bR^{\text{E}}_L$ and
one for $\bR^{\text{O}}_L$)
\begin{align}
\xy {\ar(0.05,-13)*{};(0.05,-14)*{}}; 
\xymatrix@R=8mm@C=-1mm{ 
0 \ar[r] &H^0_{\left(\bR^{\text{E},\text{O}}_{L+2}\right)_+}(B,\cO(L+3)) \ar[r] & H_{\bR^{\text{E},\text{O}}_{L}}^0 (\bY,\cE') \ar[r] &H_{\left(\bR^{\text{E},\text{O}}_L\right)_+}^0(B,\cO(L+2)) 
\ar@{-} `d[l]`[llld]  \\
H^1_{\left(\bR^{\text{E},\text{O}}_{L+2}\right)_+}(B,\cO(L+3)) \ar[r] & H^1_{\bR^{\text{E},\text{O}}_L} (\bY,\cE') \ar[r] & H^1_{\left(\bR^{\text{E},\text{O}}_L\right)_+}(B,\cO(L+2)) \ar[r] & 0 ~,
}
\endxy
\end{align}
we have for $L\geq0$ that
\begin{align}
\label{eq:cohmso10ex1}
H^0_{\bR_L^{\text{E}}}(\bY,\cE')&=\C^{L+4+(2L+7)\left\lceil\frac{L+1}2\right\rceil}~,		
&H^0_{\bR_L^{\text{O}}}(\bY,\cE')&=\C^{L+4+(2L+7)\left\lfloor\frac{L+1}2\right\rfloor}~,
\end{align}
while if $L=-2,-1$ we obtain
\begin{align}
\label{eq:cohmso10ex2}
H^0_{\bR_L^{\text{E}}}(\bY,\cE')&=\C^{(L+4)\left\lceil\frac{L+3}2\right\rceil}~,		
&H^0_{\bR_L^{\text{O}}}(\bY,\cE')&=\C^{(L+4)\left\lfloor\frac{L+3}2\right\rfloor}~.
\end{align}
Finally, $H^1_{\bR_L^{\text{E,O}}}(\bY,\cE')=0$ $\forall L$ and the cohomology is trivial everywhere if $L<2$.

\subsubsection*{Massless spectrum}

Let us now turn to the analysis of the massless spectrum, applying the methods we developed in section \ref{s:spectrum}.
As can be seen from the 
quantum numbers for each twisted sector listed in table \ref{t:quantnumbsso10}, the vacuum state energy of the (R,R) sectors is always zero, 
thus in this case it is enough to reduce to zero modes. Since we are primarily interested in highlighting 
novel features of (0,2) models, we restrict our exposition to the three (NS,R) sectors $k=1,3,5$. 
Here, $l_{|k\ra}\in\Z$ determines according to \eqref{eq:bundlevaceven} and $\eqref{eq:bundlevacodd}$ the line bundle for the vacuum as $|k\ra\in\Gamma(\cO(l_{|k\ra}))$.

\begin{table}[t!]
\begin{center}
\begin{tabular}{| c | c | c | c | c | c | c | c | c |}
\hline
$k$ 		&$l_{|k\ra}$		&$E_{|k\ra}$		&$q_{|k\ra}$		&$\qb_{|k\ra}$		&$\nu_{1,2,3}$		&$\nu_4$			&$\nut_{3,4,5}$			&$\nut_6$
\\\hline
0		&0				&0				&$-2$			&$-\ff32$			&0				&0				&0					&0\\\hline
1		&0				&$-1$			&$0$				&$-\ff32$			&$\ff1{10}$		&$\frac2{10}$		&$-\ff4{10}$			&$-\ff3{10}$\\\hline
2		&3				&0				&$0$				&$-\ff32$			&$\ff2{10}$		&$\frac4{10}$		&$-\ff{8}{10}$			&$-\ff6{10}$\\\hline
3		&$-1$			&$-\ff35$			&$-\ff25$			&$-\ff9{10}$		&$\ff3{10}$		&$\frac6{10}$		&$-\ff{2}{10}$			&$-\ff9{10}$\\\hline
4		&3				&$0$				&$-1$			&$-\ff12$			&$\ff4{10}$		&$\frac{8}{10}$		&$-\ff{6}{10}$			&$-\ff2{10}$\\\hline
5		&$-1$			&$-\ff12$			&$-1$			&$-\ff12$			&$\ff5{10}$		&0				&0					&$-\ff5{10}$\\\hline
\end{tabular}\vspace{2mm}
\begin{tabular}{| c | c | c | c | c | c | c | c | c |}
\hline
field 		&$\phi^{1,2,3}$		&$\phi^4$			&$\rho_{1,2,3}$		&$\rho_4$			&$\chi^{3,4,5}$		&$\chi^6$			&$\chib_{3,4,5}$	&$\chib_6$	\\\hline
$\bq$	&$\ff15$			&$\ff25$			&$-\ff15$			&$-\ff25$			&$-\ff45$			&$-\ff35$			&$\ff45$			&$\ff35$		\\\hline
$\bqb$	&$\ff15$			&$\ff25$			&$-\ff15$			&$-\ff25$			&$\ff15$			&$\ff25$			&$-\ff15$			&$-\ff25$		\\\hline
\end{tabular}
\end{center}
\caption{Quantum numbers for the $\SO(10)$ hybrid.}
\label{t:quantnumbsso10}
\end{table}

\subsubsection*{$k=1$ sector}

In this sector the vacuum state transforms trivially and the geometry is governed by the full $\cE\rightarrow\bY$. 
In particular, the vacuum $|1\ra$ has weight $h=-1$ and charges $(\bq_{|1\ra},\bqb_{|1\ra})=(0,-3/2)$, thus it is associated to the $\rep{28}_0$ of $\so(8)$.

At $h=-\half$ and $\bq=-1$ we find one chiral $\rep{8^v}_{-1}$, while at $\bq=1$ 
we have 
 \begin{equation}
\begin{matrix}\vspace{5mm}\\E_1^{p,q}:\end{matrix}
\begin{xy}
\xymatrix@C=10mm@R=5mm{
   H^1(\bY,\cE) & H^1(\bY,\cO_{\bY})  \\
   H^0(\bY,\cE) & H^0(\bY,\cO_{\bY})  
}
\save="x"!LD+<00mm,0pt>;"x"!RD+<20pt,0pt>**\dir{-}?>*\dir{>}\restore
\save="x"!LD+<50mm,-3mm>;"x"!LU+<50mm,2mm>**\dir{-}?>*\dir{>}\restore
\save!RD+<-40mm,-4mm>*{-\ff32}\restore
\save!RD+<-12mm,-4mm>*{-\half}\restore
\save!RD+<5mm,-4mm>*{p}\restore
\save!CL+<53mm,10mm>*{u}\restore
\end{xy}
\end{equation}
It is clear from \eqref{eq:cohmso10ex1} and \eqref{eq:cohmso10ex2} that $\cE'$ cannot contribute to the first row. 
In fact, the only contribution is of the form $(S_{[-2]}\phi_a+S_{[-3]}\phi_3)\chib_5\etab|1\ra$ at $(\bq,\bqb)=(1,-\half)$, which accounts for
4 states. The contribution to $H^0(\bY,\cE)$ at the relevant grade is 9 dimensional from the pullback factor of the bundle, 
while $\cE'$ contributes for $L=0,-1$, which yields
$21$ states, and twice for $L=-2$, since this section must be tensored with either $(\phi_3)^2$ or $\phi_4$, accounting for additional 4 states. 
Thus, the first stage of the spectral sequence reads
 \begin{equation}
\begin{matrix}\vspace{5mm}\\E_1^{p,q}:\end{matrix}
\begin{xy}
\xymatrix@C=10mm@R=5mm{
 \C^4 & 0 \\
\C^{34} \ar[r]^-{\bQb_J} & \C^{126}
}
\save="x"!LD+<00mm,0pt>;"x"!RD+<15pt,0pt>**\dir{-}?>*\dir{>}\restore
\save="x"!LD+<28mm,-3mm>;"x"!LU+<28mm,2mm>**\dir{-}?>*\dir{>}\restore
\save!RD+<-26mm,-4mm>*{-\ff32}\restore
\save!RD+<-5mm,-4mm>*{-\half}\restore
\save!RD+<5mm,-4mm>*{p}\restore
\save!CL+<31mm,8mm>*{u}\restore
\end{xy}
\end{equation}
It is straightforward to show that $\dim \ker \bQb_J=0$  and we count 92+4=96 chiral $\rep{8^s}_ 1$. 

Now onto the internal energy zero states ($h=0$). 
At $\bq=-2$ we have $H^1(\bY,\wedge^2\cE^\ast)=\C^2$
which yield two chiral $\rep{1}_{-2}$. At $\bq=2$ instead
 \begin{equation}
\begin{matrix}\vspace{5mm}\\E_1^{p,q}:\end{matrix}
\begin{xy}
\xymatrix@C=10mm@R=5mm{
H^1(\bY,\wedge^2\cE)_{22}\ar[r]^-{d_1}	& H^1(\bY,\cE)_{14}		&H^1(\bY,\cO_{\bY})_0 \\
H^0(\bY,\wedge^2\cE)_{205}\ar[r]^-{d_1}& H^0(\bY,\cE)_{1198}\ar[r]^-{d_1}	&H^0(\bY,\cO_{\bY}) _{1001}
}
\save="x"!LD+<0mm,0pt>;"x"!RD+<10pt,0pt>**\dir{-}?>*\dir{>}\restore
\save="x"!LD+<63mm,-3mm>;"x"!LU+<63mm,2mm>**\dir{-}?>*\dir{>}\restore
\save!RD+<-86mm,-4mm>*{-\ff32}\restore
\save!RD+<-54mm,-4mm>*{-\half}\restore
\save!RD+<-15mm,-4mm>*{\half}\restore
\save!RD+<2mm,-4mm>*{p}\restore
\save!CL+<66mm,10mm>*{u}\restore
\end{xy}
\end{equation}
where the subscripts indicate the dimension of the corresponding group.  
It might be worthwhile checking explicitly the map in the first row. 
Omitting the $\etab$ dependence, the complex at $u=1$ is
\begin{align}
\label{eq:firstrowd1map}
\xymatrix@C=15mm@R=0mm{
{\begin{matrix}
S_{[-3]}\phi_a\phi_3\chib_4\chib_5|1\ra_4\oplus
S_{[-2]}\phi_a\phi_3\chib_5\chib_3|1\ra_2
\end{matrix}} \ar[r]^-{d_1^1} & S_{[-2]} \phi_a G_{[5]}\chib_5 |1\ra_6
\\\oplus\\
{\begin{matrix}
S_{[-4]}G_{[2]}\chib_4\chib_5|1\ra_6\oplus
S_{[-3]}G_{[2]}\chib_5\chib_3|1\ra_4
\end{matrix}} \ar[r]^-{d_1^2} & 
S_{[-3]}  G_{[7]}\chib_5 |1\ra_8 
\\\oplus\\
{\begin{matrix}
S_{[-2]}\phi_a\phi_b\chib_4\chib_5|1\ra_3\\\oplus\\
S_{[-2]}G_{[3]}\chib_5\chib_6|1\ra_2\oplus
S_{[-2]}\phi_3\chib_2\chib_5|1\ra_1
\end{matrix}} \ar[r]^-{d_1^3} & 0
}
\end{align}
where $S_{[-d]}\in H^1(\P^1,\cO(-d))$ and $G_{[d]}$ is a homogeneous polynomial of degree $d$ in $\phi_3$ and $\phi_4$ with degree $1$ and $2$ respectively. 
Moreover, the only contribution from sections involving $\cE'$ happens at $L=-2$, which only have a $\chib_6$ component.
We verify explicitly that $\dim \ker d_1^1=\dim \ker d_1^2=2$, and $\dim \ker d_1^3=6$ . 
Although technically slightly more involved, the cohomology of the bottom row can be similarly computed and the result is
\begin{equation}
\begin{matrix}\vspace{5mm}\\E_2^{p,q}:\end{matrix}
\begin{xy}
\xymatrix@C=10mm@R=10mm{
&\C^{10} \ar[rrd]^-{d_2} & \C^2	&0 \\
&0	& \C^{54}	& \C^{62}
}
\save="x"!LD+<00mm,0pt>;"x"!RD+<25pt,0pt>**\dir{-}?>*\dir{>}\restore
\save="x"!LD+<43mm,-3mm>;"x"!LU+<43mm,2mm>**\dir{-}?>*\dir{>}\restore
\save!RD+<-41mm,-4mm>*{-\ff32}\restore
\save!RD+<-24mm,-4mm>*{-\half}\restore
\save!RD+<-5mm,-4mm>*{\half}\restore
\save!RD+<5mm,-4mm>*{p}\restore
\save!CL+<45mm,10mm>*{u}\restore
\end{xy}
\end{equation}
We need to compute a higher order map. Given a section $S^{AB}$, antisymmetric in its indices, representing an element in the kernel of \eqref{eq:firstrowd1map}, we have
\begin{align}
\label{eq:d2map}
d_2 S^{AB} \chib_A \chib_B = d_1 (T^A \chib_A -T^B \chib_B) = T^A J_A - T^B J_B~,
\end{align}
where the coefficients satisfy
\begin{align}
\pb T^A &= S^{AB} J_B~,		&\pb T^B &= S^{AB} J_A~.
\end{align}
As an example, we have for the states $-\frac1{(1+u\ub)^2}\phi_a\phi_b\chib_4\chib_5|1\ra$ 
\begin{align}
T^4&=\frac{1-\ub}{1+u\ub}\phi_a\phi_b\phi_3^4~,		&T^5&=\frac{u^2-2\ub}{1+u\ub}\phi_a\phi_b\phi_3^4~,
\end{align} 
which yield
\begin{align}
d_2 \left( S_{[-2]}\phi_a\phi_b\chib_4\chib_5|1\ra \right) = (2-u^2)\phi_a\phi_b \phi_3^8 |1\ra~.
\end{align}
Each of these three states is not in the image of $d_1$ since $2-u^2$ is not divisible by neither $1+u$ nor $2+u^3$.
Similarly, it can be verified explicitly that $\dim \ker d_2=0$, so that   
\begin{equation}
\begin{matrix}\vspace{5mm}\\E_3^{p,q}:\end{matrix}
\begin{xy}
\xymatrix@C=10mm@R=10mm{
&0 & \C^2	&0 \\
&0	& \C^{54}	& \C^{52}
}
\save="x"!LD+<00mm,0pt>;"x"!RD+<25pt,0pt>**\dir{-}?>*\dir{>}\restore
\save="x"!LD+<43mm,-3mm>;"x"!LU+<43mm,2mm>**\dir{-}?>*\dir{>}\restore
\save!RD+<-39mm,-4mm>*{-\ff32}\restore
\save!RD+<-24mm,-4mm>*{-\half}\restore
\save!RD+<-5mm,-4mm>*{\half}\restore
\save!RD+<5mm,-4mm>*{p}\restore
\save!CL+<45mm,10mm>*{u}\restore
\end{xy}
\end{equation}
We count 54 chiral $\rep{1}_2$, and the same number of anti-chirals multiplets, in agreement with CPT invariance. 

Finally, we focus on the $\bq=0$ states. 
Here we have
 \begin{equation}
\begin{matrix}\vspace{5mm}\\E_1^{p,q}:\end{matrix}
\begin{xy}
\xymatrix@C=15mm@R=2mm{
{\begin{matrix}
H^1(\bY,\cE\otimes\cE^\ast)_{26}\\\oplus\\
H^1(\bY,T_{\bY})_0\\\oplus\\
H^1(B,T^\ast_B)_1
\end{matrix}}\ar[r]^-{\bQb_J}	& H^1(\bY,\cE^\ast)_3 \\
{\begin{matrix}
H^0(\bY,\cE\otimes\cE^\ast)_{78}\\\oplus\\
H^0(\bY,T_{\bY})_{27}
\end{matrix}}\ar[r]^-{\bQb_J}	& H^0(\bY,\cE^\ast)_{473} 
}
\save="x"!LD+<00mm,0pt>;"x"!RD+<25pt,0pt>**\dir{-}?>*\dir{>}\restore
\save="x"!LD+<70mm,-3mm>;"x"!LU+<70mm,2mm>**\dir{-}?>*\dir{>}\restore
\save!RD+<-56mm,-4mm>*{-\ff32}\restore
\save!RD+<-15mm,-4mm>*{-\half}\restore
\save!RD+<5mm,-4mm>*{p}\restore
\save!CL+<73mm,20mm>*{u}\restore
\end{xy}
\end{equation}
In agreement with the general features of our models, for a generic superpotential the map in the bottom row has a one dimensional 
kernel, which corresponds to the gaugino for the global $\GU(1)_L$ symmetry. In the first row, one can show that $\dim \coker\ \bQb_J=1$
for \eqref{eq:so10supchioc}, and that the map becomes surjective for a more generic superpotential.

\subsubsection*{$k=3$ sector}

The moding in table \ref{t:quantnumbs} shows that $\phi_4$ is a heavy field in this sector, and the geometry is governed by
\begin{align}
\bY_3 = \tot\left( \oplus_{i=1}^3X_i \xrightarrow{\pi_3} \P^1\right)~.
\end{align}
For the left-moving fermions, it follows from \eqref{eq:so10Echio} that we need to analyze both monomials $\phi_{1,2}^2$ in $E^7$, 
which we can consider at once as $\phi_{1,2}$ have the same weight. In this case
\begin{align}
\nut_6-2\nu_i =-\frac9{10}- \frac{6}{10}=-\frac32~,
\end{align}
does not satisfy \eqref{eq:condtwistkodd}, thus the bundle splits, and in particular
\begin{align}
\cE|_{k=3}=\cE|_B=\oplus_{\Gammat=3}^7\pi_3^\ast \cF_{\Gammat} \oplus \pi^\ast_3 \cE_B ~.
\end{align}
The cohomology groups
\begin{align}
&H^0(\bY_3,\pi^\ast_3(\cF_6\otimes L_{|3\ra}^\ast))=\C~,		&H^1\left(\bY_3,\pi^\ast_3(\cF^\ast_3\wedge\cE^\ast_B\otimes L_{|3\ra}^\ast)\right)&=\C~,\nonumber\\
&H^1\left((\bY_3,\pi^\ast_3(\cF_4\wedge\cF_5\wedge\cF_3\otimes L_{|3\ra}^\ast)\right)=\C^2
\end{align}
correspond to a chiral $\rep{8^v}_{-1}$, an anti-chiral $\rep{1}_{-2}$ and two chiral $\rep{1}_{2}$.
At $\bq=0$ we have instead the spectral sequence
 \begin{equation}
\begin{matrix}\vspace{5mm}\\E_1^{p,q}:\end{matrix}
\begin{xy}
\xymatrix@C=15mm@R=8mm{
H^1(\bY_3,\pi_3^\ast(\cF_5\otimes X_4 \otimes L_{|3\ra}^\ast))_1
 \ar[r]^-{\bQb_J}	& H^1(\bY_3,\pi_3^\ast(\cF_5\otimes\cF^\ast_6\otimes L_{|3\ra}^\ast))_{4} \\
H^0(\bY_3,\pi_3^\ast(\cF_{3,4}\otimes X_4 \otimes L_{|3\ra}^\ast))_3  \ar[r]^-{\bQb_J}	& 
{\begin{matrix}
H^0(\bY_3,\pi_3^\ast(\cF_{3,4}\otimes\cF^\ast_6\otimes L_{|3\ra}^\ast))_{7}\\\oplus\\
H^0(\bY_3,\pi_3^\ast(\cE_B\otimes\cF^\ast_6\otimes L_{|3\ra}^\ast))_{5}\\\oplus\\
H^0(\bY_3,\pi_3^\ast L_{|3\ra}^\ast)_{22}
\end{matrix}}
}
\save="x"!LD+<00mm,0pt>;"x"!RD+<25pt,0pt>**\dir{-}?>*\dir{>}\restore
\save="x"!LD+<121mm,-3mm>;"x"!LU+<121mm,2mm>**\dir{-}?>*\dir{>}\restore
\save!RD+<-90mm,-4mm>*{-\ff32}\restore
\save!RD+<-26mm,-4mm>*{-\half}\restore
\save!RD+<5mm,-4mm>*{p}\restore
\save!CL+<124mm,22mm>*{u}\restore
\end{xy}
\end{equation}
It is straightforward to verify that for a generic superpotential both maps have zero dimensional kernels, 
thus we count 31 chiral and 3 anti-chiral singlets.

\subsubsection*{$k=5$ sector}

Finally, we are left with the last sector to analyze. 
Here all the coordinates have $\nu_i\leq1/2$, therefore the geometry is characterized by the full $\bY$. 
For the left-moving bundle we have that again
\begin{align}
\nut_6-2\nu_1 = -\frac12-1=-\frac32~,
\end{align}
thus the bundle again splits $\cE|_{k=5}=\cE|_B$.
The only effect of $\bQb_J$ in this sector is $\bQb_J \chib_3=\phi_4^2$. 
Hence, we can simply ignore the zero-mode $\chib_3$ and we restrict our attention to $\phi_4^n$ for $n=0,1$. 
With this in mind, the cohomology groups
\begin{align}
H^0(\bY,\pi^\ast L_{|5\ra})&=H^1(\bY,\wedge^2\cE\otimes \pi^\ast L_{|5\ra})=\C^2~,	&&(h=-1/2,\bq=\pm1)~,\nonumber\\
H^0(\bY,\cE^\ast\otimes\pi^\ast L_{|5\ra})&=H^0(\bY,\wedge^3\cE\otimes\pi^\ast L_{|5\ra})=\C~, && (h=0,\bq=\pm2)~,
\end{align}
yield two chiral $\rep{8^v}_{-1}$ and a chiral $\rep{1}_{-2}$ together with their CPT conjugate states. Finally at $h=\bq=0$ we obtain
\begin{align}
H^\bullet(\bY,\cE\otimes\pi^\ast L_{|5\ra}) \oplus H^\bullet(\bY,\cE\otimes T^\ast_{\bY}\otimes \pi^\ast L_{|5\ra})
\oplus H^\bullet(\bY,\wedge^2\cE\otimes\cE^\ast\otimes\pi^\ast L_{|5\ra}) =\C^{16}~,
\end{align}
for $\bullet=0,1$, thus leading to 16 chiral and anti-chiral singlets.

\subsection{An $\GE_7$ model}

We begin by listing the quantum numbers for each twisted sector introduced by the orbifold by $\Z_{14}$ in table \ref{t:quantnumbs}. 
We then turn to the massless spectrum analysis in each 
sector $k=0,\dots,7$. We point out that since the bundle $\cE$ is a sum of line bundles which pull back from the base 
the fine grading $\br$ is suitable for the relevant sheaf cohomology computations.

\begin{table}[h!]
\begin{center}
\begin{tabular}{| c | c | c | c | c | c | c | c |}
\hline
$k$ 		&$l_{|k\ra}$		&$E_{|k\ra}$		&$q_{|k\ra}$		&$\qb_{|k\ra}$		&$\nu_1$		&$\nu_2$			&$\nut_I$	
\\\hline
0		&0				&0				&$-1$			&$-1$			&0			&0				&0\\\hline
1		&0				&$-1$			&$0$				&$-1$			&$\ff1{14}$	&$\frac3{14}$		&$-\ff6{14}$\\\hline
2		&0				&0				&$1$				&$-1$			&$\ff2{14}$	&$\frac6{14}$		&$-\ff{12}{14}$\\\hline
3		&2				&$-\ff67$			&$-\ff47$			&$-\ff47$			&$\ff3{14}$	&$\frac9{14}$		&$-\ff{4}{14}$\\\hline
4		&2				&$-\ff17$			&$\ff37$			&$-\ff47$			&$\ff4{14}$	&$\frac{12}{14}$	&$-\ff{10}{14}$\\\hline
5		&0				&$-\ff{11}{14}$		&$-\ff57$			&$-\ff57$			&$\ff5{14}$	&$\frac{1}{14}$		&$-\ff{2}{14}$\\\hline
6		&0				&$-\ff{1}{7}$		&$\ff27$			&$-\ff57$			&$\ff6{14}$	&$\frac{4}{14}$		&$-\ff{8}{14}$\\\hline
7		&2				&$-\ff{1}{2}$		&$-\ff97$			&$-\ff27$			&$\ff7{14}$	&$\frac{7}{14}$		&$0$\\\hline
\end{tabular}\ \ \ \
\begin{tabular}{| c | c | c | c | c | c | c |}
\hline
field 		&$\phi^1$			&$\phi^2$			&$\rho_1$			&$\rho_2$			&$\chi^I$		&$\chib_I$	\\\hline
$\bq$	&$\ff17$			&$\ff37$			&$-\ff17$			&$-\ff37$			&$-\ff67$		&$\ff67$		\\\hline
$\bqb$	&$\ff17$			&$\ff37$			&$-\ff17$			&$-\ff37$			&$\ff17$		&$-\ff17$		\\\hline
\end{tabular}
\end{center}
\caption{Quantum numbers for the $\GE_7$ model.}
\label{t:quantnumbs}
\end{table}

\subsubsection*{Even $k$ sectors}

In the untwisted $k=0$ sector always $E_{|0\ra}=0$, thus we just restrict to zero modes. 
The GSO projection restricts us to double complexes defined by the charges of the states
$F_{[7m]}|0\ra$ for $m=0,1,2$. 
For $m=0$ the only possibility is the vacuum\footnote{We will often indicate the charges $\bq(\cO)$ and $\bqb(\cO)$ of a state $\cO$ as subscripts $\cO_{\bq,\bqb}$.}
 $|0\ra_{-1,-1}$ while at $m=2$ we find its CPT conjugate state at $(\bq,\bqb)=(1,1)$, both being associated to a $\rep{32}\in\so(12)$.
At $m=1$ the only non trivial row of the spectral sequence after taking horizontal Dolbeault cohomology is at $\bq=0$ and $U=0$ where we have 
\begin{align}
\xymatrix@C=25mm@R=3mm{
H^0(\bY,\cE)_5
\ar[r]^-{\bQb_J} &
H^0(\bY,\cO_{\bY})_{15} ~,\\
\bqb=-1 & \bqb=0~
}
\end{align}
where the subscript indicates the dimensions of the groups corresponding to the relevant grades. 
It is easy to verify that $\dim\ker \bQb_J=0$ from the potential \eqref{eq:02suppot} as $u^6$ does not divide $u^8+1$.  
The contribution is then 10 $\half$-hypers in the $(\rep{32'},\rep1,\rep1$)$\in\so(12)\oplus\mathfrak{u}(1)_L\oplus\mathfrak{u}(1)_R$.

In the twisted $k=2$ sector the vacuum has again zero energy and there are no zero modes, hence the only contribution is from the vacuum $|2\ra_{1,-1}$, which together with the CPT conjugate 
state to be found in $k=12$, teams up with the states describe above to complete the representation $(\rep{32},\rep{2},\rep{2})$.

At $k=4$ we cannot simply restrict to zero modes since $E_{|4\ra}<0$, and since there are no
additional fiber zero modes we have simply $\bY_4=B$. Keeping 
into consideration that the vacuum transforms non-trivially we find that the only well-defined zero-energy states are of the form
\begin{align}
H^1(B, X_2^\ast\otimes  L_{|4\ra}^\ast)=\C^5~.
\end{align}
These are found at $(\bq,\bqb)=(0,0)$, that is, they yield $\half$-hypers in the $(\rep{32'},\rep{1},\rep1)$.

The last sector to analyze is $k=6$, and it is easy to see that it is not possible to engineer any zero-energy states with the twisted moding.

\subsubsection*{Odd $k$ sectors}

The untwisted $k=1$ sector is as usual the richest. The vacuum has $E_{|1\ra}=-1$ and charges $(\bq,\bqb)=(0,-1)$ 
and is part of a doublet under $\su(2)_R$ associated to the $\rep{66}$ of $\so(12)$, while
internal states at $E=-\half$ are paired up with the $\rep{12}$ of $\so(12)$ and complete the  10 $\half$-hyper from the $k=0$ sector.

We now turn to energy-zero internal states
at $\bq=0$, where the first stage of the spectral sequence reads
\begin{equation}
\begin{matrix}\vspace{10mm}\\E_1^{p,u}:\end{matrix}
\begin{xy}
\xymatrix@C=15mm@R=5mm{
{\begin{matrix}
H^1\left(\bY,B_{0,0,1} \right)_3
\\\oplus\\ 
H^1\left(\bY, B_{1,1,0} \right)_2
\\\oplus\\
H^1\left(B, T_B^\ast\right)_{1}
\end{matrix}}   \ar[r]^{\bQb_J}  &
H^1\left(\bY, B_{0,1,0} \right)_{1}   \\
{\begin{matrix}
H^0\left(\bY,B_{0,0,1}\right)_{5}
\\\oplus\\ 
H^0\left(\bY,B_{1,1,0} \right)_{11}
\end{matrix}}   \ar[r]^{\bQb_J}  
  &
{\begin{matrix}
H^0\left(\bY, B_{0,1,0}\right)_{40}
\end{matrix}}
}
\save="x"!LD+<-6mm,0pt>;"x"!RD+<40pt,0pt>**\dir{-}?>*\dir{>}\restore
\save="x"!LD+<72mm,-3mm>;"x"!LU+<72mm,2mm>**\dir{-}?>*\dir{>}\restore
\save!RD+<-58mm,-4mm>*{-1}\restore
\save!RD+<-15mm,-4mm>*{0}\restore
\save!RD+<8mm,-3mm>*{p}\restore
\save!CL+<75mm,25mm>*{u}\restore
\end{xy}
\end{equation}
In computing the dimensions of these cohomology groups
we used the fact that 
$T_{\bY}=\cO\oplus T_{\bY'}$ where $\bY'=\text{tot}\left(\cO(-4)\rightarrow \P^1\right)$,
and 
\begin{align}
H_{r\geq0}^0(\bY',T_{\bY'})&=\C^{8r+4}~,		&H_{r=-1}^1(\bY',T_{\bY'})&=\C^3~,
\end{align}
and zero otherwise, where $r$ is the fine grading on $\bY'$ which simply keeps track of the power of $\phi_2$.
In the bottom row we find that again for a generic superpotential $\dim \ker \bQb_J=1$, 
corresponding to the unbroken left-moving $\GU(1)_L$, while for a more
specific superpotential this number can be higher. 
For example, for the superpotential \eqref{eq:02suppot} we have $\dim \ker \bQb_J=2$. 
Moreover, the first row map is surjective, 
thus we count 30 neutral $\half$-hypers and one $\bqb=-1$ component of the $\su(2)_R$-doublet for a vector multiplet.

In the $k=3$ sector, $\nu_2>\half$ and the geometry is $\bY_3=\tot(X_1 \xrightarrow{\pi_3} \P^1)$. 
The contribution at $h=\bq=0$ is computed by
 \begin{equation}
\begin{matrix}\vspace{5mm}\\E_1^{p,u}:\end{matrix}
\begin{xy}
\xymatrix@C=10mm@R=5mm{
 H^1(\bY_3, \cE \otimes X_2^\ast\otimes L_{|3\ra})_{13}
\ar[r]^-{\bQb_J} & 
H^1(\bY_3, \pi_3^\ast L_{|3\ra})_1 \\
0 & 
H^0(\bY_3,\pi_3^\ast X_2 \otimes \pi_3^\ast L_{|3\ra})_3
}
\save="x"!LD+<1mm,0pt>;"x"!RD+<30pt,0pt>**\dir{-}?>*\dir{>}\restore
\save="x"!LD+<98mm,-3mm>;"x"!LU+<98mm,5mm>**\dir{-}?>*\dir{>}\restore
\save!RD+<-76mm,-4mm>*{-1}\restore
\save!RD+<-20mm,-4mm>*{0}\restore
\save!RD+<9mm,-3mm>*{p}\restore
\save!CL+<101mm,12mm>*{u}\restore
\end{xy}
\end{equation}
It is easy to see that the map in the first row is surjective, so we find $12+3=15$ gauge-neutral $\ff12$-hypers.

In the $k=5$ sector there are no heavy fields, 
and only the $u=0$ row produces a non-trivial contribution
\begin{align}
\xymatrix@R=-2mm@C=15mm{
H^0(\bY,\cE\otimes T_{\bY}\otimes \pi^\ast L_{|5\ra}^\ast)_5
\ar[r]^-{\bQb_J} &
H^0(\bY,T_{\bY}\otimes \pi^\ast L_{|5\ra}^\ast)_9\\
&
{\begin{matrix}
\oplus\\
H^0(\bY, \pi^\ast L_{|5\ra}^\ast)_5
\end{matrix}} 
}
\end{align}
For a generic superpotential this map is injective, and we are left with $9$ neutral $\ff12$-hypers.

Finally, for $k=7$ at $h=\bq=0$, $\bQb_J$ is trivial and the relevant states are
counted by
\begin{align}
H^0(\bY,\cE\otimes\pi^\ast L_{|7\ra}^\ast) = H^1(\bY,\wedge^2\cE\otimes T_{\bY}\otimes\pi^\ast L_{|7\ra}^\ast)=\C^{11}~,
\end{align}
which shows that this sector is its own CPT conjugate and it contributes 22 neutral $\ff12$-hypers.

\section{An accidental hybrid}
\label{app:acchybr}

In this appendix we show that accidents do arise in non-trivial hybrid theories. Let us consider the hybrid model
defined by the data
\begin{align}
\bY&=\text{tot}\left( \cO\oplus\cO(-2) \rightarrow \P^1 \right)~,		&\cE&=\pi^\ast\cO(2)\oplus\pi^\ast\cO(-2)\oplus\pi^\ast\cO^{\oplus2}~.\\\nonumber
&\bq:\qquad	\ff17	\qquad\quad \ff37						&&\bQ:	\quad -1	\qquad \ \ -\ff67	\qquad\  -\ff67
\end{align}
In particular, the fiber LG theory coincides with the one in the example in section \ref{ss:disaccident}, 
but the fibration structure is different. This data gives $c=7$, $\cb=6$ and $r=3$.
However, the following non-singular superpotential
\begin{align}
\label{eq:02hybaccsup}
J_1&=S_{[12]}\phi_1^7~,		&J_2&=S_{[14]}\phi_1^6~,		&J_3&=J_4=\phi_2^2~,
\end{align}
is equivalent, by a field redefinition between $\cX_3$ and $\cX_4$, to $J_3=\phi_2^2$ and $J_4=0$. That is, the theory is a sum of a free Weyl fermion and 
an interacting theory, where anomaly cancellation determines
\begin{align}
\bY&=\text{tot}\left( \cO\oplus\cO(-2) \rightarrow \P^1 \right)~,		&\cE&=\pi^\ast\cO(2)\oplus\pi^\ast\cO(-2)\oplus \pi^\ast\cO~.\\\nonumber
&\bq:\qquad	\ff3{21}	\quad\quad \ff7{21}					&&\bQ:	\quad -1	\qquad -\ff{14}{21}	\qquad -\ff{18}{21}
\end{align}
By tuning $S_{[12]}$ appropriately in \eqref{eq:02hybaccsup}, this theory is 
identified as a (0,2) deformation of the (2,2) theory with target space $\bY$ and superpotential
\begin{align}
W_{(2,2)}&= S_{[14]}\phi_1^7 +\phi_2^3~.
\end{align}
This theory identifies a IR fixed point with central charges $c=\cb=3(2+ \frac1{21})$. Thus, the orbit of field redefinitions corresponding to the superpotential
\eqref{eq:02hybaccsup} is accidental and it does not contribute to the conformal manifold of the theory we are interested in.

Finally, one might ask whether there exists at all a sensible theory defined by the UV data above. We can easily construct a more general superpotential
that will achieve this. Consider 
\begin{align}
\label{eq:02hybnonaccsup}
J_1&=S_{[12]}\phi_1^7~,		&J_2&=S_{[14]}\phi_1^6~,		&J_3&=\phi_2^2~,		&J_4&=S_{[6]}\phi_2\phi_1^3+S_{[12]}\phi_1^6~,
\end{align}
where all the sections $S_{[d]}$ do not have common roots. This choice of superpotential defines a non-singular model and 
a moment of thought shows that no field redefinitions are possible which are associated to enhanced symmetries.
Hence, \eqref{eq:02hybnonaccsup} defines a RG flow to a theory with the expected $c=7$, $\cb=6$.

\bibliographystyle{./utphys}
\bibliography{./bigrefJHEP}

\providecommand{\href}[2]{#2}\begingroup\raggedright\begin{thebibliography}{10}

\bibitem{McOrist:2010ae}
J.~McOrist, ``{The Revival of (0,2) Linear Sigma Models},''
  \href{http://dx.doi.org/10.1142/S0217751X11051366}{{\em Int. J. Mod. Phys.}
  {\bf A26} (2011)  1--41},
\href{http://arxiv.org/abs/1010.4667}{{\tt arXiv:1010.4667 [hep-th]}}.

\bibitem{Sharpe:2015vza}
E.~Sharpe, ``{A few recent developments in 2d (2,2) and (0,2) theories},'' {\em
  Proc. Symp. Pure Math.} {\bf 93} (2015)  67,
\href{http://arxiv.org/abs/1501.01628}{{\tt arXiv:1501.01628 [hep-th]}}.

\bibitem{McOrist:2008ji}
J.~McOrist and I.~V. Melnikov, ``{Summing the instantons in half-twisted linear
  sigma models},'' {\em JHEP} {\bf 02} (2009)  026,
\href{http://arxiv.org/abs/0810.0012}{{\tt arXiv:0810.0012 [hep-th]}}.

\bibitem{Melnikov:2010sa}
I.~V. Melnikov and M.~R. Plesser, ``{A (0,2) mirror map},''
  \href{http://dx.doi.org/10.1007/JHEP02(2011)001}{{\em JHEP} {\bf 1102} (2011)
   001},
\href{http://arxiv.org/abs/1003.1303}{{\tt arXiv:1003.1303 [hep-th]}}.

\bibitem{Aspinwall:2010ve}
P.~S. Aspinwall, I.~V. Melnikov, and M.~R. Plesser, ``{(0,2) Elephants},'' {\em
  JHEP} {\bf 1201} (2012)  060, \href{http://arxiv.org/abs/1008.2156}{{\tt
  arXiv:1008.2156 [hep-th]}}.

\bibitem{Melnikov:2012hk}
I.~Melnikov, S.~Sethi, and E.~Sharpe, ``{Recent Developments in (0,2) Mirror
  Symmetry},'' \href{http://dx.doi.org/10.3842/SIGMA.2012.068}{{\em SIGMA} {\bf
  8} (2012)  068},
\href{http://arxiv.org/abs/1209.1134}{{\tt arXiv:1209.1134 [hep-th]}}.

\bibitem{Aspinwall:2014ava}
P.~S. Aspinwall and B.~Gaines, ``{Rational Curves and (0,2)-Deformations},''
\href{http://arxiv.org/abs/1404.7802}{{\tt arXiv:1404.7802 [hep-th]}}.

\bibitem{Aspinwall:2011vp}
P.~S. Aspinwall, ``{A McKay-like correspondence for (0,2)-deformations},''
\href{http://arxiv.org/abs/1110.2524}{{\tt arXiv:1110.2524 [hep-th]}}.

\bibitem{Closset:2015ohf}
C.~Closset, W.~Gu, B.~Jia, and E.~Sharpe, ``{Localization of twisted $
  \mathcal{N}=\left(0,\;2\right) $ gauged linear sigma models in two
  dimensions},'' \href{http://dx.doi.org/10.1007/JHEP03(2016)070}{{\em JHEP}
  {\bf 03} (2016)  070},
\href{http://arxiv.org/abs/1512.08058}{{\tt arXiv:1512.08058 [hep-th]}}.

\bibitem{Bouchard:2006dn}
V.~Bouchard, M.~Cvetic, and R.~Donagi, ``{Tri-linear couplings in an heterotic
  minimal supersymmetric standard model},'' {\em Nucl. Phys.} {\bf B745} (2006)
   62--83,
\href{http://arxiv.org/abs/hep-th/0602096}{{\tt arXiv:hep-th/0602096}}.

\bibitem{Donagi:2006yf}
R.~Donagi, R.~Reinbacher, and S.-T. Yau, ``{Yukawa couplings on quintic
  threefolds},''
\href{http://arxiv.org/abs/hep-th/0605203}{{\tt hep-th/0605203}}.

\bibitem{Greene:1990du}
B.~R. Greene, ``{Superconformal compactifications in weighted projective
  space},''
\href{http://dx.doi.org/10.1007/BF02473356}{{\em Commun. Math. Phys.} {\bf 130}
  (1990)  335--355}.

\bibitem{Distler:1993mk}
J.~Distler and S.~Kachru, ``(0,2) {L}andau-{G}inzburg theory,'' {\em Nucl.
  Phys.} {\bf B413} (1994)  213--243,
\href{http://arxiv.org/abs/hep-th/9309110}{{\tt hep-th/9309110}}.

\bibitem{Melnikov:2009nh}
I.~V. Melnikov, ``{(0,2) Landau-Ginzburg models and residues},'' {\em JHEP}
  {\bf 09} (2009)  118,
\href{http://arxiv.org/abs/0902.3908}{{\tt arXiv:0902.3908 [hep-th]}}.

\bibitem{Kawai:1994np}
T.~Kawai and K.~Mohri, ``{Geometry of (0,2) Landau-Ginzburg orbifolds},''
  \href{http://dx.doi.org/10.1016/0550-3213(94)90178-3}{{\em Nucl. Phys.} {\bf
  B425} (1994)  191--216},
\href{http://arxiv.org/abs/hep-th/9402148}{{\tt arXiv:hep-th/9402148
  [hep-th]}}.

\bibitem{Melnikov:2007xi}
I.~V. Melnikov and S.~Sethi, ``{Half-twisted (0,2) Landau-Ginzburg models},''
  {\em JHEP} {\bf 03} (2008)  040,
\href{http://arxiv.org/abs/0712.1058}{{\tt arXiv:0712.1058 [hep-th]}}.

\bibitem{Gadde:2016khg}
A.~Gadde and P.~Putrov, ``{Exact solutions of (0,2) Landau-Ginzburg models},''
\href{http://arxiv.org/abs/1608.07753}{{\tt arXiv:1608.07753 [hep-th]}}.

\bibitem{Benini:2013cda}
F.~Benini and N.~Bobev, ``{Two-dimensional SCFTs from wrapped branes and
  c-extremization},'' \href{http://dx.doi.org/10.1007/JHEP06(2013)005}{{\em
  JHEP} {\bf 06} (2013)  005},
\href{http://arxiv.org/abs/1302.4451}{{\tt arXiv:1302.4451 [hep-th]}}.

\bibitem{Melnikov:2016dnx}
I.~V. Melnikov, ``{Relevant deformations and c-extremization},''
  \href{http://dx.doi.org/10.1007/JHEP09(2016)169}{{\em JHEP} {\bf 09} (2016)
  169},
\href{http://arxiv.org/abs/1603.08935}{{\tt arXiv:1603.08935 [hep-th]}}.

\bibitem{Bertolini:2014ela}
M.~Bertolini, I.~V. Melnikov, and M.~R. Plesser, ``{Accidents in (0,2)
  Landau-Ginzburg theories},''
  \href{http://dx.doi.org/10.1007/JHEP12(2014)157}{{\em JHEP} {\bf 12} (2014)
  157},
\href{http://arxiv.org/abs/1405.4266}{{\tt arXiv:1405.4266 [hep-th]}}.

\bibitem{Witten:1993yc}
E.~Witten, ``{Phases of N = 2 theories in two dimensions},'' {\em Nucl. Phys.}
  {\bf B403} (1993)  159--222,
\href{http://arxiv.org/abs/hep-th/9301042}{{\tt arXiv:hep-th/9301042}}.

\bibitem{Beasley:2003fx}
C.~Beasley and E.~Witten, ``{Residues and world-sheet instantons},'' {\em JHEP}
  {\bf 10} (2003)  065,
\href{http://arxiv.org/abs/hep-th/0304115}{{\tt arXiv:hep-th/0304115}}.

\bibitem{Bertolini:2014dna}
M.~Bertolini and M.~R. Plesser, ``{Worldsheet instantons and (0,2) linear
  models},'' \href{http://dx.doi.org/10.1007/JHEP08(2015)081}{{\em JHEP} {\bf
  08} (2015)  081},
\href{http://arxiv.org/abs/1410.4541}{{\tt arXiv:1410.4541 [hep-th]}}.

\bibitem{Bertolini:2013xga}
M.~Bertolini, I.~V. Melnikov, and M.~R. Plesser, ``{Hybrid conformal field
  theories},'' \href{http://dx.doi.org/10.1007/JHEP05(2014)043}{{\em JHEP} {\bf
  1405} (2014)  043},
\href{http://arxiv.org/abs/1307.7063}{{\tt arXiv:1307.7063}}.

\bibitem{Babalic:2016mbw}
M.~Babalic, D.~Doryn, C.~I. Lazaroiu, and M.~Tavakol, ``{Differential models
  for B-type open-closed topological Landau-Ginzburg theories},''
\href{http://arxiv.org/abs/1610.09103}{{\tt arXiv:1610.09103 [math.DG]}}.

\bibitem{Babalic:2016hjh}
M.~Babalic, D.~Doryn, C.~I. Lazaroiu, and M.~Tavakol, ``{On B-type open-closed
  Landau-Ginzburg theories defined on Calabi-Yau Stein manifolds},''
\href{http://arxiv.org/abs/1610.09813}{{\tt arXiv:1610.09813 [math.DG]}}.

\bibitem{2016arXiv160808962B}
A.~{Basalaev}, A.~{Takahashi}, and E.~{Werner}, ``{Orbifold Jacobian algebras
  for invertible polynomials},'' {\em ArXiv e-prints} (2016)  ,
  \href{http://arxiv.org/abs/1608.08962}{{\tt arXiv:1608.08962 [math.AG]}}.

\bibitem{2015arXiv150204872S}
K.~{Saito}, ``{Coherence of direct images of the De Rham complex},'' {\em ArXiv
  e-prints} (2015)  , \href{http://arxiv.org/abs/1502.04872}{{\tt
  arXiv:1502.04872 [math.AG]}}.

\bibitem{Chiang:1997kt}
T.-M. Chiang, J.~Distler, and B.~R. Greene, ``{Some features of (0,2) moduli
  space},'' {\em Nucl. Phys.} {\bf B496} (1997)  590--616,
\href{http://arxiv.org/abs/hep-th/9702030}{{\tt arXiv:hep-th/9702030}}.

\bibitem{Guffin:2008pi}
J.~Guffin and E.~Sharpe, ``{A-twisted heterotic Landau-Ginzburg models},''
\href{http://arxiv.org/abs/0801.3955}{{\tt arXiv:0801.3955 [hep-th]}}.

\bibitem{Distler:1995mi}
J.~Distler, ``Notes on (0,2) superconformal field theories,''
\href{http://arxiv.org/abs/hep-th/9502012}{{\tt hep-th/9502012}}.

\bibitem{Bertolini:2018now}
M.~Bertolini and M.~Romo, ``{Aspects of (2,2) and (0,2) hybrid models},''
\href{http://arxiv.org/abs/1801.04100}{{\tt arXiv:1801.04100 [hep-th]}}.

\bibitem{Kachru:1993pg}
S.~Kachru and E.~Witten, ``{Computing the complete massless spectrum of a
  Landau- Ginzburg orbifold},'' {\em Nucl. Phys.} {\bf B407} (1993)  637--666,
\href{http://arxiv.org/abs/hep-th/9307038}{{\tt arXiv:hep-th/9307038}}.

\bibitem{Melnikov:2011ez}
I.~V. Melnikov and E.~Sharpe, ``{On marginal deformations of (0,2) non-linear
  sigma models},'' \href{http://dx.doi.org/10.1016/j.physletb.2011.10.055}{{\em
  Phys.Lett.} {\bf B705} (2011)  529--534},
\href{http://arxiv.org/abs/1110.1886}{{\tt arXiv:1110.1886 [hep-th]}}.

\bibitem{Katz:2004nn}
S.~H. Katz and E.~Sharpe, ``Notes on certain (0,2) correlation functions,''
  {\em Commun. Math. Phys.} {\bf 262} (2006)  611--644,
\href{http://arxiv.org/abs/hep-th/0406226}{{\tt hep-th/0406226}}.

\bibitem{Silverstein:1994ih}
E.~Silverstein and E.~Witten, ``Global {U(1)} {R} symmetry and conformal
  invariance of (0,2) models,'' {\em Phys. Lett.} {\bf B328} (1994)  307--311,
\href{http://arxiv.org/abs/hep-th/9403054}{{\tt hep-th/9403054}}.

\bibitem{Gepner:1987vz}
D.~Gepner, ``{Exactly solvable string compactifications on manifolds of SU(N)
  holonomy},'' {\em Phys.Lett.} {\bf B199} (1987)  380--388.

\bibitem{Adams:2005tc}
A.~Adams, J.~Distler, and M.~Ernebjerg, ``Topological heterotic rings,'' {\em
  Adv. Theor. Math. Phys.} {\bf 10} (2006)  657--682,
\href{http://arxiv.org/abs/hep-th/0506263}{{\tt hep-th/0506263}}.

\bibitem{Vafa:1989xc}
C.~Vafa, ``{String Vacua and Orbifoldized L-G Models},''
{\em Mod. Phys. Lett.} {\bf A4} (1989)  1169.

\bibitem{Aspinwall:2009qy}
P.~S. Aspinwall and M.~Ronen~Plesser, ``{Decompactifications and Massless
  D-Branes in Hybrid Models},''
\href{http://arxiv.org/abs/0909.0252}{{\tt arXiv:0909.0252 [hep-th]}}.

\bibitem{Hori:2013gga}
K.~Hori and J.~Knapp, ``{Linear sigma models with strongly coupled phases - one
  parameter models},'' \href{http://dx.doi.org/10.1007/JHEP11(2013)070}{{\em
  JHEP} {\bf 11} (2013)  070},
\href{http://arxiv.org/abs/1308.6265}{{\tt arXiv:1308.6265 [hep-th]}}.

\bibitem{Bott:1982df}
R.~Bott and L.~W. Tu, {\em Differential forms in algebraic topology}, vol.~82
  of {\em Graduate Texts in Mathematics}.
\newblock Springer-Verlag, New York, 1982.

\bibitem{Sharpe:1998zu}
E.~R. Sharpe, ``{Kahler cone substructure},''
  \href{http://dx.doi.org/10.4310/ATMP.1998.v2.n6.a7}{{\em Adv. Theor. Math.
  Phys.} {\bf 2} (1999)  1441--1462},
\href{http://arxiv.org/abs/hep-th/9810064}{{\tt arXiv:hep-th/9810064
  [hep-th]}}.

\bibitem{Atiyah:1957xx}
M.~F. Atiyah, ``Complex analytic connections in fibre bundles,'' {\em Trans.
  AMS} {\bf 85} (1957) no.~1, 181--207.

\bibitem{Anderson:2011ty}
L.~B. Anderson, J.~Gray, A.~Lukas, and B.~Ovrut, ``{The Atiyah class and
  complex structure stabilization in heterotic Calabi-Yau compactifications},''
  \href{http://arxiv.org/abs/1107.5076}{{\tt arXiv:1107.5076 [hep-th]}}.

\bibitem{Benini:2012cz}
F.~Benini and N.~Bobev, ``{Exact two-dimensional superconformal R-symmetry and
  c-extremization},''
  \href{http://dx.doi.org/10.1103/PhysRevLett.110.061601}{{\em Phys.Rev.Lett.}
  {\bf 110} (2013)  061601},
\href{http://arxiv.org/abs/1211.4030}{{\tt arXiv:1211.4030 [hep-th]}}.

\bibitem{Distler:1996tj}
J.~Distler, B.~R. Greene, and D.~R. Morrison, ``{Resolving singularities in
  (0,2) models},'' {\em Nucl. Phys.} {\bf B481} (1996)  289--312,
\href{http://arxiv.org/abs/hep-th/9605222}{{\tt arXiv:hep-th/9605222}}.

\bibitem{Aspinwall:1994cf}
P.~S. Aspinwall and B.~R. Greene, ``{On the geometric interpretation of N=2
  superconformal theories},''
  \href{http://dx.doi.org/10.1016/0550-3213(94)00571-U}{{\em Nucl.Phys.} {\bf
  B437} (1995)  205--230},
\href{http://arxiv.org/abs/hep-th/9409110}{{\tt arXiv:hep-th/9409110
  [hep-th]}}.

\bibitem{McOrist:2007kp}
J.~McOrist and I.~V. Melnikov, ``{Half-twisted correlators from the Coulomb
  branch},'' {\em JHEP} {\bf 04} (2008)  071,
\href{http://arxiv.org/abs/0712.3272}{{\tt arXiv:0712.3272 [hep-th]}}.

\bibitem{Kreuzer:2010ph}
M.~Kreuzer, J.~McOrist, I.~V. Melnikov, and M.~Plesser, ``{(0,2) deformations
  of linear sigma models},''
  \href{http://dx.doi.org/10.1007/JHEP07(2011)044}{{\em JHEP} {\bf 1107} (2011)
   044}, \href{http://arxiv.org/abs/1001.2104}{{\tt arXiv:1001.2104 [hep-th]}}.

\bibitem{McOrist:2011bn}
J.~McOrist and I.~V. Melnikov, ``{Old issues and linear sigma models},'' {\em
  Adv.Theor.Math.Phys.} {\bf 16} (2012)  251--288,
\href{http://arxiv.org/abs/1103.1322}{{\tt arXiv:1103.1322 [hep-th]}}.

\bibitem{Aspinwall:2011us}
P.~S. Aspinwall and M.~R. Plesser, ``{Elusive worldsheet instantons in
  heterotic string compactifications},''
  \href{http://arxiv.org/abs/1106.2998}{{\tt arXiv:1106.2998 [hep-th]}}.

\end{thebibliography}\endgroup

\end{document}